\documentclass[11pt]{article}
\pdfoutput=1
\usepackage{amsfonts}
\usepackage{amssymb}
\usepackage{mathrsfs}
\usepackage{graphicx}
\usepackage{amsmath,amsthm,amssymb,amscd}
\usepackage[all]{xy}
\usepackage{hyperref}
\usepackage{multirow}
\usepackage{color}
\usepackage{float}
\usepackage{mathtools,slashed}
\usepackage{bbold}
\usepackage{bbm}
\usepackage{slashed}

\usepackage{subcaption}

\usepackage{array}

\usepackage{amsxtra,graphics,epsfig,bm,tikz,xfrac,lscape}
\usetikzlibrary{decorations.pathmorphing}
\usetikzlibrary{decorations.markings}
\usetikzlibrary{arrows, decorations.markings, calc, fadings, decorations.pathreplacing, patterns, decorations.pathmorphing, positioning}

\usetikzlibrary{positioning,shapes}
\usetikzlibrary{chains}
\usetikzlibrary{arrows,fit,decorations.pathreplacing}
\tikzstyle{every picture}+=[remember picture]
\tikzstyle{na} = [baseline=-.5ex]

\newcommand{\cv}{{\cal V}}

\newcommand{\nn}{\nonumber}

\def\eqa{\begin{eqnarray}}
\def\eqae{\end{eqnarray}}
\def\eq{\begin{equation}}
\def\eqe{\end{equation}}
\def\be{\begin{equation}}
\def\ee{\end{equation}}
\def\bea{\begin{eqnarray}}
\def\eea{\end{eqnarray}}
\def\ba{\begin{array}}
\def\ea{\end{array}}
\def\bd{\begin{displaymath}}
\def\ed{\end{displaymath}}

\def\Tr{{\rm Tr}}

\def\>{\rangle}
\def\<{\langle}
\def\a{\alpha}

\def\tQ{\tilde{Q}}

\newcommand{\fft}[2]{\frac{#1}{#2}}
\newcommand{\ft}[2]{{\textstyle\frac{#1}{#2}}}

\numberwithin{equation}{section}

\newcommand{\tx}{\tilde{x}}

\newcommand{\tabincell}[2]{\begin{tabular}{@{}#1@{}}#2\end{tabular}}

\definecolor{darkblue}{rgb}{0,0,0.5}
\definecolor{darkred}{rgb}{0.5,0,0}
\definecolor{darkgreen}{rgb}{0,0.5,0}
\definecolor{orange}{rgb}{0.9,0.58,0}

\begin{document}

\begin{titlepage}
\hfill LCTP-20-18

\vskip 1 cm

\begin{center}
{\Large \bf Universal Logarithmic Behavior in Microstate Counting   }\\

\vskip .7cm

{\Large \bf  and the Dual One-loop Entropy  of AdS$_4$ Black Holes}\\

\end{center}

\vskip .7 cm

\vskip 1 cm
\begin{center}
{ \large Leopoldo A. Pando Zayas${}^{a,b}$ and Yu Xin${}^c$}
\end{center}

\vskip .4cm \centerline{\it ${}^a$ Leinweber Center for Theoretical
Physics,   Randall Laboratory of Physics}
\centerline{ \it The University of
Michigan, Ann Arbor, MI 48109-1120}

\bigskip 
\centerline{\it ${}^b$ The Abdus Salam International Centre for Theoretical Physics}
\centerline{\it Strada Costiera 11,  34151 Trieste, Italy}

\bigskip\bigskip

%\bigskip\bigskip

\centerline{\it ${}^c$ Department of Applied Mathematics and Theoretical Physics,}
\centerline{\it University of Cambridge, Cambridge, CB3 0WA, UK}

\vskip 1 cm

\vskip 1.5 cm
\begin{abstract}

We numerically study the topologically twisted index of several  three-dimensional supersymmetric field theories on a genus $g$ Riemann surface times a circle, $\Sigma_g\times S^1$.  We show that for a large class of theories with leading term of the order $N^{3/2}$, where $N$ is generically the rank of the gauge group, there is a universal logarithmic correction of the form  $\frac{g-1}{2} \log N$.  We explain how this logarithmic subleading correction can be obtained as a one-loop effect on the dual supergravity theory for magnetically charged, asymptotically AdS$_4\times M^7$ black holes for a large class of Sasaki-Einstein manifolds,  $M^7$. The matching of the logarithmic correction relies on a generic cohomological property of $M^7$ and it is independent of the black hole charges.  We argue that our supergravity results apply also to rotating, electrically charged asymptotically AdS$_4\times M^7$ black holes. We present explicitly  the quiver gauge theories and the gravity side corresponding to $M^7=N^{0,1,0}, V^{5,2}$ and $Q^{1,1,1}$.  
\end{abstract}

\vskip  1.5 cm

{\tt lpandoz@umich.edu, yx328@cam.ac.uk}
\end{titlepage}

\tableofcontents
%%%%%%%%%%%%%%%%%%%%%%%%%%%%%%%%%%%%%%%%%%%%%%%%%%%%%%%%%%%%%%%%%%%%%%%%%%%%%%%%%%%%%%%%%%%%%%
\section{Introduction}
%%%%%%%%%%%%%%%%%%%%%

One remarkable recent result in the context of the AdS/CFT correspondence is the microscopic understanding of the Bekenstein-Hawking entropy of a class of asymptotically AdS$_4$ black holes.  Benini, Hristov and Zaffaroni demonstrated in \cite{Benini:2015eyy} that the topologically twisted index of ABJM theory reproduces the entropy of the dual magnetically charged, asymptotically AdS$_4\times S^7$ black holes.

Similar microscopic foundations via the topologically twisted index were provided for the corresponding macroscopic black hole entropy in different situations, including: dyonic black holes \cite{Benini:2016rke}, black holes with hyperbolic horizons \cite{Cabo-Bizet:2017jsl} and asymptotically AdS$_4$ black holes in massive IIA supergravity \cite{Benini:2017oxt,Hosseini:2017fjo}. Some interesting progress has also been reported in the higher dimensional context \cite{Hosseini:2018uzp,Crichigno:2018adf,Suh:2018tul,Hosseini:2018usu,Suh:2018szn} and for AdS$_4$ black holes embeddable in certain universal sectors of M2 \cite{Azzurli:2017kxo}  and M5 backgrounds \cite{Gang:2018hjd,Gang:2019uay} (see  \cite{Hosseini:2018qsx,Zaffaroni:2019dhb}  for reviews and a complete list of references for  those developments). An analogous microscopic description, rooted in the superconformal index, has recently  been presented for rotating, electrically charged AdS$_4$ black holes  \cite{Choi:2019zpz,Nian:2019pxj}, including in a  universal sector arising from wrapped M5 branes \cite{Bobev:2019zmz,Benini:2019dyp}.

The robust agreement at the leading order inspired attempts to understand the topologically twisted index beyond the large $N$ limit with focus on the logarithmic corrections to the entropy on both sides of the correspondence  \cite{Liu:2017vll,Jeon:2017aif}. The initial conclusion, however,  was that more work was required and that Sen's quantum entropy formalism in its current formulation needed to be amended to also account for hair degrees of freedom away from the near-horizon region. Ultimately, precise agreement was found in \cite{Liu:2017vbl} whose computation  focused on the asymptotically AdS$_4$ region of the black hole solution.  Further successful matches  of the logarithmic contributions were provided in the case of universally embedded black holes \cite{Gang:2019uay,Benini:2019dyp}. This subleading agreement motivates us to embark on a systematic exploration of a large class of models with the aim of demonstrating that the logarithmic correction is, indeed, quite universal. This is precisely one of the main results of this paper:  {\it an expression for the logarithmic corrections of the topologically twisted index for a large class of  field theories on $\Sigma_g\times S^1$, which we find to be $\frac{g-1}{2}\log N$}.

Let us describe two important previous results  that make the journey to a universal logarithmic correction plausible. The first  precedent pointing to the fact that the coefficient of $\log N$ in the topologically twisted index could  be universal comes from a subleading analysis of the free energy of a large class of 3d field theories. The exact partition function for a large class of Chern-Simons matter theories on $S^3$ can be computed using field theory localization and certain matrix model techniques, the answer can be  succinctly written in terms of an Airy function  \cite{Marino:2011eh}.  In some cases the supergravity dual is known to be a background of M-theory on AdS$_4\times M^7$, where $M^7$ is a   Sasaki-Einstein seven-dimensional manifold.  The universality of the logarithmic term in the free energy on $S^3$ established in \cite{Marino:2011eh} was beautifully elucidated from the dual supergravity point of view  in \cite{Bhattacharyya:2012ye}  and shown to depend on some mild cohomological properties of the seven-dimensional manifold $M^7$.   Some of our arguments in this manuscript mimic that analysis closely. The other important source of inspiration for us is a group of works  that established a leading order in $N$ relationship between the free energy on $S^3$ and the topologically twisted index on $S^2\times S^1$ presented and developed in  \cite{Hosseini:2016tor,Hosseini:2016ume}. There is a formal background that arguably provides a rigorous basis for relations among the free energy on $S^3$ and the topologically twisted index in $\Sigma_g \times S^1$   \cite{Closset:2017zgf,Toldo:2017qsh,Gang:2018hjd,Closset:2018ghr} but we were particularly inspired by the two developments mentioned above.  In this manuscript we effectively ask the questions of whether there is a relationship between the free energy on $S^3$ and the topologically twisted index in $\Sigma_g\times S^1$ beyond the leading order in $N$ and, in particular, whether we can establish  the universality of the logarithmic in $N$ correction.   We are not able to answer the broader question of the relationship between the free energy on $S^3$  and the topologically twisted index beyond the large $N$ limit but we  present strong numerical evidence  in favor of a universal logarithmic correction in the topologically twisted index very similar to the universality of the logarithmic term for the free energy.

The rest of the manuscript is organized as follows. We start in section \ref{Sec:TTI} by briefly reviewing the topologically twisted  index in general and its form for the ABJM theory. Sections \ref{Sec:N010}, \ref{Sec:V52} and \ref{Sec:Q111} are devoted to extensive numerical evaluations of the topologically twisted index for the 3d Chern-Simons matter theories dual to M-theory on AdS$_4\times M^7$ for $M^7=N^{0,1,0}, V^{5,2}, Q^{1,1,1}$, respectively.  We discuss the one-loop gravity computation dual to the universal result in section \ref{Sec:Sugra}. We conclude in section \ref{Sec:Conclusions} where we also point to a number of interesting, in our opinion, open problems. 

%%%%%%%%%%%%%%%%%%%%%%%%%%%%%%%%%%%%%%%%%%%%%%%%%%%%%%%%%%%%%%%%%%%%%%%%%%%%%%%%%%%%%%%%%%%%%%
\section{The topologically twisted index for  generic ${\cal N}=2$ theories}\label{Sec:TTI}
%%%%%%%%%%%%%%%%%%%%%%%%%%%%%%%%%%%%%%%%%%%%%%%%%

%%%%%%%%%%%%%%%%%%%%%%%%%%%%%%%%%%%%%%%%%%%%%%%%%%%%%%%
%\subsection{The Index for theories with leading  $N^{3/2}$ behavior}
%%%%%%%%%%%%%%%%%%%%%%%%%%%%%%%%%%%%%%%%%%%%%

In this section we will briefly review the construction and structure of the topologically twisted index for 3d ${\cal N}=2$ supersymmetric theories.   The topologically twisted index for three dimensional ${\cal N}=2$ field theories was defined in \cite{Benini:2015noa} (see other related works \cite{Honda:2015yha,Closset:2015rna,Hosseini:2016tor,Hosseini:2016ume,Closset:2016arn}) by evaluating the supersymmetric partition function on $S^1\times S^2$ with a topological twist on $S^2$.   One considers a 3d theory, usually containing Yang-Mills,  ${\cal L}_{YM}$,  and Chern-Simons, ${\cal L}_{CS}$, interactions on  $S^2\times S^1$ with metric and background field given as
\be
ds^2 =R^2 (d\theta^2+\sin^2\theta d\phi^2)+\beta^2dt^2, \qquad A^R=\frac{1}{2}\cos\theta d\phi.
\ee
 There is typically a set of flavor symmetries characterized by Cartan-valued magnetic fluxes:
\be
J^f=\frac{1}{2\pi}\int_{S^2}F^f = \mathfrak{n}.
\ee
With these magnetic fluxes one associates flavor fugacities $y=\exp\left[ i \left(A_t^f+i\beta \sigma^f\right)\right]$, where the constant potential $A_t^f$ is a flat connection for the flavor symmetry and $\sigma^f$ is a real mass for the three-dimensional field theory. Similarly the fugacities for the dynamical fields are $x=\exp\left[ i \left(A_t+i\beta \sigma\right)\right]$, where $A_t$ runs over the maximal torus of the gauge group and $\sigma$ over the corresponding Cartan subalgebra.  

The topologically twisted index generically takes the from 
\be
Z(\mathfrak{n}, y)=\frac{1}{|W|}\sum\limits_{\mathfrak{m}\in \Gamma_{\mathfrak h}}\oint_{\cal C}Z_{int}(x,y;\mathfrak{m},\mathfrak{n}).
\ee
There is an algorithmic way of constructing $Z_{int}$ depending on the field content of the theory. Let us define the building blocks that go into $Z_{int}$. For a chiral multiplet
\be
Z_{1-loop}^{chiral}=\prod\limits_{\rho \in {\mathfrak R}}\left(\frac{x^{\rho/2}y^{\rho_f/2}}{1-x^\rho y^{\rho_f}}\right)^{\rho(\mathfrak{m})+\rho_f(\mathfrak{n})-q+1},
\ee
where ${\mathfrak R}$ is the representation of the gauge group $G$, $\rho$ denote the corresponding weights, $q$ is the R-charge of the field and $\rho_f$ is the weight of the multiplet under the flavor symmetry group. For the gauge multiplet one has 
\be
Z_{1-loop}^{gauge}=\prod\limits_{\alpha \in G}(1-x^\alpha)\left(id u\right)^r,
\ee
where $r$ is the rank of the gauge group and $\alpha$ donate the roots of $G$.  We also use  $u=A_t+i\beta \sigma $ which lives on the complexified Cartan subalgebra, essentially, $x=e^{iu}$.

The only classical contribution to $Z_{int}$ comes from the Chern-Simons term and takes the form 
\be
Z_{class}^{CS}=x^{k\mathfrak{m}}, 
\ee
where $k$ is the Chern-Simons level and $\mathfrak{m}$ is the magnetic flux taking values in the co-root latice $\Gamma_{\mathfrak h}$ of the gauge group. There is also the contribution of a $U(1)$ topological symmetry with holonomy $\xi=e^{iz}$ and flux $\mathfrak{t}$:
\be
Z^{top}_{class}=x^{\mathfrak{t}}\xi^{\mathfrak{m}}.
\ee

With these ingredients one has that the index takes the general form 
\be\label{Eq:ZTTI}
Z(\mathfrak{n}, y) =\frac{1}{|W|}\sum\limits_{\mathfrak{m}\in \Gamma_{\mathfrak h}}\oint_{\cal C}\prod\limits_{{\rm Cartan}}\left(\frac{dx}{2\pi i x}x^{k\mathfrak{m}}\right)
\prod\limits_{\alpha\in G}(1-x^\alpha)\prod\limits_{I}\prod\limits_{\rho_I\in \mathfrak{R}_I}\left(\frac{x^{\rho_I/2}y_I^{1/2}}{1-x^{\rho_I} y_I}\right)^{\rho_I(\frak{m})-\frak{n}_I+1},
\ee
where $\alpha$ are the roots of $G$, $\rho_I$ are the weights of the representation $\mathfrak{R}_I$ and $\frak{m}$ are gauge magnetic fluxes living in the co-root lattice $\Gamma_{\mathfrak h}$.

The index depends on a choice of fugacities $y_I$ for the flavor group and a choice of integer magnetic charges $\mathfrak{n}_I$ for the R-symmetry of the theory. Both $y_I$ and $\mathfrak{n}_I$ are parameterized by the global symmetries of the theory. Each monomial term $W$ in the superpotential imposes a constraint:

\be
\prod\limits_{I \in W}y_I=1,\qquad \prod\limits_{I \in W}\mathfrak{n}_I=2,
\ee
where the product and sum are restricted to the fields entering in $W$. These constraints are called the marginality conditions of the superpotential.

After summing over the magnetic fluxes, $\mathfrak{m}$,  in Eq.~(\ref{Eq:ZTTI}), one obtains an expression for the index whose poles are located at positions determined by  the following Bethe-Ansatz like expression
\be
\exp\left(i\,{\rm sign}(k_a)B^{(a)}_i\right)=1.
\ee

%%%%%%%%%%%%%%%%%%%%%%%%%%%%%%
For the class of theories we are interested in this manuscript  it is convenient to consider some representations explicitly. The ingredients in the topologically twisted index that we will require are: 

\begin{itemize}

\item The Vandermonde determinant contributes to the logarithm of the index as
\bea
\log \prod\limits_{i\neq j}\left(1-\frac{x_i^{(a)}}{x_j^{(a)}}\right)&=& \log \prod\limits_{i<j}\left(1-\frac{x_j^{(a)}}{x_i^{(a)}}\right)^2\left(-\frac{x_i^{(a)}}{x_j^{(a)}}\right) \nonumber \\
&=& i\sum\limits_{i<j}^N(u_i^{(a)}-u_j^{(a)}+\pi)-2\sum\limits_{i<j}^N{\rm Li}_1\left(e^{i(u_j^{(a)}-u_i^{(a)})}\right).
\eea

\item The topological symmetry contributes as
\be
i\sum\limits_{i=1}^N u_i^{(a)}\mathfrak{t}_a,
\ee
where $\mathfrak{t}_a$ is the flux of the $U(1)_a$ topological symmetry. 

\item A bi-fundamental chiral multiplet transforming in $(\bar{\bf N}, {\bf N})$ of $U(N)_a\times U(N)_b$ with magnetic flux $\mathfrak{n}_{(b,a)}$ and chemical potential $\Delta_{(b,a)}$  contributes as

\bea
&&\prod\limits_{i=1}^N\left(\frac{x_i^{(a)}}{x_i^{(b)}}\right)^{\frac{1}{2}(\mathfrak{n}_{(b,a)}-1)}
\left(1-y_{(b,a)}\frac{x_i^{(b)}}{x_i^{(a)}} \right)^{\mathfrak{n}_{(b,a)}-1} \nonumber \\
&\times &\prod\limits_{i<j}^N(-1)^{\mathfrak{n}_{(b,a)}-1}\left(\frac{x_i^{(a)}x_i^{(b)}}{x_j^{(a)}x_j^{(b)}}  \right)^{\frac{1}{2}(\mathfrak{n}_{(b,a)}-1)} \left(1-y_{(b,a)}\frac{x_j^{(b)}}{x_i^{(a)}} \right)^{\mathfrak{n}_{(b,a)}-1} 
\left(1-y_{(b,a)}^{-1}\frac{x_j^{(a)}}{x_i^{(b)}} \right)^{\mathfrak{n}_{(b,a)}-1}. \nonumber \\
\eea

\item Fundamental and anti-fundamental fields contribute as
\bea
&&\log \prod\limits_{i=1}^N\prod_{\substack{{\rm anti-fundamental}\\ a}}\left(x_i^{(a)}\right)^{\frac{1}{2}(\tilde{\mathfrak{n}}_{a}-1)}
\big[1-\tilde{y}_a\left(x_i^{(a)}\right)^{-1}\big]^{\tilde{\mathfrak{n}}_{a}-1} \nonumber \\
&\times& \prod_{\substack{{\rm fundamental}\\ a}}\left(x_i^{(a)}\right)^{\frac{1}{2}(\mathfrak{n}_{a}-1)}
\big[1-y_a^{-1}\left(x_i^{(a)}\right)^{-1}\big]^{\mathfrak{n}_{a}-1}.
\eea

\item As anticipated above, the contour integration is best expressed in terms of the position of the poles, given by the BA equation in terms of $B_i$.  In taking the residues from one set of variables $x_i$ to another $B_i$, we incur a Jacobian  denoted by $\mathbb{B}$:

\be
\mathbb{B}=\frac{\partial(e^{iB^{(a)}_j},e^{iB_j^{(b)}})}{\partial (\log x_l^{(a)}, \log x_l^{(b)})}=
\left(
\begin{array}{cc}
x_l^{(a)}\frac{e^{iB_j^{(a)}}}{\partial x_l^{(a)}}& x_l^{(b)}\frac{e^{iB_j^{(a)}}}{\partial x_l^{(b)}}\\
x_l^{(a)}\frac{e^{iB_j^{(b)}}}{\partial x_l^{(a)}}& x_l^{(b)}\frac{e^{iB_j^{(b)}}}{\partial x_l^{(b)}}
\end{array}
\right)_{2N\times 2N},
\ee
where 
\bea
\exp\left(i\,{\rm sign}(k_a)B^{(a)}_i\right)&=&(\xi^{(a)})^{{\rm sign}(k_a)}(x_i^{(a)})^{k_a}\prod\limits_{\substack{{\rm bi-fundamentals} \\ (b,a)\, {\rm and}\, (a,b)}} \,\,\prod\limits_{j=1}^N\frac{\sqrt{\frac{x_i^{(a)}}{x_j^{(b)}}y_{(a,b)}}}{1-\frac{x_i^{(a)}}{x_j^{(b)}}y_{(a,b)}}
\frac{1-\frac{x_j^{(b)}}{x_i^{(a)}}y_{(b,a)}}{\sqrt{\frac{x_j^{(b)}}{x_i^{(a)}}y_{(b,a)}}} \nonumber \\
&&\times \prod\limits_{\substack{{\rm fundamentals} \\a}} \frac{\sqrt{x_i^{(a)}y_{a}}}{1-x_i^{(a)}y_{a}}
\prod\limits_{\substack{{\rm anti-fundamentals} \\a}} \frac{1-\frac{1}{x_i^{(a)}}\tilde{y}_{a}}{\sqrt{\frac{1}{x_i^{(a)}}\tilde{y}_{a}}}.
\eea
\end{itemize}

%%%%%%%%%%%%%%%%%%%%%%%%%%%%%%%%%%%%%%%%%%
\subsubsection*{The Topologically Twisted Index}
%%%%%%%%%%%%%%%%%%%%%%%%%%%%%%%%%%%%%%%%%%%%
Explicitly, the general expression of the index is
\bea
Z&=&\frac{1}{(N!)^{|G|}}\sum\limits_{\left\{\mathfrak{m};\,\mathfrak{m}\in \mathbb{Z}^N\right\}}\int_{\cal C} \prod\limits_{a=1}^{|G|} \left[ \prod\limits_{i=1}^N
\frac{dx_i^{(a)}}{2\pi i x_i^{(a)}} \left(x_i^{(a)}\right)^{k_a\mathfrak{m}_i^{(a)}+\mathfrak{t}^{(a)}}\left(\xi^{(a)}\right)^{{\rm sign}(k_a)\mathfrak{m}_i^{(a)}} \times \prod\limits_{i\neq j}^N\left(1-\frac{x_i^{(a)}}{x_j^{(a)}}\right)\right] \nonumber\\
&&\times \prod\limits_{i,j=1}^N\left[\prod\limits_{\substack{{\rm bi-fundamentals} \\ (b,a)\, {\rm and}\,(a,b)}} \left(\frac{\sqrt{\frac{x_i^{(a)}}{x_j^{(b)}}y_{(a,b)}}}{1-\frac{x_i^{(a)}}{x_j^{(b)}}y_{(a,b)}}\right)^{\mathfrak{m}_i^{(a)}-\mathfrak{m}_j^{(b)}-\mathfrak{n}_{(a,b)}+1}
\left(\frac{\sqrt{\frac{x_j^{(b)}}{x_i^{(a)}}y_{(b,a)}}}{1-\frac{x_j^{(b)}}{x_i^{(a)}}y_{(b,a)}}\right)^{\mathfrak{m}_j^{(b)}-\mathfrak{m}_i^{(a)}-\mathfrak{n}_{(b,a)}+1} \right. \nonumber\\
&&\times 
\left.\prod\limits_{\substack{{\rm adjoints} \\ (a,a)}} \left(\frac{\sqrt{\frac{x_i^{(a)}}{x_j^{(a)}}y_{(a,a)}}}{1-\frac{x_i^{(a)}}{x_j^{(a)}}y_{(a,a)}}\right)^{\mathfrak{m}_i^{(a)}-\frac{1}{2}\mathfrak{n}_{(a,a)}+\frac{1}{2}}\left(\frac{\sqrt{\frac{x_j^{(a)}}{x_i^{(a)}}y_{(a,a)}}}{1-\frac{x_j^{(a)}}{x_i^{(a)}}y_{(a,a)}}\right)^{-\mathfrak{m}_i^{(a)}-\frac{1}{2}\mathfrak{n}_{(a,a)}+\frac{1}{2}}\right] \nonumber\\
&&\times \prod\limits_{i=1}^N\left[\prod\limits_{\substack{{\rm fundamentals} \\a}}\Bigg(\frac{\sqrt{x_i^{(a)}y_{a}}}{1-x_i^{(a)}y_{a}}\Bigg)^{\mathfrak{m}_i^{(a)}-\mathfrak{n}_a+1} \prod\limits_{\substack{{\rm anti-fundamentals} \\a}}\Bigg(\frac{\sqrt{\frac{1}{x_i^{(a)}}\tilde{y}_{a}}}{1-\frac{1}{x_i^{(a)}}\tilde{y}_{a}}\Bigg)^{-\mathfrak{m}_i^{(a)}-\tilde{\mathfrak{n}}_a+1}\right].
\nonumber\\
\eea

The sum over magnetic fluxes is effectively a geometric sum introducing a large cut-off $M$ and the index takes the form

\bea
Z&=&\frac{1}{(N!)^{|G|}} \int_{\cal C} \prod\limits_{a=1}^{|G|} \left[ \prod\limits_{i=1}^N
\frac{dx_i^{(a)}}{2\pi i x_i^{(a)}} \left(x_i^{(a)}\right)^{\mathfrak{t}^{(a)}} \times \prod\limits_{i\neq j}^N\left(1-\frac{x_i^{(a)}}{x_j^{(a)}}\right) \times \prod\limits_{i=1}^N\frac{\big(e^{iB_i^{(a)}}\big)^M}{e^{iB_i^{(a)}}-1}\right] \nonumber\\
&&\times \prod\limits_{i,j=1}^N\left[\prod\limits_{\substack{{\rm bi-fundamentals} \\ (b,a)\, {\rm and}\,(a,b)}} \left(\frac{\sqrt{\frac{x_i^{(a)}}{x_j^{(b)}}y_{(a,b)}}}{1-\frac{x_i^{(a)}}{x_j^{(b)}}y_{(a,b)}}\right)^{1-\mathfrak{n}_{(a,b)}}
\left(\frac{\sqrt{\frac{x_j^{(b)}}{x_i^{(a)}}y_{(b,a)}}}{1-\frac{x_j^{(b)}}{x_i^{(a)}}y_{(b,a)}}\right)^{1-\mathfrak{n}_{(b,a)}} \right. \nonumber\\
&&\times 
\left.\prod\limits_{\substack{{\rm adjoints} \\ (a,a)}} \left(\frac{\sqrt{y_{(a,a)}}}{1-\frac{x_j^{(a)}}{x_i^{(a)}}y_{(a,a)}}\right)^{1-\mathfrak{n}_{(a,a)}} \right] \nonumber\\
&&\times \prod\limits_{i=1}^N\left[\prod\limits_{\substack{{\rm fundamentals} \\a}}\Bigg(\frac{\sqrt{x_i^{(a)}y_{a}}}{1-x_i^{(a)}y_{a}}\Bigg)^{(1-\mathfrak{n}_a)} \prod\limits_{\substack{{\rm anti-fundamentals} \\a}}\Bigg(\frac{\sqrt{\frac{1}{x_i^{(a)}}\tilde{y}_{a}}}{1-\frac{1}{x_i^{(a)}}\tilde{y}_{a}}\Bigg)^{(1-\tilde{\mathfrak{n}}_a)}\right].
\eea
This is precisely the main expression we will consider. 
%%%%%%%%%%%%%%%%%%%%%%%%%%%%%%%%%%%%%%%%%%
\subsubsection*{The Bethe Ansatz Potential}
%%%%%%%%%%%%%%%%%%%%%%%%%%%%%%%%%%%%%%%%%%%%
An  alternative way to package the information in the index is to consider the so-called Bethe-Ansatz potential, ${\cal V}$. The Bethe-Ansatz potential succinctly summarizes the Bethe-Ansatz equations.  For the representations we will consider in this manuscript it is possible to write
\be
\cv=\cv^{CS}+\cv^{\rm bi-fund}+\cv^{\rm adjoint}+\cv^{\rm (anti-)fund}.
\ee
Introducing chemical potentials:
\be
y_I=e^{i\Delta_I}, \qquad \xi^{(a)}=e^{i\Delta_m^{(a)}},
\ee
the Bethe potential is given by 
\bea
\label{Eq:BAPotential_CS}
\cv^{CS}&=&\sum\limits_{i=1}^N\bigg[-\frac{k_a}{2}(u_i^{(a)})^2-{\rm sign}(k_a)\Delta_m^{(a)}u_i^{(a)}\bigg], \\
\label{Eq:BAPotential_bifund}
\cv^{\rm bi-fund}&=& \sum\limits_{\substack{{\rm bi-fundamentals} \\ (b,a) \,{\rm and} \,(a,b)}}\,\,\sum\limits_{i,j=1}^N
\bigg[{\rm Li}_2\left(e^{i(u_j^{(b)}-u_i^{(a)}+\Delta_{(b,a)})}\right)- {\rm Li}_2\left(e^{i(u_j^{(b)}-u_i^{(a)}-\Delta_{(a,b)})}\right)\bigg] \nonumber \\
&&+{\rm Arg}\Bigg[{\rm exp}\Bigg(i\bigg(-\frac{1}{2}{\rm Arg}\bigg[{\rm exp}\Big(i \sum\limits_{\substack{{\rm bi-fundamentals} \\ (b,a) \,{\rm and} \,(a,b)}}\,\, \left(\Delta_{(b,a)}+\Delta_{(a,b)}\right)\Big)\bigg]\nonumber \\
&&+\sum\limits_{\substack{{\rm bi-fundamentals} \\ (b,a) \,{\rm and} \,(a,b)}}\,\, \pi\bigg)\Bigg)\Bigg]\sum\limits_{i,j=1}^N \left(u_j^{(b)}-u_i^{(a)}\right),
\eea
and
\bea
\label{Eq:BAPotential_fund}
\cv^{\rm (anti-)fund}&=&\sum\limits_{i=1}^N\bigg[\sum\limits_{\substack{{\rm anti-fundamental}\\a}}{\rm Li}_2\left(e^{i(-u_i^{(a)}+\tilde{\Delta}_a)}\right) - \sum\limits_{\substack{{\rm fundamental}\\a}}{\rm Li}_2\left(e^{i(-u_i^{(a)}-\Delta_a)}\right)\bigg] \nonumber \\
&&+\frac{1}{2}\sum\limits_{i=1}^N\bigg[\sum\limits_{\substack{{\rm anti-fundamental}\\a}}\left(\tilde{\Delta}_a-\pi\right)u_i^{(a)} + \sum\limits_{\substack{{\rm fundamental}\\a}}\left(\Delta_a-\pi\right)u_i^{(a)}\bigg] \nonumber \\
&&-\frac{1}{4}\sum\limits_{i=1}^N\bigg[\sum\limits_{\substack{{\rm anti-fundamental}\\a}}\left(u_i^{(a)}\right)^2 - \sum\limits_{\substack{{\rm fundamental}\\a}}\left(u_i^{(a)}\right)^2\bigg].
\eea

Adjoint fields are treated as a special case of bi-fundamentals with $\Delta_{(b,a)}=\Delta_{(a,b)}=\Delta_{(a,a)}$ and an explicit factor of $1/2$. The second term in the bi-fundamental potential Eq.~(\ref{Eq:BAPotential_bifund}) is a little different from (A.10) in \cite{Hosseini:2016tor} for consistency with the potential in the ABJM theory in \cite{Benini:2015eyy}. This difference will only translate all of the eigenvalues along the real axis by a constant depending on $N$, which has no effect on the final result of ${\rm Re}\log Z$. Under the choice of the second term in Eq.~(\ref{Eq:BAPotential_bifund}), the eigenvalues for different values of $N$ will be concentrated without any translation along the real axis.

%%%%%%%%%%%%%%%%%%%%%%%%%%%%%%%%%%%%%%%%%%%%%%%%%%%%%%%%%%
\subsection{The topologically twisted index of ABJM beyond the large $N$ limit}\label{Sec:FieldTheory}
%%%%%%%%%%%%%%%%%%%%%%%%%%%%%%%%%%%%%%%%%%%%%%%%%%%%%%%%%%%%%%%%%%%%%%%%%%%%%%%%%%%%%%%%%%%%%%
As a way of giving the above general description of the topologically twisted index some concrete context, let us consider the ABJM theory \cite{Aharony:2008ug} which is a three-dimensional supersymmetric Chern-Simons-matter theory with gauge group
$U(N)_k \times U(N)_{-k}$  (the subscripts denote the Chern-Simons levels) and matter in bifundamental representations. A simple representation of the theory is via standard  ${\cal N}=2$ notation in terms of the quiver diagram below:
\bea
\begin{tikzpicture}[baseline, font=\footnotesize, scale=0.8]
\begin{scope}[auto,%
  every node/.style={draw, minimum size=0.5cm}, node distance=2cm];
  % the vertices
\node[circle] (USp2k) at (-0.1, 0) {$N_{+k}$};
\node[circle, right=of USp2k] (BN)  {$N_{-k}$};
\end{scope}
  % the edges
\draw[solid,line width=0.2mm,<-]  (USp2k) to[bend right=15] node[midway,above] {$B_2 $}node[midway,above] {}  (BN) ;
\draw[solid,line width=0.2mm,->]  (USp2k) to[bend right=50] node[midway,above] {$A_1$}node[midway,above] {}  (BN) ; 
\draw[solid,line width=0.2mm,<-]  (USp2k) to[bend left=15] node[midway,above] {$B_1$} node[midway,above] {} (BN) ;  
\draw[solid,line width=0.2mm,->]  (USp2k) to[bend left=50] node[midway,above] {$A_2$} node[midway,above] {} (BN) ;    
%\draw[black,-> ] (USp2k) edge [out={-150},in={150},loop,looseness=10] (USp2k) node at (-2,1) {$\phi_1$} ;
%\draw[black,-> ] (BN) edge [out={-30},in={30},loop,looseness=10] (BN) node at (5.8,1) {$\phi_2$};
\end{tikzpicture}
\eea
The superpotential of the theory is 
\begin{equation}
 W = \Tr\left[A_1 B_1 A_2 B_2 -A_1B_2 A_2 B_1 \right] \, .
\end{equation}

There are a total of four $U(1)$ gauge fields from the Cartan of the $SO(8)$ R-symmetry, with corresponding charges $\mathfrak{n}_a$ satisfying the supersymmetry constraint $\sum \mathfrak{n}_a=2$. The expression for the topologically twisted index, before summing over the magnetic fluxes $\mathfrak{m}$ takes the form
\bea
Z&=&\frac{1}{(N!)^2}\sum\limits_{\mathfrak{m},\tilde{\mathfrak{m}}\in \mathbb{Z}^N}\int_{\cal C}\prod\limits_{i=1}^N
\frac{dx_i}{2\pi i x_i}\frac{d\tx_i}{2\pi i \tx_i} x_i^{k\mathfrak{m}_i +\mathfrak{t} }\tx_i^{-k\tilde{\mathfrak{m}}_i+\tilde{\mathfrak{t}}}
\xi^{\mathfrak{m}_i}\tilde{\xi}^{-\tilde{\mathfrak{m}}_i} \times \prod\limits_{i\neq j}^N(1-\frac{x_i}{x_j})(1-\frac{\tx_i}{\tx_j}) \nonumber\\
&&\times \prod\limits_{i,j=1}^N\prod\limits_{a=1,2}
\left(\frac{\sqrt{\frac{x_i}{\tx_j}y_a}}{1-\frac{x_i}{\tx_j}y_a}\right)^{\mathfrak{m}_i-\tilde{\mathfrak{m}}_j-\mathfrak{n}_a+1}
\prod\limits_{b=3,4}
\left(\frac{\sqrt{\frac{\tx_j}{x_i}y_b}}{1-\frac{\tx_j}{x_i}y_b}\right)^{\tilde{\mathfrak{m}}_j-\mathfrak{m}_i-\mathfrak{n}_b+1}.
\eea

Performing the summation over magnetic fluxes introducing a large cut-off $M$ we get
\bea
Z&=&\frac{1}{(N!)^2}\int_{\cal C}\prod\limits_{i=1}^N
\frac{dx_i}{2\pi i x_i}\frac{d\tx_i}{2\pi i \tx_i}\prod\limits_{i\neq j}^N(1-\frac{x_i}{x_j})(1-\frac{\tx_i}{\tx_j}) \nonumber\\
&&\times \prod\limits_{i,j=1}^N\prod\limits_{a=1,2}
\left(\frac{\sqrt{\frac{x_i}{\tx_j}y_a}}{1-\frac{x_i}{\tx_j}y_a}\right)^{1-\mathfrak{n}_a}
\prod\limits_{b=3,4}
\left(\frac{\sqrt{\frac{\tx_j}{x_i}y_b}}{1-\frac{\tx_j}{x_i}y_b}\right)^{1-\mathfrak{n}_b} \nonumber \\
&&\times \prod\limits_{i=1}^N\frac{\big(e^{iB_i}\big)^M}{e^{iB_i}-1} \prod\limits_{j=1}^N\frac{\big(e^{i\tilde{B}_j}\big)^M}{e^{i\tilde{B}_j}-1}.
\eea

The topologically twisted index for ABJM theory was worked out in \cite{Benini:2015eyy}, and reduces to the evaluation of
the partition function
\be
Z(y_a,\mathfrak{n}_a)=\prod_{a=1}^4 y_a^{-\frac{1}{2}N^2 \mathfrak{n}_a}\sum_{I\in BAE}\frac{1}{\det\mathbb{B}}
\frac{\prod_{i=1}^N x_i^N \tilde{x}_i^N\prod_{i\neq j}\left(1-\frac{x_i}{x_j}\right)\left(1-\frac{\tilde{x}_i}{\tilde{x}_j}\right)}{\prod_{i,j=1}^N\prod_{a=1,2}(\tilde{x}_j-y_ax_i)^{1-\mathfrak{n}_a}\prod_{a=3,4}(x_i-y_a\tilde{x}_j)^{1-\mathfrak{n}_a}},
\label{eq:logZ}
\ee
where $y_a$ are the corresponding fugacities.  The summation is over all solutions $I$ of the ``Bethe Ansatz Equations" (BAE) $e^{iB_i}=e^{i\tilde{B}_i}=1$ modulo permutations, where
\begin{align}
e^{iB_i}&=x_i^k\prod_{j=1}^N\frac{(1-y_3 \frac{\tilde{x}_j}{x_i})(1-y_4 \frac{\tilde{x}_j}{x_i})}{(1-y_1^{-1} \frac{\tilde{x}_j}{x_i})(1-y_2^{-1} \frac{\tilde{x}_j}{x_i})},\nn\\
e^{i\tilde{B}_j}&=\tilde{x}_j^k\prod_{i=1}^N\frac{(1-y_3 \frac{\tilde{x}_j}{x_i})(1-y_4 \frac{\tilde{x}_j}{x_i})}{(1-y_1^{-1} \frac{\tilde{x}_j}{x_i})(1-y_2^{-1} \frac{\tilde{x}_j}{x_i})}.
\end{align}
Here $k$ is the Chern-Simons level, and the two sets of variables $\{x_i\}$ and $\{\tilde x_j\}$ arise from the $U(N)_k\times U(N)_{-k}$ structure of ABJM theory.  Finally, the $2N\times 2N$ matrix $\mathbb{B}$ is the Jacobian relating the $\{x_i,\tilde x_j\}$ variables to the $\{e^{iB_i},e^{i\tilde B_j}\}$ variables
\begin{equation}
\mathbb B=\begin{pmatrix}x_l\fft{\partial e^{iB_j}}{\partial x_l}&\tilde x_l\fft{\partial e^{iB_j}}{\partial\tilde x_l}\\[4pt]
x_l\fft{\partial e^{i\tilde B_j}}{\partial x_l}&\tilde x_l\fft{\partial e^{i\tilde B_j}}{\partial\tilde x_l}\end{pmatrix}.
\end{equation}

It is convenient to introduce the chemical potentials $\Delta_a$ according to $y_a=e^{i\Delta_a}$ and furthermore perform a change of variables $x_i=e^{iu_i}$, $\tilde{x}_j=e^{i\tilde{u}_j}$.  In this case, the BAE become
\begin{align}
0&=ku_i -i\sum_{j=1}^N\left[\sum_{a=3,4}\log\left(1-e^{i(\tilde{u}_j-u_i+\Delta_a)}\right)-
\sum_{a=1,2}\log\left(1-e^{i(\tilde{u}_j-u_i-\Delta_a)}\right)\right]-2\pi n_i, \nonumber \\
0&=k\tilde{u}_j -i\sum_{i=1}^N\left[\sum\limits_{a=3,4}\log\left(1-e^{i(\tilde{u}_j-u_i+\Delta_a)}\right)-
\sum_{a=1,2}\log\left(1-e^{i(\tilde{u}_j-u_i-\Delta_a)}\right)\right]-2\pi \tilde{n}_j.
\label{eq:BAE}
\end{align}
The topologically twisted index is evaluated by first solving these equations for $\{u_i,\tilde u_j\}$, and then inserting the resulting solution into the partition function Eq.~(\ref{eq:logZ}).  This procedure was carried out in \cite{Benini:2015eyy} in the large $N$ limit with $k=1$ by introducing the parametrization:
\begin{equation}
u_i=iN^{1/2}\, t_i +\pi-\ft12\delta v(t_i), \qquad \tilde{u}_i=iN^{1/2}\,t_i+\pi+\ft12\delta v(t_i),
\end{equation}
where we have further made use of reflection symmetry about $\pi$ along the real axis.  In the large $N$ limit, the eigenvalue distribution becomes continuous, and the set $\{t_i\}$ may be described by an eigenvalue density $\rho(t)$.

The leading order solution for $\rho(t)$ and $\delta v(t)$ was worked out in \cite{Benini:2015eyy}, and the resulting partition function exhibits the expected $N^{3/2}$ scaling of ABJM theory:
\be
{\rm Re}\log Z_0=-\frac{N^{3/2}}{3}\sqrt{2\Delta_1\Delta_2\Delta_3\Delta_4}\sum\limits_a\frac{\mathfrak{n}_a}{\Delta_a}.
\label{eq:logZ0}
\ee
%
%%%%%%%%%%%%%%%%%%%%%%%%%%%%%%%%%%%%%%%%%%%
\subsection*{Evaluation of the index beyond the leading order in $N$}

Given that the BA approach provides the exact answer in $N$ a numerical study of this topologically twisted index was performed in \cite{Liu:2017vll} and established that presence of a logarithmic correction of the form $-\frac{1}{2}\log N$. In this manuscript we take \cite{Liu:2017vll} as a blueprint and extend that analysis to  a number of models  with the goal of determining whether this logarithmic contribution is universal; we find that, indeed, it is. Let us thus briefly review the main result and some of the techniques of \cite{Liu:2017vll}.

In the ABJM context, one expects the subleading behavior of the index to have the form
\begin{equation}
{\rm Re}\log Z={\rm Re}\log Z_0+f_1(\Delta_a,\mathfrak{n}_a)N^{1/2}+f_2(\Delta_a,\mathfrak{n}_a)\log N+f_3(\Delta_a,\mathfrak{n}_a)+\mathcal O(N^{-1/2}),
\label{eq:ABJMrlz}
\end{equation}
where the functions $f_1$, $f_2$ and $f_3$ are linear in the magnetic fluxes $n_a$. 

\begin{figure}[h]
\centering
\includegraphics[width=0.7\linewidth]{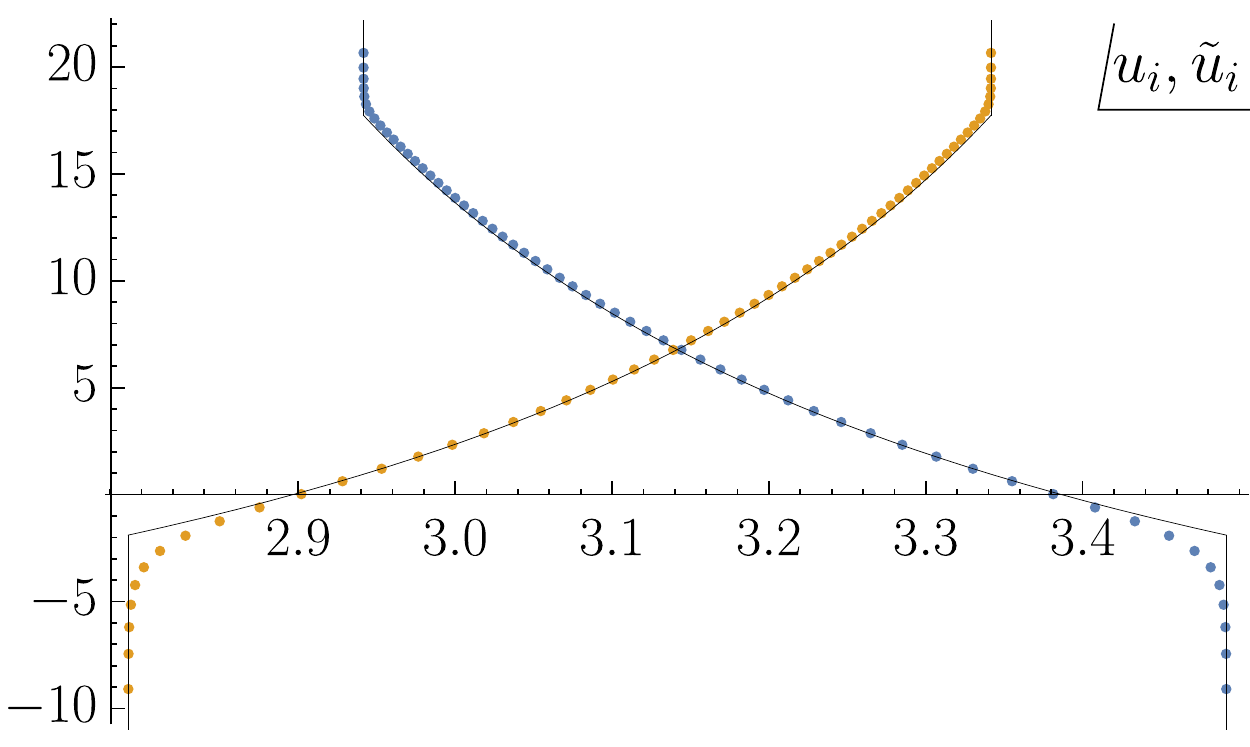}
\caption{The solution to the BAE for $\Delta_a=\{0.4,0.5,0.7,2\pi-1.6\}$ and $N=60$.  The solid lines correspond to the leading order expression obtained in \cite{Benini:2015eyy}.}
\label{fig:evals}
\end{figure}

\begin{figure}[h]
\centering
\hbox to \linewidth{\includegraphics[width=0.48\linewidth]{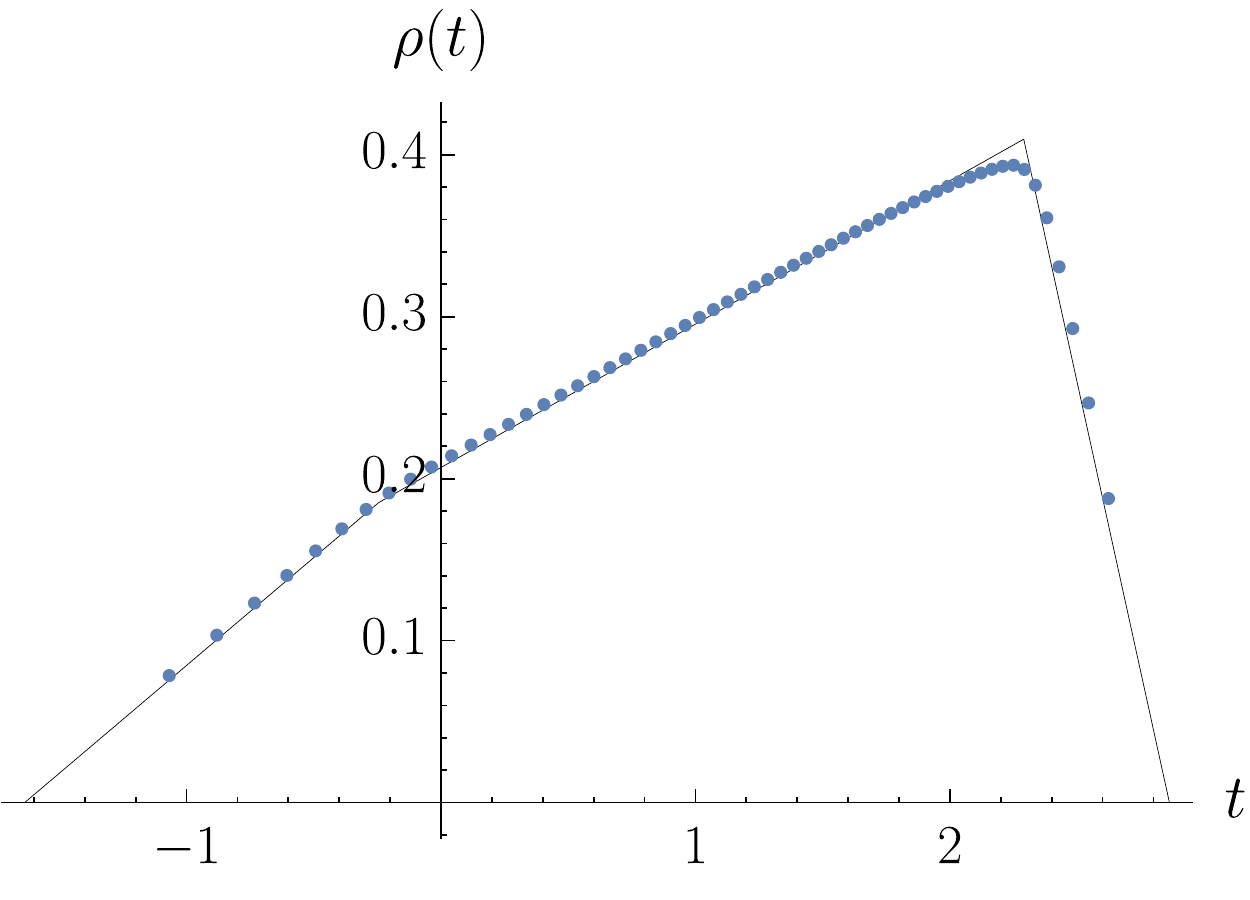}\hss\includegraphics[width=0.48\linewidth]{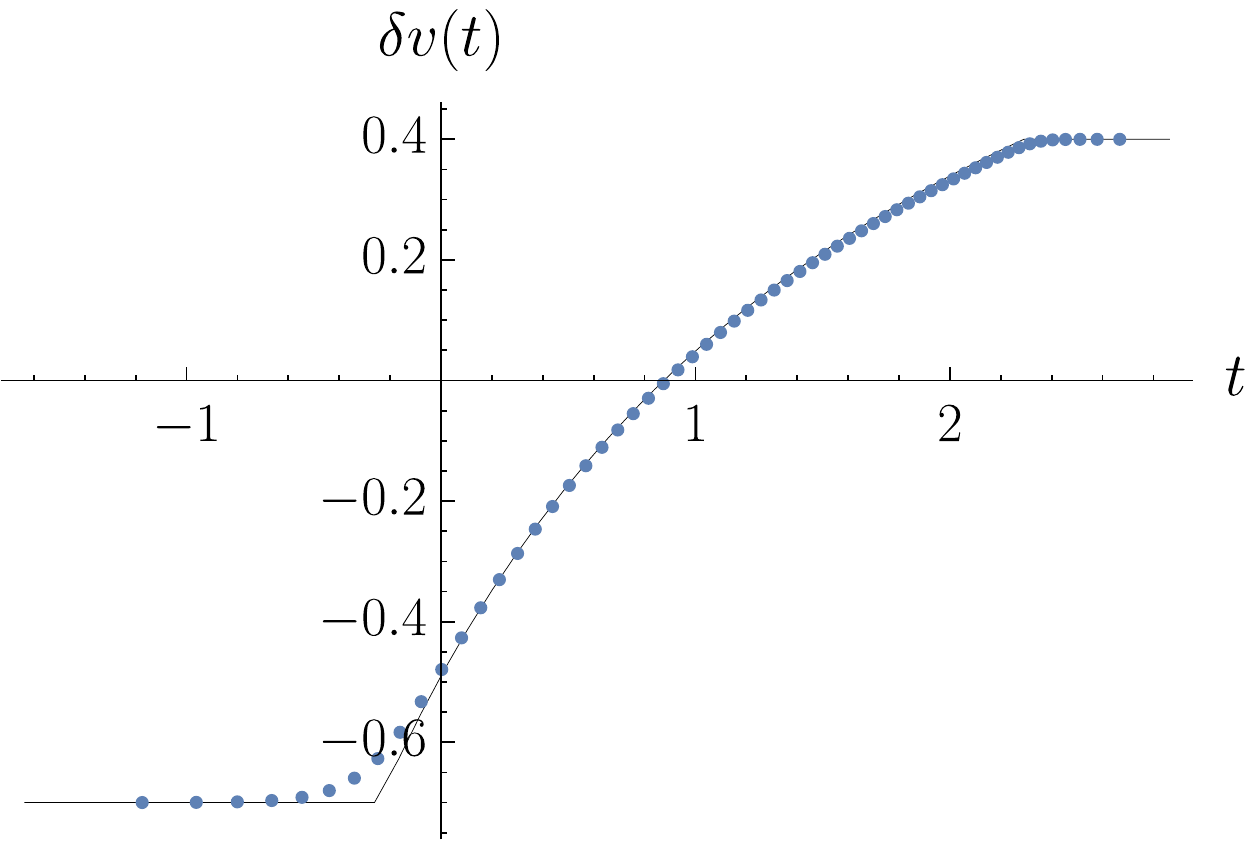}}
\caption{The eigenvalue density $\rho(t)$ and the function $\delta v(t)$ for $\Delta_a=\{0.4,0.5,0.7,2\pi-1.6\}$ and $N=60$, compared with the leading order expression.}
\label{fig:rhodv}
\end{figure}

Let us quote some results from \cite{Liu:2017vll} where the numerical solution for the eigenvalues $u_i$ and $\tilde u_i$ for $\Delta_a=\{0.4, 0.5, 0.7,2\pi-1.6\}$ and $N=60$ is shown in Figure~\ref{fig:evals} quoted from \cite{Liu:2017vll}.  The corresponding eigenvalue density $\rho(t)$ and function $\delta v(t)$ are shown in Figure~\ref{fig:rhodv} quoted from \cite{Liu:2017vll}.  

%%%%
%%%%

%%%%
%%%%

Once the eigenvalues are obtained, it is then simply a matter of numerically evaluating the index Eq.~(\ref{eq:logZ}) on the solution to the BAE.   For a given set of chemical potentials $\Delta_a$, we compute $\log Z$ for a range of $N$.  We then subtract out the leading behavior Eq.~(\ref{eq:logZ0}) and decompose the residuals into a sum of four independent terms:
\begin{equation}
{\rm Re}\log Z={\rm Re}\log Z_0+A+B_1\mathfrak{n}_1+B_2\mathfrak{n}_2+B_3\mathfrak{n}_3,
\end{equation}
where we have used the condition $\sum_a \mathfrak{n}_a=2$.  At this stage, we then perform a linear least-squares fit of $A$ and $B_a$ to the function
\begin{equation}
f(N)=f_1N^{1/2}+f_2\log N+f_3+f_4N^{-1/2}+f_5N^{-1}+f_6N^{-3/2}.
\end{equation}
%%%%

The results of the numerical fit are presented in Table~\ref{tbl:dat} quoted from \cite{Liu:2017vll} whose main result is that the numerical evidence points to the coefficient of the $\log N$ term being exactly $-1/2$.  We thus have
\begin{equation}
{\rm Re}\log Z=-\frac{N^{3/2}}{3}\sqrt{2\Delta_1\Delta_2\Delta_3\Delta_4}\sum\limits_a\frac{\mathfrak{n}_a}{\Delta_a}
+N^{1/2}f_1(\Delta_a,\mathfrak{n}_a)-\fft12\log N+f_3(\Delta_a,\mathfrak{n}_a)+\mathcal O(N^{-1/2}),
\label{eq:tti}
\end{equation}
where $f_1$ and $f_3$ remain to be determined.  

\begin{table}[t]
\centering
\begin{tabular}{|l l l||l|l|l|}
\hline
$\Delta_1$&$\Delta_2$&$\Delta_3$&$f_1$&$f_2$&$f_3$\\
\hline
$\pi/2$&$\pi/2$&$\pi/2$&$+3.0545$&$-0.4999$&$-3.0466$\\
\hline
$\pi/4$&$\pi/2$&$\pi/4$&\tabincell{l}{$+4.2215-0.0491\mathfrak{n}_1$\\$-0.1473\mathfrak{n}_2-0.0491\mathfrak{n}_3$}&\tabincell{l}{$-0.4996+0.0000\mathfrak{n}_1$\\$+0.0000\mathfrak{n}_2+0.0000\mathfrak{n}_3$}&\tabincell{l}{$-4.1710-0.2943\mathfrak{n}_1$\\$+0.0645\mathfrak{n}_2-0.2943\mathfrak{n}_3$}\\
\hline
$0.3$&$0.4$&$0.5$&\tabincell{l}{$+7.9855-0.2597\mathfrak{n}_1$\\$-0.5833\mathfrak{n}_2-0.6411\mathfrak{n}_3$}&\tabincell{l}{$-0.4994-0.0061\mathfrak{n}_1$\\$-0.0020\mathfrak{n}_2-0.0007\mathfrak{n}_3$}&\tabincell{l}{$-9.8404-0.9312\mathfrak{n}_1$\\$-0.0293\mathfrak{n}_2+0.3739\mathfrak{n}_3$}\\
\hline
$0.4$&$0.5$&$0.7$&\tabincell{l}{$+6.6696-0.1904\mathfrak{n}_1$\\$-0.4166\mathfrak{n}_2-0.4915\mathfrak{n}_3$}&\tabincell{l}{$-0.4986-0.0016\mathfrak{n}_1$\\$-0.0008\mathfrak{n}_2-0.0001\mathfrak{n}_3$}&\tabincell{l}{$-7.5313-0.6893\mathfrak{n}_1$\\$-0.1581\mathfrak{n}_2+0.2767\mathfrak{n}_3$}\\
\hline
\end{tabular}
\caption{(ABJM) Numerical fit for ${\rm Re}\log Z={\rm Re}\log Z_0+f_1N^{1/2}+f_2\log N+f_3+\cdots$.  The values of $N$ used in the fit range from $50$ to $N_{\rm max}$ in steps of $10$ where $N_{\rm max}=290, 150, 190, 120$ for the four cases, respectively. We made use of the fact that the index is independent of the magnetic fluxes when performing the fit for the special case ($\Delta_a=\{\pi/2,\pi/2,\pi/2,\pi/2\}$).}
\label{tbl:dat}
\end{table}

%%%%%%%%%%%%%%%%%%%%%%%%%%%%%%%%%%%%%%%%%%%%%%%%%%%%%%%
\section{The topologically twisted index of $N^{0,1,0}$}\label{Sec:N010}
%%%%%%%%%%

 In this section we study the Chern-Simons matter theory whose holographic dual is described by M-theory on ${\rm AdS}_4 \times N^{0,1,0}/\mathbb{Z}_k$ \cite{Fabbri:1999hw, Billo:2000zr, Yee:2006ba}. The space $N^{0,1,0}$ is a homogeneous Sasaki-Einstein manifold of dimension seven and defined as the coset $SU(3)/U(1)$. The manifold has the isometry $SU(3) \times SU(2)$; the latter $SU(2)$ is identified with the R-symmetry.

The  field theory was discussed in \cite{Gaiotto:2009tk, Imamura:2011uj, Cheon:2011th} and shown to be described by the following quiver diagram:
\bea
\begin{tikzpicture}[font=\footnotesize, scale=0.8]
\begin{scope}[auto,%
  every node/.style={draw, minimum size=0.5cm}, node distance=2cm];
  % the vertices
\node[circle] (USp2k) at (0., 0) {$N_{+k}$};
\node[circle, right=of USp2k] (BN)  {$N_{-k}$};
\node[rectangle, below=of USp2k] (Ur)  {$r$};
\end{scope}
  % the edges
\draw[solid,line width=0.2mm,<-]  (USp2k) to[bend right=15] node[midway,above] {$B_2 $}node[midway,above] {}  (BN) ;
\draw[solid,line width=0.2mm,->]  (USp2k) to[bend right=50] node[midway,above] {$A_1$}node[midway,above] {}  (BN) ; 
\draw[solid,line width=0.2mm,<-]  (USp2k) to[bend left=15] node[midway,above] {$B_1$} node[midway,above] {} (BN) ;  
\draw[solid,line width=0.2mm,->]  (USp2k) to[bend left=50] node[midway,above] {$A_2$} node[midway,above] {} (BN) ;    
\draw[black,-> ] (USp2k) edge [out={-150},in={150},loop,looseness=10] (USp2k) node at (-2.,1) {$\phi_1$} ;
\draw[black,-> ] (BN) edge [out={-30},in={30},loop,looseness=10] (BN) node at (5.8,1) {$\phi_2$};
%\draw (0,0.55) arc (4:280:0.75cm) node at (-2,1) {$\phi_1$} ;
%\draw(4,0.55) arc (0:-300:-0.75cm) node at (5.8,1) {$\phi_2$} ;
\draw[draw=black,solid,line width=0.2mm,<-]  (USp2k) to[bend left=20] node[midway,right] {$\tilde{q}$} node[midway,above] {} (Ur) ;  
\draw[draw=black,solid,line width=0.2mm,->]  (USp2k) to[bend right=20] node[midway,left] {$q$} node[midway,above] {} (Ur) ;    
\end{tikzpicture}
\eea
The superpotential is 
\bea 
W = \Tr \left(A_1 \phi_2 B_2 - B_2 \phi_1 A_1 - A_2 \phi_2 B_1 + B_1 \phi_1 A_2 + \frac{k}{2} \phi_1^2 - \frac{k}{2} \phi_2^2 + \tilde{q} \phi_1 q \right) \, .
\eea

The free energy on $S^3$ has been shown to match  the gravity computation \cite{Gaiotto:2009tk}; a discussion of the superconformal index was presented in  \cite{Imamura:2011uj,Cheon:2011th}.  In the context of the topologically twisted index, this theory was recently considered  by Hosseini and Mekareeya in   \cite{Hosseini:2016ume} from which we borrow  much, including the notation and the leading order analysis.

%%%%%%%%%%%%%%%%%%%%%%%%%%%%%%%%%%%%%%%%%%%
\subsection{Numerical solutions to the system of BAEs}\label{Sec:N010_BAEs}
The topologically twisted index can be algorithmically assembled from the field theory content and the result is 
\bea
Z&=&\frac{1}{(N!)^2}\sum\limits_{\mathfrak{m},\tilde{\mathfrak{m}}\in \mathbb{Z}^N}\int_{\cal C}\prod\limits_{i=1}^N
\frac{dx_i}{2\pi i x_i}\frac{d\tx_i}{2\pi i \tx_i} x_i^{k\mathfrak{m}_i}\tx_i^{-k\tilde{\mathfrak{m}}_i} \times \prod\limits_{i\neq j}^N\left(1-\frac{x_i}{x_j}\right)\left(1-\frac{\tx_i}{\tx_j}\right) \nonumber\\
&&\times \prod\limits_{i,j=1}^N\prod\limits_{a=1,2}
\left(\frac{\sqrt{\frac{x_i}{\tx_j}y_a}}{1-\frac{x_i}{\tx_j}y_a}\right)^{\mathfrak{m}_i-\tilde{\mathfrak{m}}_j-\mathfrak{n}_a+1}
\prod\limits_{b=3,4}
\left(\frac{\sqrt{\frac{\tx_j}{x_i}y_b}}{1-\frac{\tx_j}{x_i}y_b}\right)^{\tilde{\mathfrak{m}}_j-\mathfrak{m}_i-\mathfrak{n}_b+1} \nonumber\\
&&\times \prod\limits_{i=1}^N\Bigg(\frac{\sqrt{x_i y_q}}{1-x_i y_q}\Bigg)^{r(\mathfrak{m}_i-\mathfrak{n}_q+1)}\Bigg(\frac{\sqrt{\frac{1}{x_i} y_{\tilde{q}}}}{1-\frac{1}{x_i} y_{\tilde{q}}}\Bigg)^{r(-\mathfrak{m}_i-\mathfrak{n}_{\tilde{q}}+1)}.
\eea

Performing the summation over magnetic fluxes by  introducing a large cut-off $M$ we get
\bea
Z&=&\frac{1}{(N!)^2}\int_{\cal C}\prod\limits_{i=1}^N
\frac{dx_i}{2\pi i x_i}\frac{d\tx_i}{2\pi i \tx_i}\prod\limits_{i\neq j}^N\left(1-\frac{x_i}{x_j}\right)\left(1-\frac{\tx_i}{\tx_j}\right) \nonumber\\
&&\times \prod\limits_{i,j=1}^N\prod\limits_{a=1,2}
\left(\frac{\sqrt{\frac{x_i}{\tx_j}y_a}}{1-\frac{x_i}{\tx_j}y_a}\right)^{1-\mathfrak{n}_a}
\prod\limits_{b=3,4}
\left(\frac{\sqrt{\frac{\tx_j}{x_i}y_b}}{1-\frac{\tx_j}{x_i}y_b}\right)^{1-\mathfrak{n}_b} \nonumber\\
&&\times \prod\limits_{i=1}^N\Bigg(\frac{\sqrt{x_i y_q}}{1-x_i y_q}\Bigg)^{r(1-\mathfrak{n}_q)}\Bigg(\frac{\sqrt{\frac{1}{x_i} y_{\tilde{q}}}}{1-\frac{1}{x_i} y_{\tilde{q}}}\Bigg)^{r(1-\mathfrak{n}_{\tilde{q}})} \times \prod\limits_{i=1}^N\frac{\big(e^{iB_i}\big)^M}{e^{iB_i}-1} \prod\limits_{j=1}^N\frac{\big(e^{i\tilde{B}_j}\big)^M}{e^{i\tilde{B}_j}-1},
\eea
where the Bethe Ansatz equations are
\bea\label{Eq:BAEN010}
1&=&e^{iB_i}=x_i^k\prod\limits_{j=1}^N\frac{\left(1-y_3 \frac{\tilde{x}_j}{x_i}\right)\left(1-y_4 \frac{\tilde{x}_j}{x_i}\right)}{\left(1-y_1^{-1} \frac{\tilde{x}_j}{x_i}\right)\left(1-y_2^{-1} \frac{\tilde{x}_j}{x_i}\right)} \times \Bigg(\frac{\sqrt{x_i y_q}}{1-x_i y_q}\Bigg)^r \Bigg(\frac{\sqrt{\frac{1}{x_i} y_{\tilde{q}}}}{1-\frac{1}{x_i} y_{\tilde{q}}}\Bigg)^{-r}, \nonumber\\
1&=&e^{i\tilde{B}_j}=\tilde{x}_j^k\prod\limits_{i=1}^N\frac{\left(1-y_3 \frac{\tilde{x}_j}{x_i}\right)\left(1-y_4 \frac{\tilde{x}_j}{x_i}\right)}{\left(1-y_1^{-1} \frac{\tilde{x}_j}{x_i}\right)\left(1-y_2^{-1} \frac{\tilde{x}_j}{x_i}\right)}.
\eea

The compact expression for the index in terms of solutions to the Bethe-Ansatz equations Eq.~(\ref{Eq:BAEN010}) takes the form
\bea
\label{Eq:N010_Index}
Z\left(y_a,\mathfrak{n}_a\right)&=&\left(-1\right)^{\frac{Nr}{2}}y_q^{-\frac{1}{2}Nr\mathfrak{n}_q}y_{\tilde{q}}^{-\frac{1}{2}Nr\mathfrak{n}_{\tilde{q}}}\prod_{a=1}^4 y_a^{-\frac{1}{2}N^2 \mathfrak{n}_a}\nonumber\\
&&\times\sum_{I\in BAE}\left[\frac{1}{\det\mathbb{B}}
\frac{\prod_{i=1}^N x_i^N \tilde{x}_i^N \prod_{i\neq j}\left(1-\frac{x_i}{x_j}\right)\left(1-\frac{\tilde{x}_i}{\tilde{x}_j}\right)}{\prod_{i,j=1}^N\prod_{a=1,2}\left(\tilde{x}_j-y_ax_i\right)^{1-\mathfrak{n}_a}\prod_{a=3,4}\left(x_i-y_a\tilde{x}_j\right)^{1-\mathfrak{n}_a}}\right. \nonumber\\
&&\left.\times\prod_{i=1}^N \frac{x_i^{\frac{1}{2}r}}{\left(1-x_i y_q\right)^{r\left(1-\mathfrak{n}_q \right)}\left(x_i-y_{\tilde{q}}\right)^{r\left(1-\mathfrak{n}_{\tilde{q}} \right)}}\right].
\eea

The transformation matrix $\mathbb{B}$ describing the change in integration variables from $x_i$ to $B_i$ is 
\be
\mathbb{B}\Big|_{\rm BAEs}=
\left(
\begin{array}{cc}
\delta_{jl}\left[k-\sum_{m=1}^N G_{jm}+rx_j\left(\frac{1}{x_j-y_{\tilde{q}}}-\frac{1}{x_j-y_q^{-1}}\right) \right] & G_{jl}\\
-G_{lj} & \delta_{jl}\left[k+\sum_{m=1}^N G_{mj} \right]
\end{array}
\right),
\ee
where
\be
D(z)=\frac{(1-zy_3)(1-zy_4)}{(1-zy_1^{-1})(1-zy_2^{-1})},\qquad G_{ij}=\frac{\partial\log D(z)}{\partial\log z}\Big|_{z=\tilde{x}_j/x_i}.
\ee

The Bethe-Ansatz equations Eq.~(\ref{Eq:BAEN010}) can be obtained from the potential which takes the form
\bea
\label{Eq:N010BethePotential}
\mathcal{V}&=&\sum\limits_{i=1}^N\left[\frac{k}{2}\left(\tilde{u}_i^2-u_i^2\right)-2\pi \left(\tilde{n}_i \tilde{u}_i-n_iu_i\right)\right]\nonumber \\
&&+ \sum\limits_{i,j=1}^N\left[\sum\limits_{a=3,4}{\rm Li}_2\left(e^{i\left(\tilde{u}_j-u_i+\Delta_a\right)}\right)-
\sum\limits_{a=1,2}{\rm Li}_2\left(e^{i\left(\tilde{u}_j-u_i-\Delta_a\right)}\right)\right]\nonumber \\
&&+ r\sum\limits_{i=1}^N\left[{\rm Li}_2\left(e^{i\left(-u_i+\Delta_{\tilde{q}}\right)}\right)-{\rm Li}_2\left(e^{i\left(-u_i-\Delta_q\right)}\right)\right] + \frac{r}{2}\sum\limits_{i=1}^N\left[\left(\Delta_{\tilde{q}}+\Delta_q-2\pi\right)u_i\right],
\eea
where
\be
\sum\limits_{i=1}^N\left[-2\pi \left(\tilde{n}_i \tilde{u}_i-n_iu_i\right)\right]=\left(4\pi-\sum\limits_{a=1}^4\Delta_a\right)\sum\limits_{i>j}^N\left(\tilde{u}_j-u_i\right)=2\pi\sum\limits_{i>j}^N\left(\tilde{u}_j-u_i\right).
\ee

We use the leading order solution to the system of BAEs in \cite{Hosseini:2016ume} as a seed for the exact numerical solution to the BAE's in Eq.~(\ref{Eq:BAEN010}). We will assume,  as in  \cite{Hosseini:2016tor}, that $0<v(t)+\Delta_q<2\pi$ and $0<-v(t)+\Delta_{\tilde{q}}<2\pi$, then we have
\be
-\Delta_q<v(t)<\Delta_{\tilde{q}},
\ee
thus we set the initial real part axis to be
\be
\label{Eq:qReAxis}
\frac{v(t)+\tilde{v}(t)}{2}=\frac{\Delta_{\tilde{q}}-\Delta_q}{2}.
\ee

The marginality condition on the superpotential requires that $\Delta_q+\Delta_{\tilde{q}}=\pi$. For comparison, recall that in the ABJM theory, the real part axis defined above is $\pi$ when $k=1$, thus we assume the range of the real part axis here should be near to $\pi$ to match with the ABJM theory when $r=0$. If we set $\left(\Delta_{\tilde{q}}-\Delta_q\right)/2=\pi$ we will get $\{\Delta_q, \Delta_{\tilde{q}}\}=\{-\pi/2, 3\pi/2\}$.  We will see the effects of the values of $\{\Delta_q, \Delta_{\tilde{q}}\}$ in the numerical solutions. 

Before entering the details of the numerical analysis, let us remark that, as compared to ABJM, the system has three new parameters which are $r$ -- the number of flavors of fundamental hypermultiplets, and the fugacities $\{\Delta_q, \Delta_{\tilde{q}}\}$ discussed above. Our goal is to explore the space of new parameters as well as those parameters already present in ABJM, namely $N$ -- the rank of the gauge group and $\Delta_a=\{\Delta_1, \Delta_2, \Delta_3, \Delta_4\}$ -- the fugacities of the bi-fundamental matter. 

\begin{figure}[H]
    \centering
    \begin{subfigure}[b]{0.75\textwidth}
	    \begin{subfigure}[b]{0.5\textwidth}
	        \includegraphics[width=\textwidth]{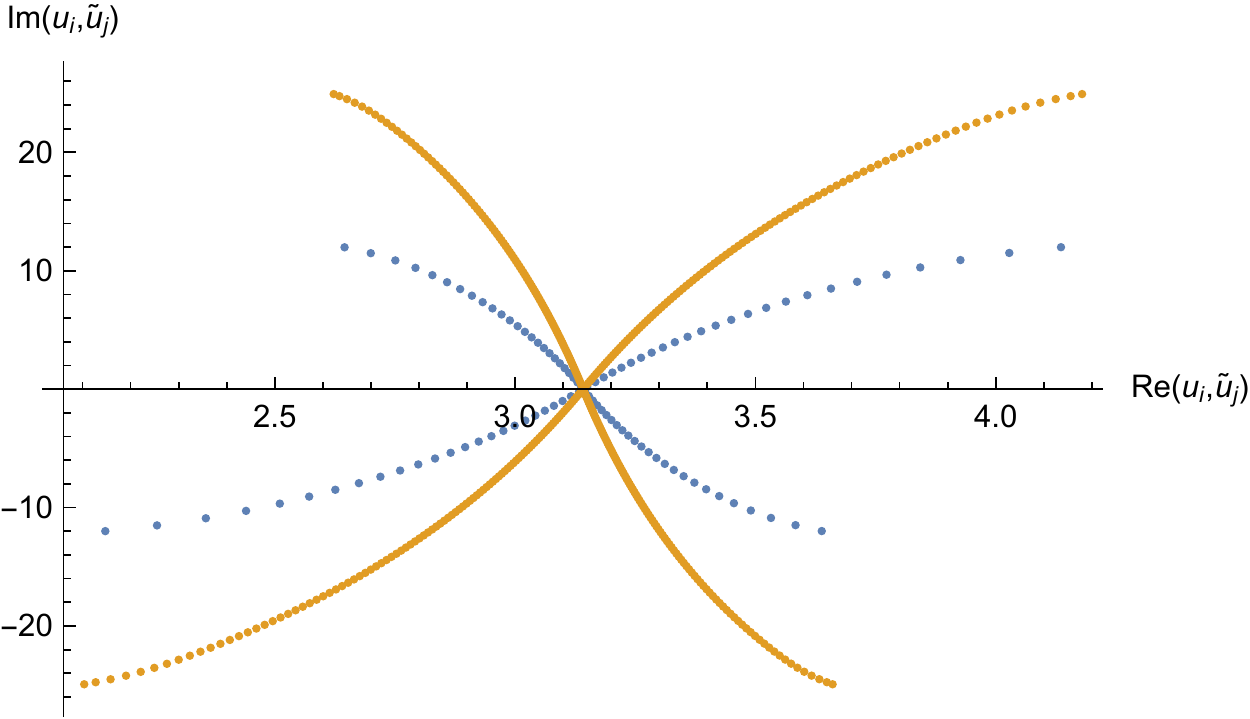}
	        \caption{Eigenvalue distribution}
	    \end{subfigure}
	    \quad %add desired spacing between images, e. g. ~, \quad, \qquad, \hfill etc. 
	    \begin{subfigure}[b]{0.5\textwidth}
	        \includegraphics[width=\textwidth]{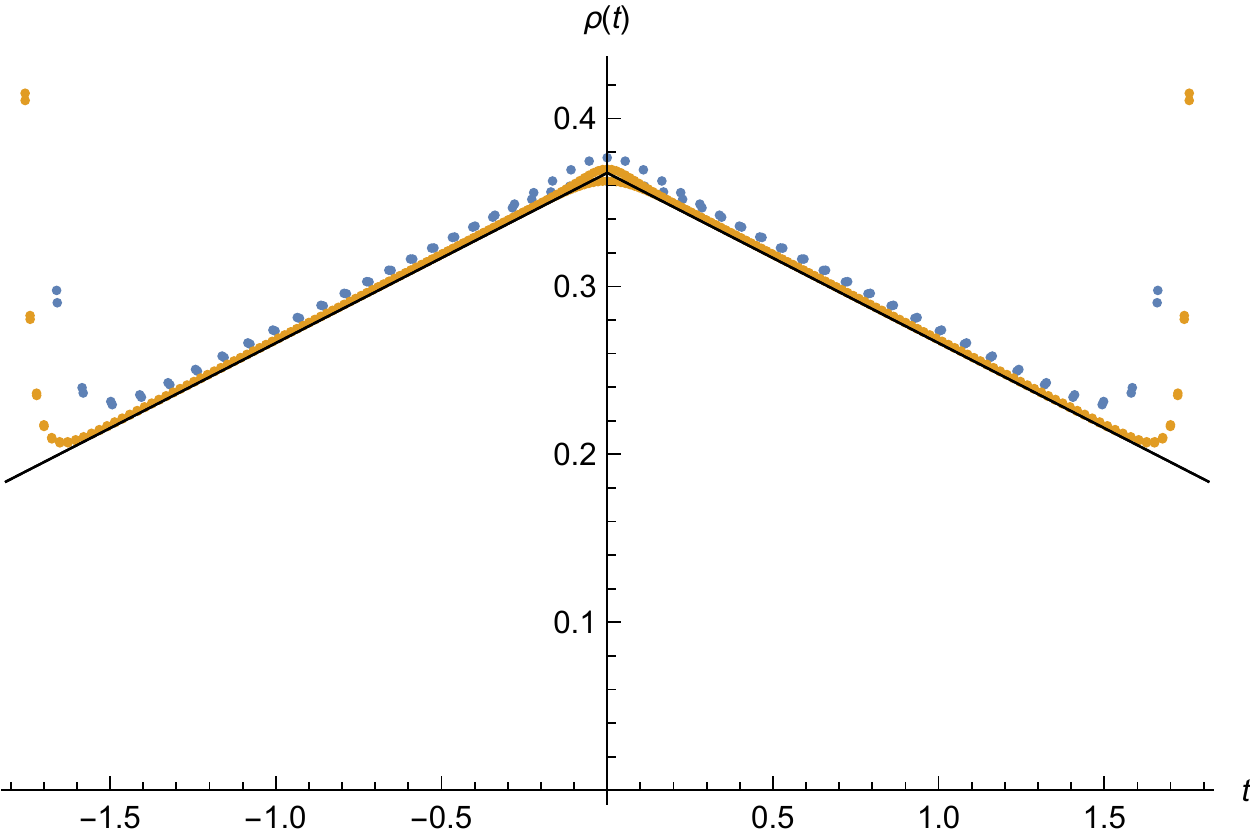}
	        \caption{Eigenvalue density $\rho(t)$}
	    \end{subfigure}\\
	    ~ %add desired spacing between images, e. g. ~, \quad, \qquad, \hfill etc. 
	    \begin{subfigure}[b]{0.5\textwidth}
	        \includegraphics[width=\textwidth]{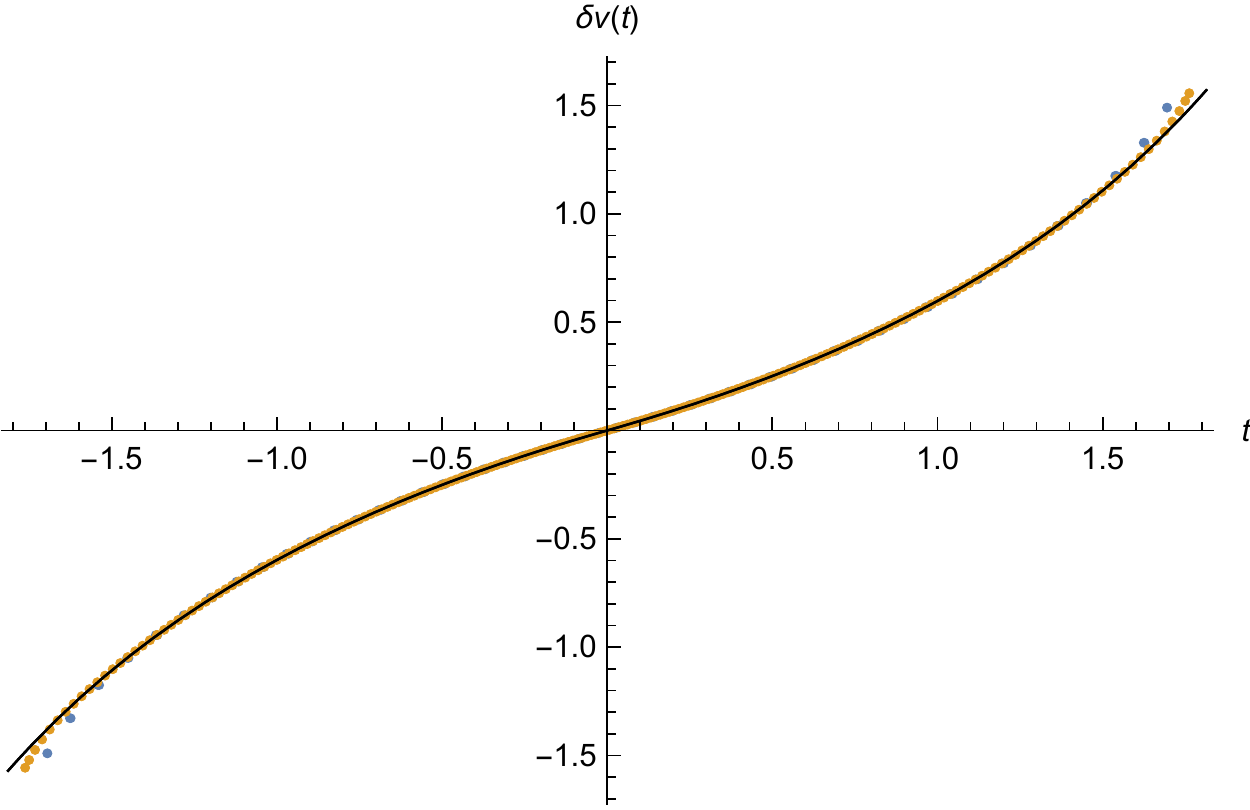}
	        \caption{Real part difference $\delta v(t)$}
	    \end{subfigure}
	    \quad %add desired spacing between images, e. g. ~, \quad, \qquad, \hfill etc. 
	    \begin{subfigure}[b]{0.5\textwidth}
	        \includegraphics[width=\textwidth]{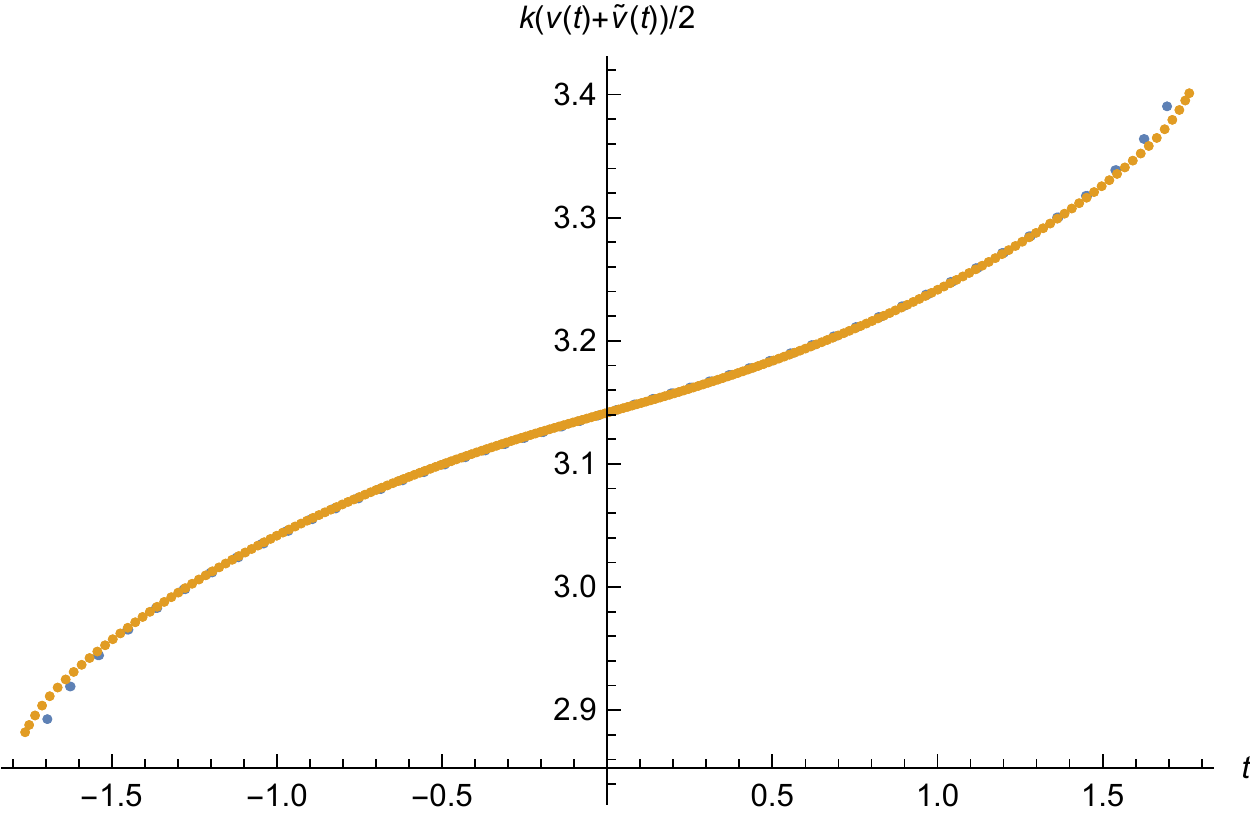}
	        \caption{Real part axis $k(v(t)+\tilde{v}(t))/2$}
	    \end{subfigure}
    \end{subfigure}
    \quad %add desired spacing between images, e. g. ~, \quad, \qquad, \hfill etc. 
    \begin{subfigure}[b]{0.2\textwidth}
        \includegraphics[width=\textwidth]{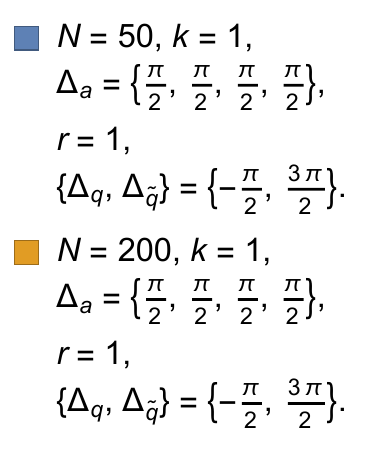}
    \end{subfigure}
    \caption{Eigenvalues for the special case $\Delta_a=\{\frac{\pi}{2}, \frac{\pi}{2}, \frac{\pi}{2}, \frac{\pi}{2}\}$ for $N=50$ (blue) and $200$ (orange) with the same other parameters.}
    \label{fig:N010_Plot_N}
\end{figure}

\begin{figure}[H]
    \centering
    \begin{subfigure}[b]{0.75\textwidth}
	    \begin{subfigure}[b]{0.5\textwidth}
	        \includegraphics[width=\textwidth]{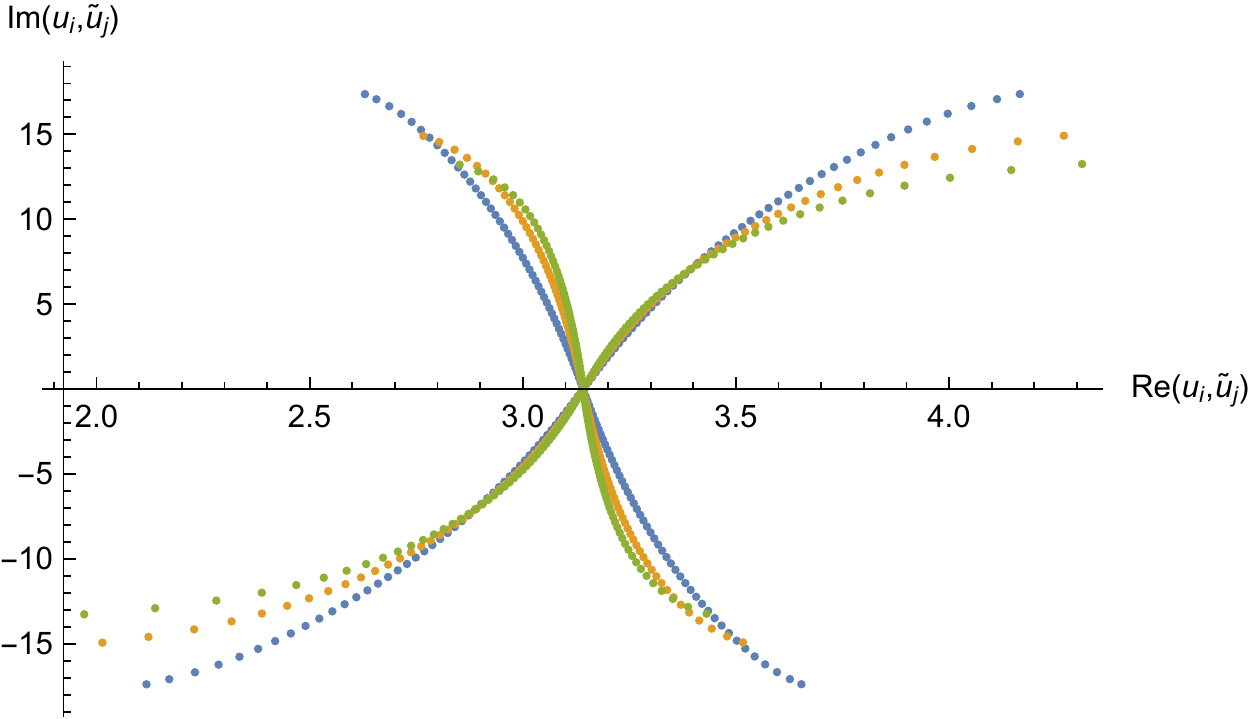}
	        \caption{Eigenvalue distribution}
	    \end{subfigure}
	    \quad %add desired spacing between images, e. g. ~, \quad, \qquad, \hfill etc. 
	    \begin{subfigure}[b]{0.5\textwidth}
	        \includegraphics[width=\textwidth]{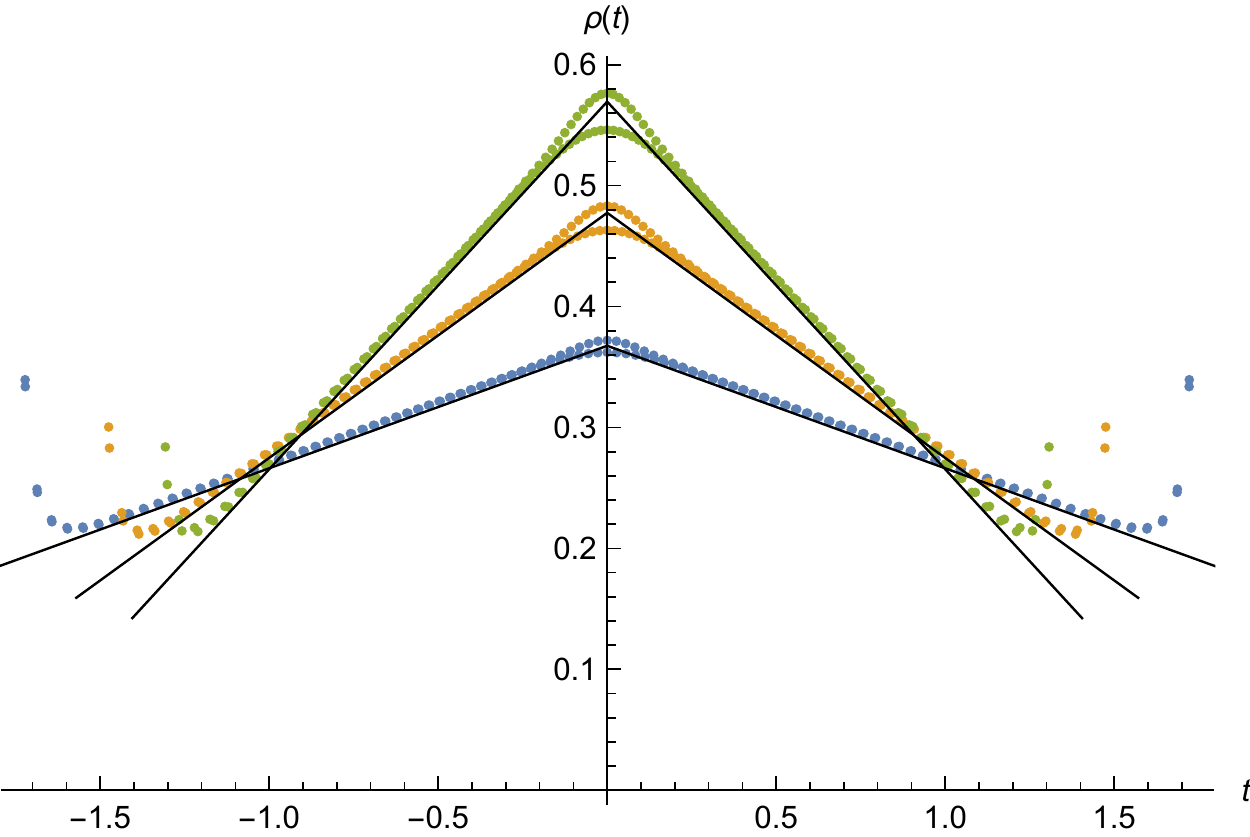}
	        \caption{Eigenvalue density $\rho(t)$}
	    \end{subfigure}\\
	    ~ %add desired spacing between images, e. g. ~, \quad, \qquad, \hfill etc. 
	    \begin{subfigure}[b]{0.5\textwidth}
	        \includegraphics[width=\textwidth]{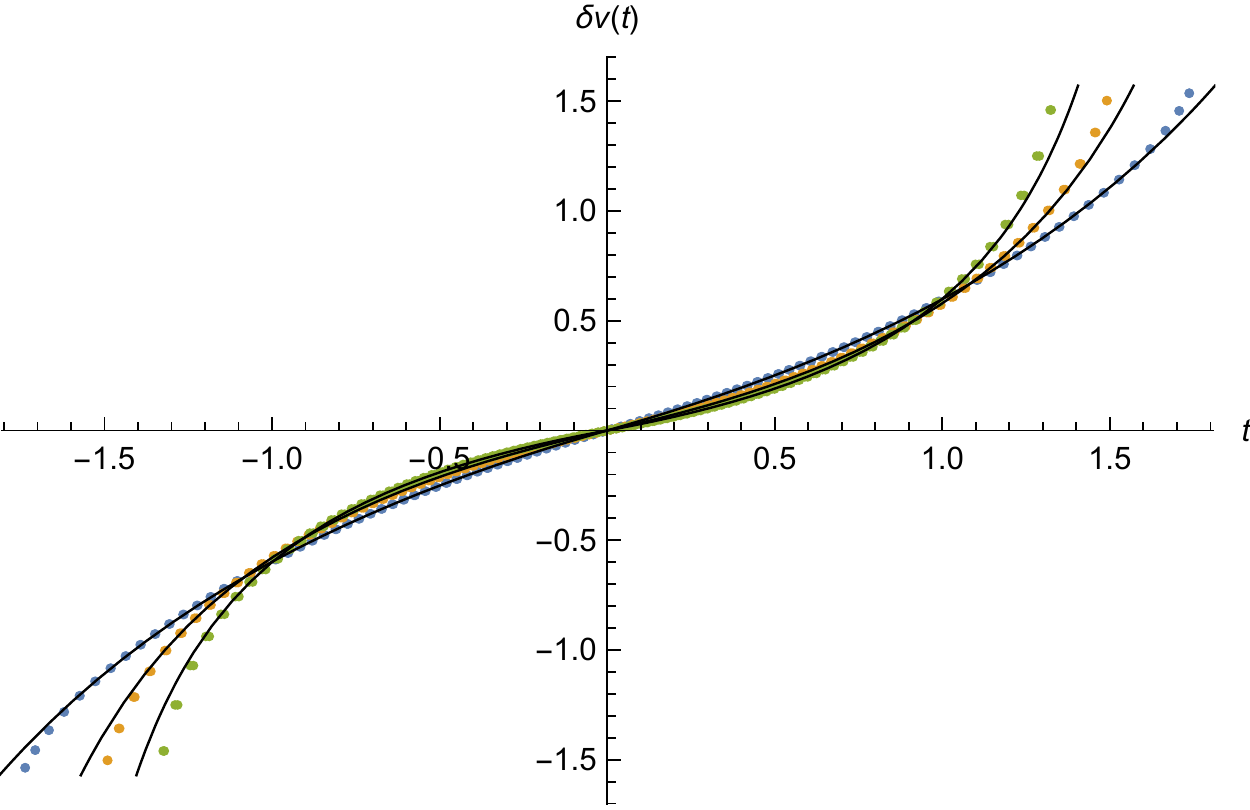}
	        \caption{Real part difference $\delta v(t)$}
	    \end{subfigure}
	    \quad %add desired spacing between images, e. g. ~, \quad, \qquad, \hfill etc. 
	    \begin{subfigure}[b]{0.5\textwidth}
	        \includegraphics[width=\textwidth]{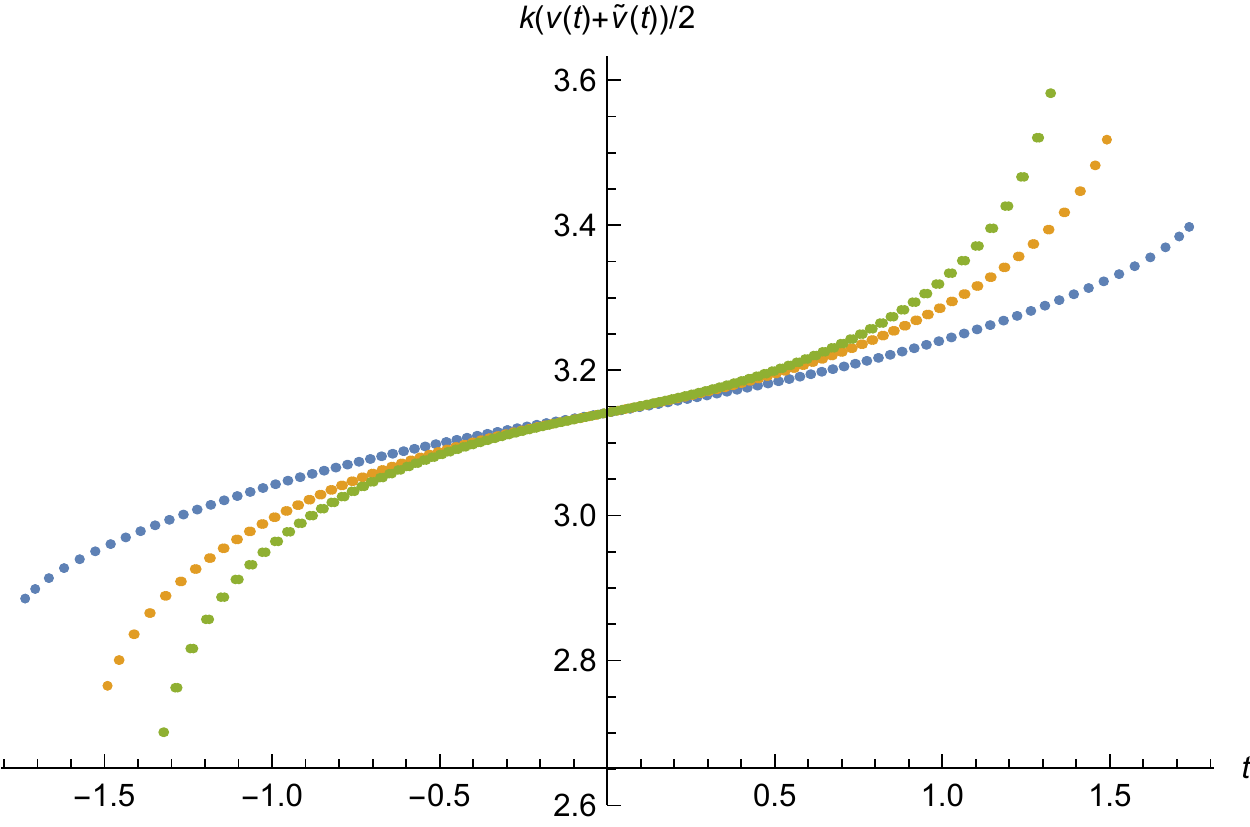}
	        \caption{Real part axis $k(v(t)+\tilde{v}(t))/2$}
	    \end{subfigure}
    \end{subfigure}
    \quad %add desired spacing between images, e. g. ~, \quad, \qquad, \hfill etc. 
    \begin{subfigure}[b]{0.2\textwidth}
        \includegraphics[width=\textwidth]{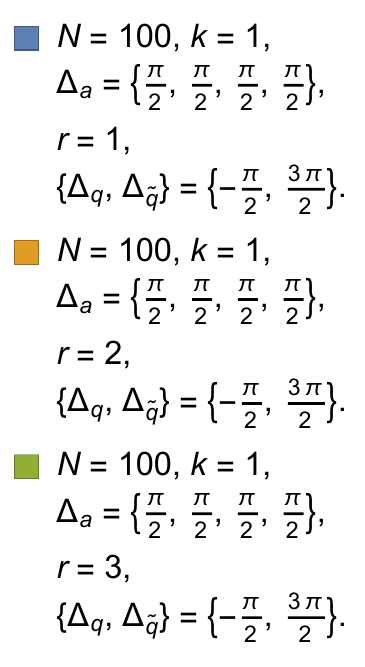}
    \end{subfigure}
    \caption{Eigenvalues for the special case $\Delta_a=\{\frac{\pi}{2}, \frac{\pi}{2}, \frac{\pi}{2}, \frac{\pi}{2}\}$ for $r=1$ (blue), $2$ (orange) and $3$ (green) with other parameters kept the same.}
    \label{fig:N010_Plot_r}
\end{figure}

\begin{figure}[H]
    \centering
    \begin{subfigure}[b]{0.75\textwidth}
	    \begin{subfigure}[b]{0.5\textwidth}
	        \includegraphics[width=\textwidth]{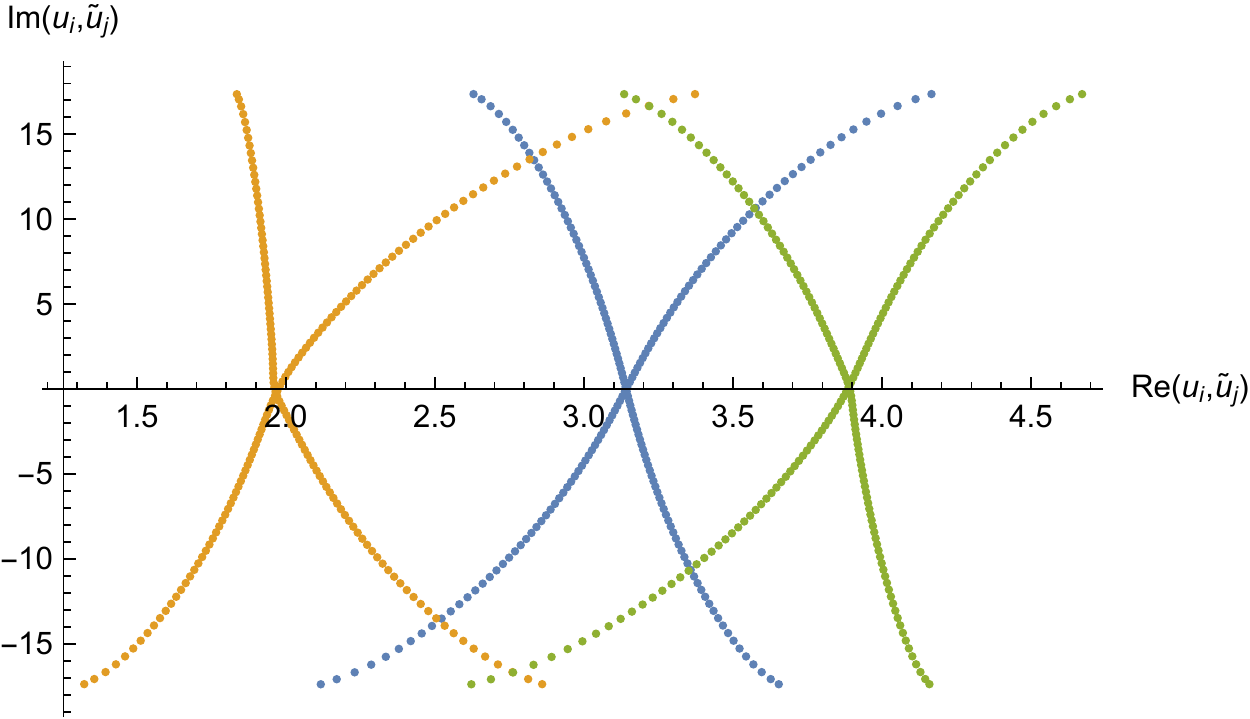}
	        \caption{Eigenvalue distribution}
	    \end{subfigure}
	    \quad %add desired spacing between images, e. g. ~, \quad, \qquad, \hfill etc. 
	    \begin{subfigure}[b]{0.5\textwidth}
	        \includegraphics[width=\textwidth]{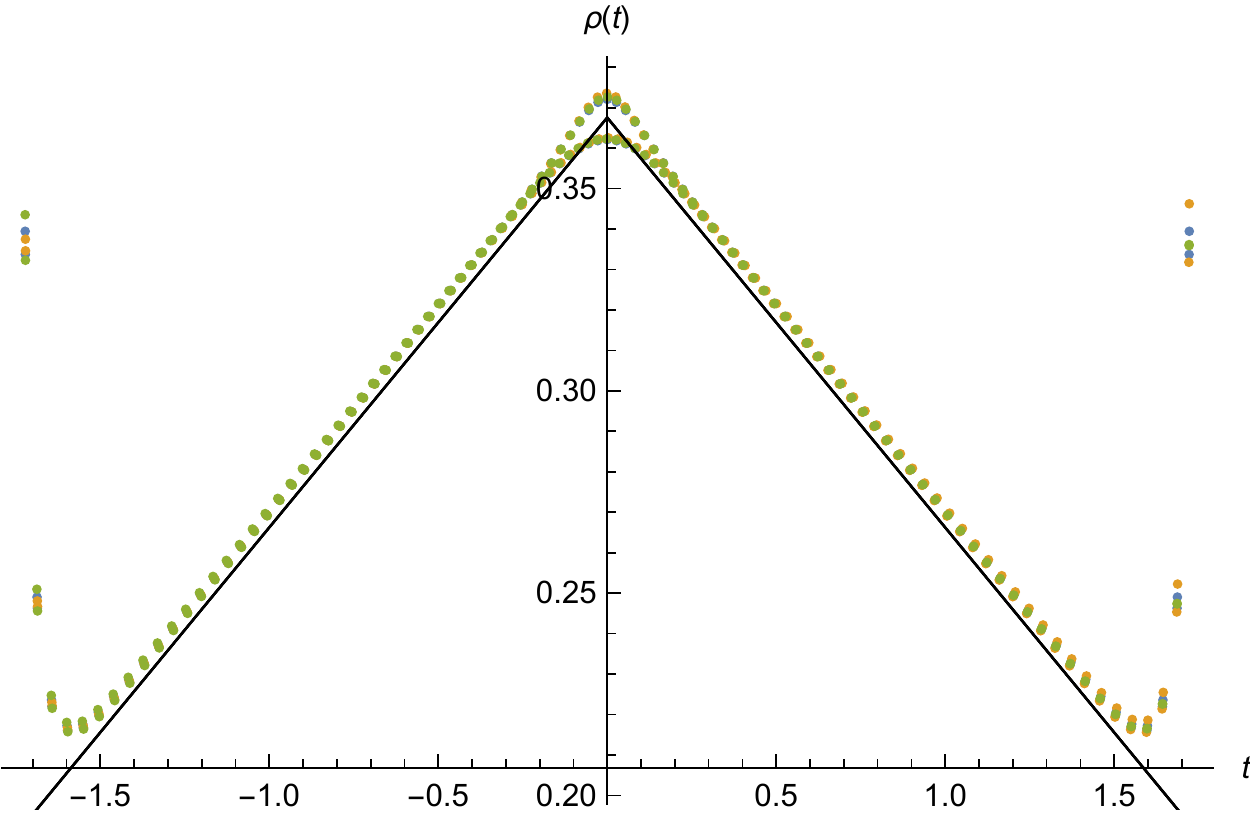}
	        \caption{Eigenvalue density $\rho(t)$}
	    \end{subfigure}\\
	    ~ %add desired spacing between images, e. g. ~, \quad, \qquad, \hfill etc. 
	    \begin{subfigure}[b]{0.5\textwidth}
	        \includegraphics[width=\textwidth]{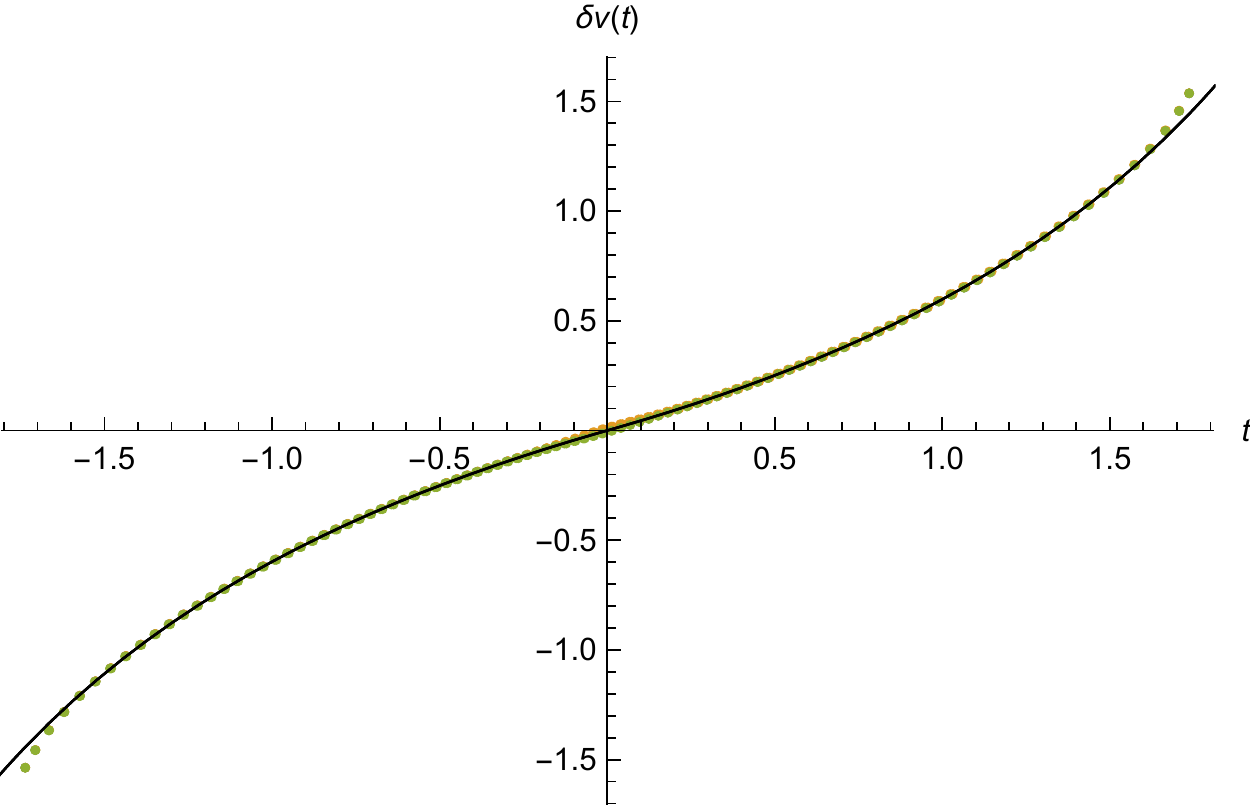}
	        \caption{Real part difference $\delta v(t)$}
	    \end{subfigure}
	    \quad %add desired spacing between images, e. g. ~, \quad, \qquad, \hfill etc. 
	    \begin{subfigure}[b]{0.5\textwidth}
	        \includegraphics[width=\textwidth]{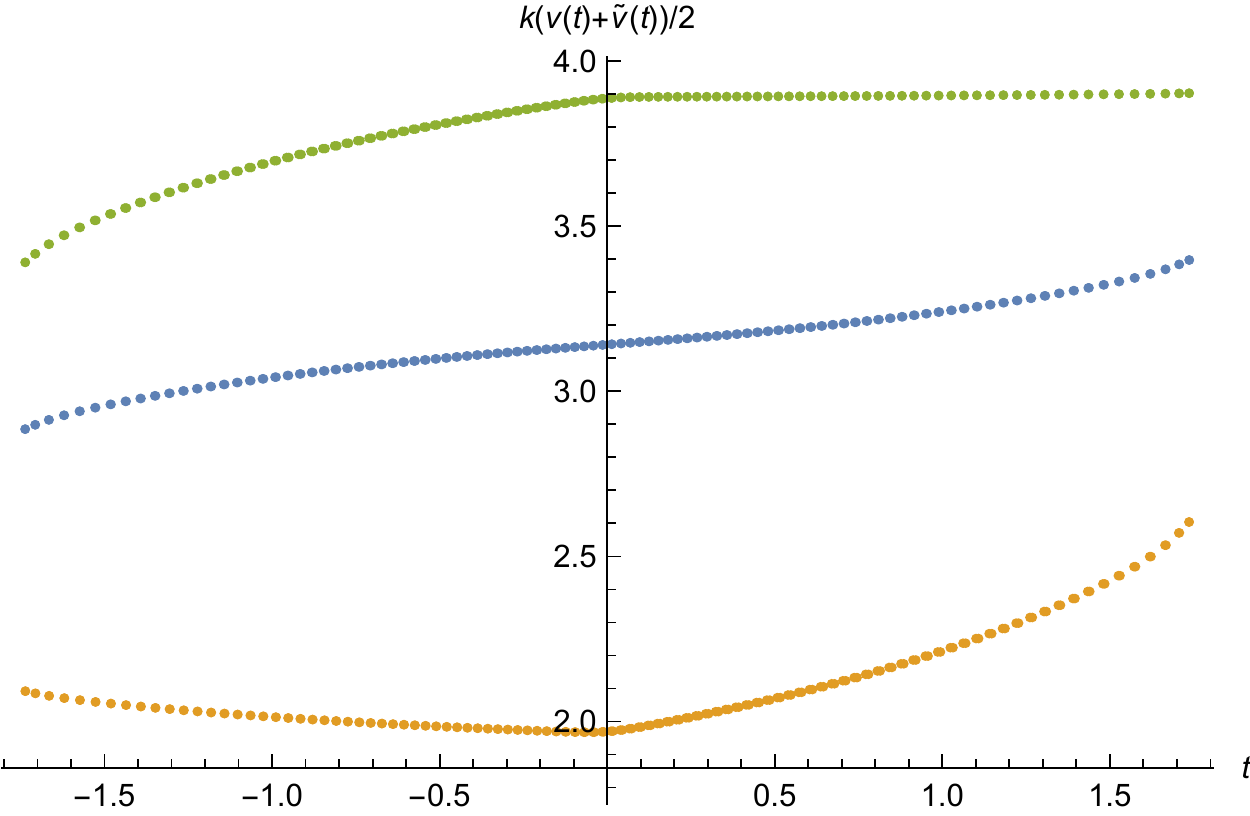}
	        \caption{Real part axis $k(v(t)+\tilde{v}(t))/2$}
	    \end{subfigure}
    \end{subfigure}
    \quad %add desired spacing between images, e. g. ~, \quad, \qquad, \hfill etc. 
    \begin{subfigure}[b]{0.2\textwidth}
        \includegraphics[width=\textwidth]{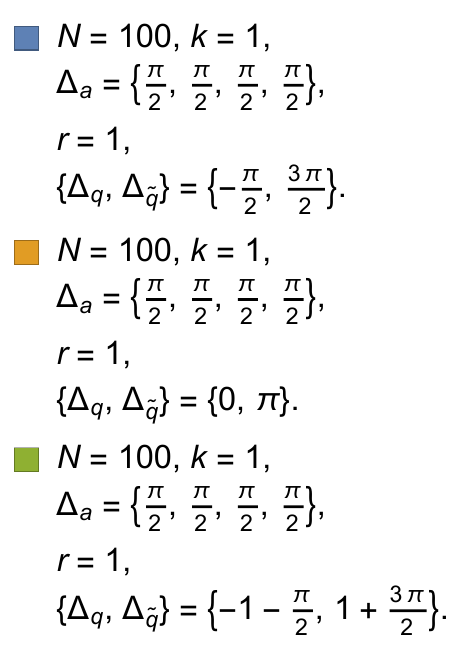}
    \end{subfigure}
    \caption{Eigenvalues for the special case $\Delta_a=\{\frac{\pi}{2}, \frac{\pi}{2}, \frac{\pi}{2}, \frac{\pi}{2}\}$ for $\{\Delta_q, \Delta_{\tilde{q}}\}=\{-\frac{\pi}{2},\frac{3\pi}{2}\}$ (blue), $\{0,\pi\}$ (orange) and $\{-1-\frac{\pi}{2},1+\frac{3\pi}{2}\}$ (green) with the same other parameters.}
    \label{fig:N010_Plot_Delq}
\end{figure}

\begin{figure}[H]
    \centering
    \begin{subfigure}[b]{0.75\textwidth}
	    \begin{subfigure}[b]{0.5\textwidth}
	        \includegraphics[width=\textwidth]{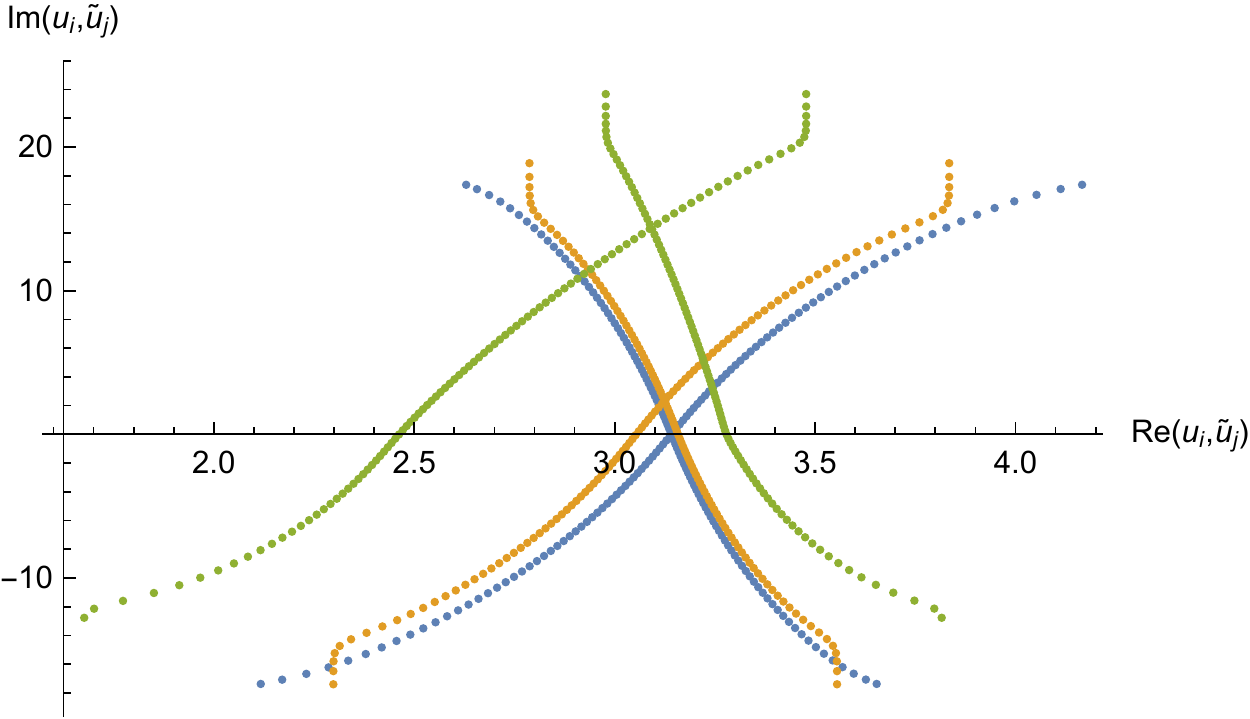}
	        \caption{Eigenvalue distribution}
	    \end{subfigure}
	    \quad %add desired spacing between images, e. g. ~, \quad, \qquad, \hfill etc. 
	    \begin{subfigure}[b]{0.5\textwidth}
	        \includegraphics[width=\textwidth]{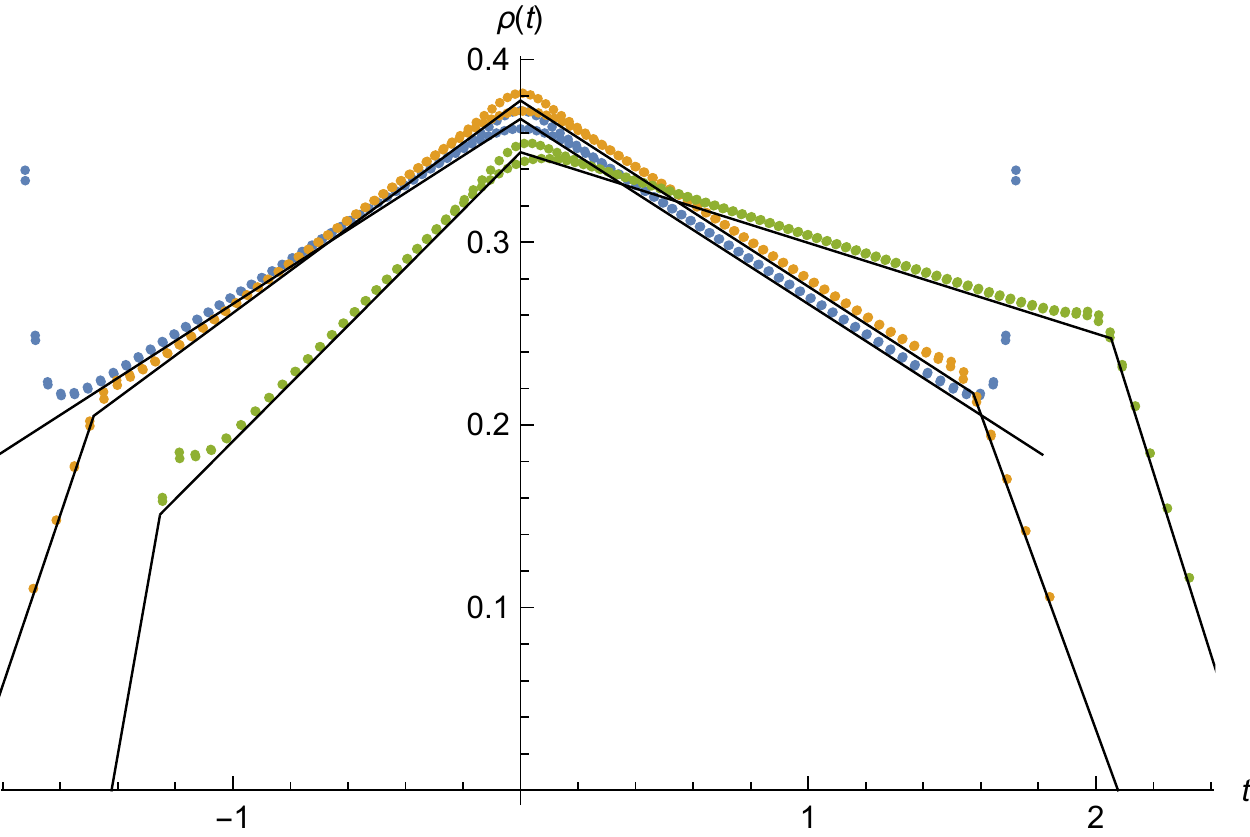}
	        \caption{Eigenvalue density $\rho(t)$}
	    \end{subfigure}\\
	    ~ %add desired spacing between images, e. g. ~, \quad, \qquad, \hfill etc. 
	    \begin{subfigure}[b]{0.5\textwidth}
	        \includegraphics[width=\textwidth]{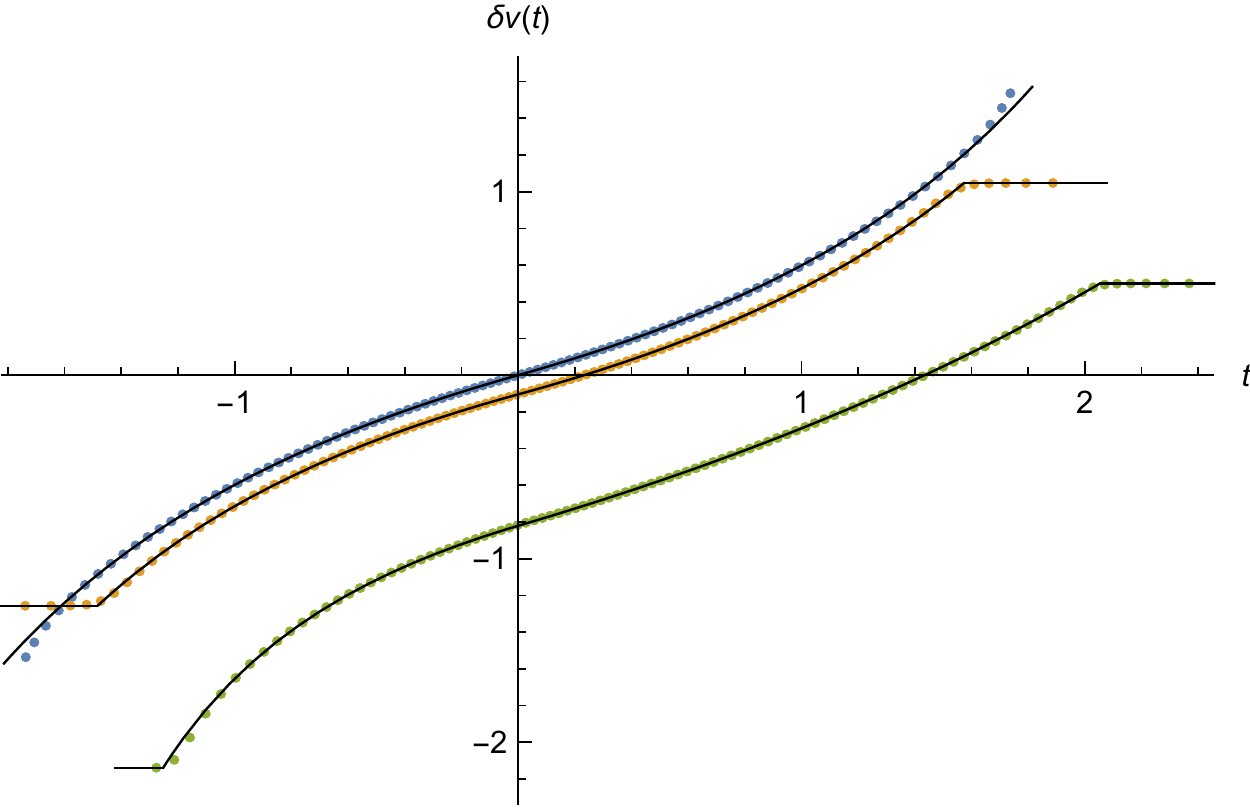}
	        \caption{Real part difference $\delta v(t)$}
	    \end{subfigure}
	    \quad %add desired spacing between images, e. g. ~, \quad, \qquad, \hfill etc. 
	    \begin{subfigure}[b]{0.5\textwidth}
	        \includegraphics[width=\textwidth]{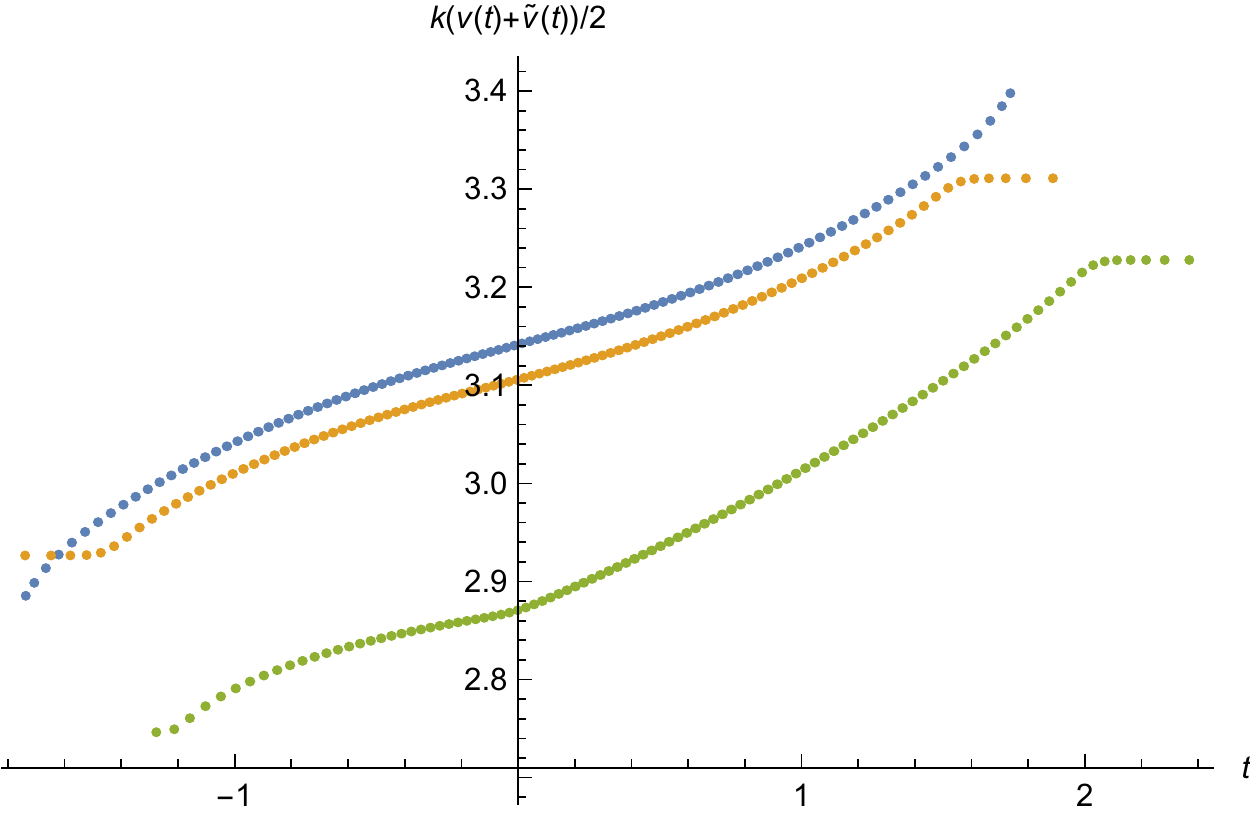}
	        \caption{Real part axis $k(v(t)+\tilde{v}(t))/2$}
	    \end{subfigure}
    \end{subfigure}
    \quad %add desired spacing between images, e. g. ~, \quad, \qquad, \hfill etc. 
    \begin{subfigure}[b]{0.2\textwidth}
        \includegraphics[width=\textwidth]{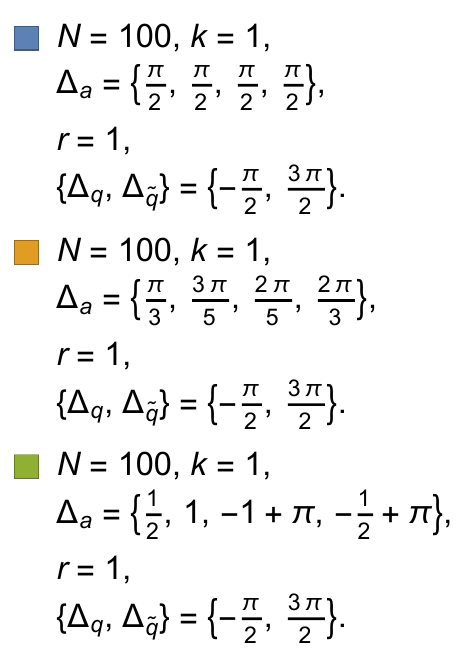}
    \end{subfigure}
    \caption{Eigenvalues for the special case $\Delta_a=\{\frac{\pi}{2}, \frac{\pi}{2}, \frac{\pi}{2}, \frac{\pi}{2}\}$ (blue), and the general cases $\Delta_a=\{\frac{\pi}{3}, \frac{3\pi}{5}, \frac{2\pi}{5}, \frac{2\pi}{3}\}$ (orange) and $\Delta_a=\{\frac{1}{2}, 1, -1+\pi, -\frac{1}{2}+\pi\}$ (green) with the same other parameters.}
    \label{fig:N010_Plot_Dela}
\end{figure}

The numerical solutions to the BAEs can be obtained using FindRoot in Mathematica as implemented in\cite{Liu:2017vll, PandoZayas:2019hdb}. In the following we focus on the case $k=1$. The numerical solutions for different values of $N$, $r$, $\{\Delta_q, \Delta_{\tilde{q}}\}$ and $\Delta_a$ are shown in Figures~\ref{fig:N010_Plot_N} - \ref{fig:N010_Plot_Dela}. The black lines are the analytical results in \cite{Hosseini:2016ume}. 

The numerical solutions show that the eigenvalues are not reflectively symmetric about $\pi$ alone the real axis as the ABJM theory. Furthermore, the imaginary part of $u_i$ is not exactly the same as $\tilde{u}_i$ so that there are two numerical results of the eigenvalue density $\rho(t)$, the real part difference $\delta v(t)$ and the real part axis $\left(v(t)+\tilde{v}(t)\right)/2$, though it is not obvious in the last two because of overlapping.

In Figure~\ref{fig:N010_Plot_N} we describe the eigenvalues as exact numerical solutions of the BAE Eq.~(\ref{Eq:BAEN010}). The plots show that the imaginary part of the eigenvalues scales as $N^{1/2}$. The eigenvalue densities are very well described by the leading analytical result of \cite{Hosseini:2016ume} except for some deviations at the edges of the intervals. 

Figure~\ref{fig:N010_Plot_r} explores the nature of the eigenvalues as one changes  the number of fundamental flavors $r$, the most prominent change is  accurately captured by the slope in the eigenvalue density. In Figure~\ref{fig:N010_Plot_Delq} we explore the effects of changing the fugacities $\{\Delta_q, \Delta_{\tilde{q}}\}$. The main effect is in the real part axis $\left(v(t)+\tilde{v}(t)\right)/2$.

Finally, in Figure~\ref{fig:N010_Plot_Dela} we explore the eigenvalues away from the symmetric point of the bi-fundamental fugacities. As in the ABJM case \cite{Liu:2017vll}, this is numerically challenging as one needs confront numerically, various numerical singularities due to branch points in the polylogarithmic functions. 

%%%%%%%%%%%%%%%%%%%%%%%%%%%%%%%%%%
\subsection{The subleading term of the index at large $N$}

Having achieved control of the eigenvalues we can proceed to analyze the index. As the ABJM theory, we expand the index beyond the leading order in $N$ and we expect the subleading behavior of the index to have the form
\be
{\rm Re}\log Z=f_1(k,r,\Delta,\mathfrak{n})N^{3/2}+f_2(k,r,\Delta,\mathfrak{n})N^{1/2}+f_3(k,r,\Delta,\mathfrak{n})\log{N}+f_4(k,r,\Delta,\mathfrak{n})+\mathcal O(N^{-1/2}),
\ee
where the functions $f_1$, $f_2$, $f_3$ and $f_4$ are linear in the magnetic fluxes $\mathfrak{n}$.

For $k=1$ and a given set of $\Delta_a$, $r$ and $\{\Delta_q, \Delta_{\tilde{q}}\}$, we can compute the index Eq.~(\ref{Eq:N010_Index}) and ${\rm Re}\log Z$ for a range of $N$ using the numerical solutions obtained in section \ref{Sec:N010_BAEs}. Then we decompose ${\rm Re}\log Z$ into a sum of four independent terms 
\be
\setlength\abovedisplayskip{10pt}
\setlength\belowdisplayskip{10pt}
{\rm Re}\log Z=A+B_1\mathfrak{n}_3+B_2\mathfrak{n}_4+B_3\mathfrak{n}_q,
\ee
where we have used the marginality condition on the superpotential $\mathfrak{n}_1+\mathfrak{n}_4=1$, $\mathfrak{n}_2+\mathfrak{n}_3=1$ and $\mathfrak{n}_q+\mathfrak{n}_{\tilde{q}}=1$. Then we perform a linear least-squares fit for $A$ and $B_a$ to the function
\be
\setlength\abovedisplayskip{10pt}
\setlength\belowdisplayskip{10pt}
f(N)=f_1N^{3/2}+f_2N^{1/2}+f_3\log{N}+f_4+\sum_{p=1}^{p_c} f_{p+4}\,\, N^{\left(1-2p\right)/2},
\label{Eq:N010FitlogZN}
\ee
where $p_c$ is the cutoff needed for the numerical fitting. Notice that the inverse powers of $N$ should be $N^{(1-2p)/2}$, instead of $N^{p/2}$, for a integer $p$ as before, because of the stability which has been checked numerically.

%%%%%%%%%%%%%%%%%%%%%%%%%%%%%%%%%%%%%%%

%%%%
\begin{table}[H]
\renewcommand{\arraystretch}{1.1}
\setlength{\tabcolsep}{5pt}
\begin{subtable}[H]{1\textwidth}
\centering
\caption{$k=1$;$\,\{\Delta_1, \Delta_2, \Delta_3, \Delta_4\}=\{\frac{\pi}{2}, \frac{\pi}{2}, \frac{\pi}{2}, \frac{\pi}{2}\}$.}
\begin{tabular}{|c|c c|c||l|l|l|l|}
\hline
$r$&$\Delta_q$&$\Delta_{\tilde{q}}$&$N(s)$&$f_1$&$f_2$&$f_3$&$f_4$\\
\hline
\multirow{9}{*}{$1$}&$-\frac{\pi}{2}$&$\frac{3\pi}{2}$&$100\sim300(10)$&$-2.41840$&$+2.11612$&$-0.50066$&$-2.29495$ \\
\cline{2-8}
&$0$&$\pi$&$100\sim300(10)$&$-2.41840$&\tabincell{l}{$+1.73825$\\$+0.15115\mathfrak{n}_3$\\$+0.15115\mathfrak{n}_4$\\$+0.45345\mathfrak{n}_q$}&$-0.50056$&$-2.29565$ \\
\cline{2-8}
&$-\frac{\pi}{2}-1$&$\frac{3\pi}{2}+1$&$100\sim300(10)$&$-2.41840$&\tabincell{l}{$+2.35669$\\$-0.09623\mathfrak{n}_3$\\$-0.09623\mathfrak{n}_4$\\$-0.28868\mathfrak{n}_q$}&$-0.50080$&$-2.29412$ \\
\hline
$2$&$-\frac{\pi}{2}$&$\frac{3\pi}{2}$&$100\sim300(10)$&$-3.14159$&$+2.08351$&$-0.50065$&$-2.49377$ \\
\hline
$3$&$-\frac{\pi}{2}$&$\frac{3\pi}{2}$&$100\sim300(10)$&$-3.74657$&$+2.38846$&$-0.50068$&$-3.10991$ \\
\hline
\end{tabular}
\end{subtable}
\vspace{3mm}

\begin{subtable}[H]{1\textwidth}
\centering
\caption{$k=1$;$\,\{\Delta_1, \Delta_2, \Delta_3, \Delta_4\}=\{\frac{\pi}{3}, \frac{3\pi}{5}, \frac{2\pi}{5}, \frac{2\pi}{3}\}$.}
\begin{tabular}{|c|c c|c||l|l|l|l|}
\hline
$r$&$\Delta_q$&$\Delta_{\tilde{q}}$&$N(s)$&$f_1$&$f_2$&$f_3$&$f_4$\\
\hline
\multirow{9}{*}{$1$}&$-\frac{\pi}{2}$&$\frac{3\pi}{2}$&$100\sim200(5)$&\tabincell{l}{$-2.45347$\\$-0.37599\mathfrak{n}_3$\\$+0.44419\mathfrak{n}_4$}&\tabincell{l}{$+2.16951$\\$-0.00208\mathfrak{n}_3$\\$+0.01685\mathfrak{n}_4$\\$+0.01959\mathfrak{n}_q$}&\tabincell{l}{$-0.50086$\\$-0.00002\mathfrak{n}_3$\\$+0.00002\mathfrak{n}_4$}&\tabincell{l}{$-2.37003$\\$-0.00489\mathfrak{n}_3$\\$+0.00569\mathfrak{n}_4$}\\
\cline{2-8}
&$0$&$\pi$&$85\sim105(1)$&\tabincell{l}{$-2.45347$\\$-0.37599\mathfrak{n}_3$\\$+0.44419\mathfrak{n}_4$}&\tabincell{l}{$+1.81164$\\$+0.16898\mathfrak{n}_3$\\$+0.13521\mathfrak{n}_4$\\$+0.46029\mathfrak{n}_q$}&\tabincell{l}{$-0.50107$\\$-0.00007\mathfrak{n}_3$\\$+0.00008\mathfrak{n}_4$\\$-0.00002\mathfrak{n}_q$}&\tabincell{l}{$-2.36898$\\$-0.00460\mathfrak{n}_3$\\$+0.00537\mathfrak{n}_4$\\$+0.00012\mathfrak{n}_q$}\\
\cline{2-8}
&$-\frac{\pi}{2}-1$&$\frac{3\pi}{2}+1$&$100\sim200(5)$&\tabincell{l}{$-2.45347$\\$-0.37599\mathfrak{n}_3$\\$+0.44419\mathfrak{n}_4$}&\tabincell{l}{$+2.40631$\\$-0.08786\mathfrak{n}_3$\\$-0.08581\mathfrak{n}_4$\\$-0.26097\mathfrak{n}_q$}&\tabincell{l}{$-0.50082$\\$-0.00001\mathfrak{n}_3$\\$+0.00003\mathfrak{n}_4$}&\tabincell{l}{$-2.37028$\\$-0.00493\mathfrak{n}_3$\\$+0.00566\mathfrak{n}_4$}\\
\hline
$2$&$-\frac{\pi}{2}$&$\frac{3\pi}{2}$&$90\sim110(1)$&\tabincell{l}{$-3.16053$\\$-0.46062\mathfrak{n}_3$\\$+0.49512\mathfrak{n}_4$}&\tabincell{l}{$+2.13744$\\$+0.19592\mathfrak{n}_3$\\$-0.16992\mathfrak{n}_4$\\$+0.02533\mathfrak{n}_q$}&\tabincell{l}{$-0.50098$\\$+0.00024\mathfrak{n}_3$\\$-0.00043\mathfrak{n}_4$\\$-0.00007\mathfrak{n}_q$}&\tabincell{l}{$-2.59551$\\$-0.12623\mathfrak{n}_3$\\$+0.12251\mathfrak{n}_4$\\$+0.00039\mathfrak{n}_q$}\\
\hline
$3$&$-\frac{\pi}{2}$&$\frac{3\pi}{2}$&$80\sim100(1)$&\tabincell{l}{$-3.75949$\\$-0.53656\mathfrak{n}_3$\\$+0.55794\mathfrak{n}_4$}&\tabincell{l}{$+2.47949$\\$+0.46309\mathfrak{n}_3$\\$-0.43345\mathfrak{n}_4$\\$+0.02703\mathfrak{n}_q$}&\tabincell{l}{$-0.50119$\\$+0.00691\mathfrak{n}_3$\\$-0.00646\mathfrak{n}_4$\\$+0.00146\mathfrak{n}_q$}&\tabincell{l}{$-3.29270$\\$-0.37592\mathfrak{n}_3$\\$+0.36784\mathfrak{n}_4$\\$-0.00802\mathfrak{n}_q$}\\
\hline
\end{tabular}
\end{subtable}
\end{table}

\clearpage

\begin{table}
\ContinuedFloat
\renewcommand{\arraystretch}{1.1}
\setlength{\tabcolsep}{5pt}
\begin{subtable}[H]{1\textwidth}
\centering
\caption{$k=1$;$\,\{\Delta_1, \Delta_2, \Delta_3, \Delta_4\}=\{\frac{1}{2}, 1, \pi-1, \pi-\frac{1}{2}\}$.}
\begin{tabular}{|c|c c|c||l|l|l|l|}
\hline
$r$&$\Delta_q$&$\Delta_{\tilde{q}}$&$N(s)$&$f_1$&$f_2$&$f_3$&$f_4$\\
\hline
\multirow{9}{*}{$1$}&$-\frac{\pi}{2}$&$\frac{3\pi}{2}$&$100\sim140(2)$&\tabincell{l}{$-2.69204$\\$+0.02322\mathfrak{n}_3$\\$+0.51870\mathfrak{n}_4$}&\tabincell{l}{$+2.10645$\\$+0.05070\mathfrak{n}_3$\\$+0.05726\mathfrak{n}_4$\\$+0.15070\mathfrak{n}_q$}&\tabincell{l}{$-0.50101$\\$-0.00001\mathfrak{n}_3$}&\tabincell{l}{$-2.42730$\\$+0.00051\mathfrak{n}_3$\\$+0.01181\mathfrak{n}_4$\\$+0.00005\mathfrak{n}_q$}\\
\cline{2-8}
&$0$&$\pi$&$100\sim120(1)$&\tabincell{l}{$-2.69204$\\$+0.02322\mathfrak{n}_3$\\$+0.51870\mathfrak{n}_4$}&\tabincell{l}{$+1.79254$\\$+0.19251\mathfrak{n}_3$\\$+0.14797\mathfrak{n}_4$\\$+0.58329\mathfrak{n}_q$}&\tabincell{l}{$-0.50079$\\$-0.00008\mathfrak{n}_3$\\$-0.00002\mathfrak{n}_q$}&\tabincell{l}{$-2.42886$\\$+0.00089\mathfrak{n}_3$\\$+0.01185\mathfrak{n}_4$\\$+0.00010\mathfrak{n}_q$}\\
\cline{2-8}
&$-\frac{\pi}{2}-1$&$\frac{3\pi}{2}+1$&$100\sim120(1)$&\tabincell{l}{$-2.69204$\\$+0.02322\mathfrak{n}_3$\\$+0.51870\mathfrak{n}_4$}&\tabincell{l}{$+2.33490$\\$-0.04105\mathfrak{n}_3$\\$-0.03333\mathfrak{n}_4$\\$-0.12470\mathfrak{n}_q$}&\tabincell{l}{$-0.50108$\\$+0.00002\mathfrak{n}_3$\\$-0.00005\mathfrak{n}_4$}&\tabincell{l}{$-2.42691$\\$+0.00031\mathfrak{n}_3$\\$+0.01213\mathfrak{n}_4$\\$-0.00003\mathfrak{n}_q$}\\
\hline
$2$&$-\frac{\pi}{2}$&$\frac{3\pi}{2}$&$80\sim180(5)$&\tabincell{l}{$-3.28009$\\$-0.14052\mathfrak{n}_3$\\$+0.41670\mathfrak{n}_4$}&\tabincell{l}{$+1.91465$\\$+0.19945\mathfrak{n}_3$\\$-0.00011\mathfrak{n}_4$\\$+0.20126\mathfrak{n}_q$}&\tabincell{l}{$-0.50031$\\$-0.00026\mathfrak{n}_3$\\$-0.00042\mathfrak{n}_4$\\$-0.00040\mathfrak{n}_q$}&\tabincell{l}{$-2.52837$\\$-0.07857\mathfrak{n}_3$\\$+0.04734\mathfrak{n}_4$\\$+0.00239\mathfrak{n}_q$}\\
\hline
$3$&$-\frac{\pi}{2}$&$\frac{3\pi}{2}$&$100\sim140(2)$&\tabincell{l}{$-3.83240$\\$-0.23232\mathfrak{n}_3$\\$+0.40375\mathfrak{n}_4$}&\tabincell{l}{$+2.18910$\\$+0.36535\mathfrak{n}_3$\\$-0.13462\mathfrak{n}_4$\\$+0.21699\mathfrak{n}_q$}&\tabincell{l}{$-0.50565$\\$+0.00191\mathfrak{n}_3$\\$+0.00215\mathfrak{n}_4$\\$+0.00343\mathfrak{n}_q$}&\tabincell{l}{$-3.11766$\\$-0.20891\mathfrak{n}_3$\\$+0.14117\mathfrak{n}_4$\\$-0.01975\mathfrak{n}_q$}\\
\hline
\end{tabular}
\end{subtable}
\caption{($N^{0,1,0}$) Numerical fit for ${\rm Re}\log Z=f_1N^{3/2}+f_2N^{1/2}+f_3\log{N}+f_4+\sum_{p=1}^{p_c=5} N^{\left(1-2p\right)/2}$. The $s$ in the bracket of $N$ is the step of $N$.}
\label{tbl:N010_FitRelogZ}
\end{table}
%%%%
The results of the numerical fit for ${\rm Re}\log Z$ with $N$ are presented in Table~\ref{tbl:N010_FitRelogZ}. The error between the analytical leading term computed by the index theorem in  \cite{Hosseini:2016ume} and the numerical leading term $f_1N^{3/2}$ is negligible. More precisely,  the analytic leading order and the numerical result match to number of significant digits present in the table. The leading term is, indeed, independent of $\{\Delta_q, \Delta_{\tilde{q}}\}$. The numerical results indicate that the coefficient $f_3$ of the $\log N$ term is precisely $-1/2$.

The main result of this section is the numerical evidence pointing to the presence of a correction of the form $-\frac{1}{2}\log N$ in the topologically twisted index. 

%%%%%%%%%%%%%%%%%%%%%%%%%%%%%%%%%%%%%%%%%%%%%%%%%%%%%%%
\section{The topologically twisted index of $V^{5,2}$}\label{Sec:V52}
%%%%%%%%%%
One particularly interesting model is the field theory dual to AdS$_4\times V^{5,2}/\mathbb{Z}_k$ becuase the manifold $V^{5,2}$ is non-toric. A simple way to visualize this seven-dimensional manifold is as a homogeneous space $V^{5,2}=SO(5)/SO(3)$. There are two models for the dual field theory. Following the literature, we call model I, the proposal of Martelli and Sparks \cite{Martelli:2009ga} and model II the proposal of Jafferis \cite{Jafferis:2009th}. The free energy on $S^3$ of the field theories was discussed in \cite{Cheon:2011vi,Martelli:2011qj,Jafferis:2011zi} and perfect agreement at leading order was found with the dual supergravity solutions. The topologically twisted index for both models was studied at leading large $N$ order in \cite{Hosseini:2016ume} where the authors established the equivalence of both models. Here we go beyond the leading order in $N$ and demonstrate the equivalence of both models at the level of the topologically twisted index up to, and including, logarithmic in $N$ terms.

\subsection{Model I}
Model I was originally proposed in \cite{Martelli:2009ga} with the following quiver diagram:
\bea
\begin{tikzpicture}[baseline, font=\footnotesize, scale=0.8]
\begin{scope}[auto,%
  every node/.style={draw, minimum size=0.5cm}, node distance=2cm];
  % the vertices
\node[circle] (USp2k) at (-0.1, 0) {$N_{+k}$};
\node[circle, right=of USp2k] (BN)  {$N_{-k}$};
\end{scope}
  % the edges
\draw[solid,line width=0.2mm,<-]  (USp2k) to[bend right=15] node[midway,above] {$B_2 $}node[midway,above] {}  (BN) ;
\draw[solid,line width=0.2mm,->]  (USp2k) to[bend right=50] node[midway,above] {$A_1$}node[midway,above] {}  (BN) ; 
\draw[solid,line width=0.2mm,<-]  (USp2k) to[bend left=15] node[midway,above] {$B_1$} node[midway,above] {} (BN) ;  
\draw[solid,line width=0.2mm,->]  (USp2k) to[bend left=50] node[midway,above] {$A_2$} node[midway,above] {} (BN) ;    
\draw[black,-> ] (USp2k) edge [out={-150},in={150},loop,looseness=10] (USp2k) node at (-2,1) {$\phi_1$} ;
\draw[black,-> ] (BN) edge [out={-30},in={30},loop,looseness=10] (BN) node at (5.8,1) {$\phi_2$};
\end{tikzpicture}
\eea
The superpotential accompanying the quiver diagram is 
\begin{equation}
 W = \Tr\left[ \phi_1^3 + \phi_2^3 +\phi_1(A_1 B_2 + A_2 B_1) + \phi_2 (B_2 A_1+ B_1 A_2) \right] \, .
\end{equation}

\subsubsection{Numerical solutions to the system of BAEs}

Collecting all the relevant building blocks following from the quiver diagram we have
\bea
Z&=&\frac{1}{(N!)^2}\sum\limits_{\mathfrak{m},\tilde{\mathfrak{m}}\in \mathbb{Z}^N}\int_{\cal C}\prod\limits_{i=1}^N
\frac{dx_i}{2\pi i x_i}\frac{d\tx_i}{2\pi i \tx_i} x_i^{k\mathfrak{m}_i}\tx_i^{-k\tilde{\mathfrak{m}}_i} \times \prod\limits_{i\neq j}^N\left(1-\frac{x_i}{x_j}\right)\left(1-\frac{\tx_i}{\tx_j}\right) \nonumber\\
&&\times \prod\limits_{i,j=1}^N\left[\prod\limits_{a=1,2}
\left(\frac{\sqrt{\frac{x_i}{\tx_j}y_a}}{1-\frac{x_i}{\tx_j}y_a}\right)^{\mathfrak{m}_i-\tilde{\mathfrak{m}}_j-\mathfrak{n}_a+1}
\prod\limits_{b=3,4}
\left(\frac{\sqrt{\frac{\tx_j}{x_i}y_b}}{1-\frac{\tx_j}{x_i}y_b}\right)^{\tilde{\mathfrak{m}}_j-\mathfrak{m}_i-\mathfrak{n}_b+1} \right.\nonumber\\
&&\left.\times 
\left(\frac{\sqrt{\frac{x_i}{x_j}y_{\phi_1}}}{1-\frac{x_i}{x_j}y_{\phi_1}}\right)^{\mathfrak{m}_i-\frac{1}{2}\mathfrak{n}_{\phi_1}+\frac{1}{2}}
\left(\frac{\sqrt{\frac{x_j}{x_i}y_{\phi_1}}}{1-\frac{x_j}{x_i}y_{\phi_1}}\right)^{-\mathfrak{m}_i-\frac{1}{2}\mathfrak{n}_{\phi_1}+\frac{1}{2}} \right. \nonumber\\
&&\left.\times 
\left(\frac{\sqrt{\frac{\tx_i}{\tx_j}y_{\phi_2}}}{1-\frac{\tx_i}{\tx_j}y_{\phi_2}}\right)^{-\tilde{\mathfrak{m}}_j-\frac{1}{2}\mathfrak{n}_{\phi_2}+\frac{1}{2}}
\left(\frac{\sqrt{\frac{\tx_j}{\tx_i}y_{\phi_2}}}{1-\frac{\tx_j}{\tx_i}y_{\phi_2}}\right)^{\tilde{\mathfrak{m}}_j-\frac{1}{2}\mathfrak{n}_{\phi_2}+\frac{1}{2}} \right].
\eea

Performing the summation over magnetic fluxes, as in previous cases, by  introducing a large cut-off $M$ we get
\bea
Z&=&\frac{1}{(N!)^2}\int_{\cal C}\prod\limits_{i=1}^N
\frac{dx_i}{2\pi i x_i}\frac{d\tx_i}{2\pi i \tx_i}\prod\limits_{i\neq j}^N\left(1-\frac{x_i}{x_j}\right)\left(1-\frac{\tx_i}{\tx_j}\right) \nonumber\\
&&\times \prod\limits_{i,j=1}^N\left[\prod\limits_{a=1,2}
\left(\frac{\sqrt{\frac{x_i}{\tx_j}y_a}}{1-\frac{x_i}{\tx_j}y_a}\right)^{1-\mathfrak{n}_a}
\prod\limits_{b=3,4}
\left(\frac{\sqrt{\frac{\tx_j}{x_i}y_b}}{1-\frac{\tx_j}{x_i}y_b}\right)^{1-\mathfrak{n}_b} \right.\nonumber\\
&&\left.\times
\left(\frac{\sqrt{y_{\phi_1}}}{1-\frac{x_j}{x_i}y_{\phi_1}}\right)^{1-\mathfrak{n}_{\phi_1}}
\left(\frac{\sqrt{y_{\phi_2}}}{1-\frac{\tx_i}{\tx_j}y_{\phi_2}}\right)^{1-\mathfrak{n}_{\phi_2}} \right] \times \prod\limits_{i=1}^N\frac{\big(e^{iB_i}\big)^M}{e^{iB_i}-1} \prod\limits_{j=1}^N\frac{\big(e^{i\tilde{B}_j}\big)^M}{e^{i\tilde{B}_j}-1},
\eea
where the Bethe Ansatz equations are
\bea
1&=&e^{iB_i}=x_i^k\prod\limits_{j=1}^N\frac{\left(1-y_3 \frac{\tilde{x}_j}{x_i}\right)\left(1-y_4 \frac{\tilde{x}_j}{x_i}\right)}{\left(1-y_1^{-1} \frac{\tilde{x}_j}{x_i}\right)\left(1-y_2^{-1} \frac{\tilde{x}_j}{x_i}\right)} \left(\frac{1}{\sqrt{y_1 y_2 y_3 y_4}}\right) \left(\frac{x_i-x_j y_{\phi_1}}{x_j-x_i y_{\phi_1}}\right), \nonumber\\
1&=&e^{i\tilde{B}_j}=\tilde{x}_j^k\prod\limits_{i=1}^N\frac{\left(1-y_3 \frac{\tilde{x}_j}{x_i}\right)\left(1-y_4 \frac{\tilde{x}_j}{x_i}\right)}{\left(1-y_1^{-1} \frac{\tilde{x}_j}{x_i}\right)\left(1-y_2^{-1} \frac{\tilde{x}_j}{x_i}\right)} \left(\frac{1}{\sqrt{y_1 y_2 y_3 y_4}}\right) \left(\frac{\tx_i-\tx_j y_{\phi_2}}{\tx_j-\tx_i y_{\phi_2}}\right).
\eea
We highlight the ambiguity of selecting a branch as discussed around Eq.~(\ref{Eq:BAPotential_bifund}) by explicitly keeping the product $\sqrt{y_1y_2y_3y_4}$ in the above expression. 

The compact expression for the index, once the solutions to the BAE are known, is

\bea
\label{Eq:V52_Model1_Index}
Z\left(y_a,\mathfrak{n}_a\right)&=&\left(\prod_{a=1}^4 y_a^{-\frac{1}{2}N^2 \mathfrak{n}_a} \right)
y_{\phi_1}^{-\frac{1}{2}N^2\mathfrak{n}_{\phi_1}}y_{\phi_2}^{-\frac{1}{2}N^2\mathfrak{n}_{\phi_2}}
\nonumber\\
&&\times\sum_{I\in BAE}\left[\frac{1}{\det\mathbb{B}}
\frac{\prod_{i=1}^N x_i^{\frac{1}{2}N\left(6-\sum_{a=1}^4\mathfrak{n}_a-2\mathfrak{n}_{\phi_1}\right)} \tilde{x}_i^{\frac{1}{2}N\left(6-\sum_{a=1}^4\mathfrak{n}_a-2\mathfrak{n}_{\phi_2}\right)}}{\prod_{i,j=1}^N\prod_{a=1,2}\left(\tilde{x}_j-y_ax_i\right)^{1-\mathfrak{n}_a}\prod_{a=3,4}\left(x_i-y_a\tilde{x}_j\right)^{1-\mathfrak{n}_a}} \right.\nonumber\\
&&\left.\times
\frac{\prod_{i\neq j}\left(1-\frac{x_i}{x_j}\right)\left(1-\frac{\tilde{x}_i}{\tilde{x}_j}\right)}{\prod_{i,j=1}^N\left(x_i-x_j y_{\phi_1}\right)^{1-\mathfrak{n}_{\phi_1}}\left(\tx_j-\tx_i y_{\phi_2}\right)^{1-\mathfrak{n}_{\phi_2}}}\right].
\eea

The transformation matrix $\mathbb{B}$ is given by

\be
\mathbb{B}\Big|_{\rm BAEs}=
\left(
\begin{array}{cc}
\mathbb{B}_{jl} & G_{jl}\\
-G_{lj} & \tilde{\mathbb{B}}_{jl}
\end{array}
\right),
\ee
where 

\bea
\mathbb{B}_{jl}&=&\delta_{jl}\left[k-\sum\limits_{m=1}^N G_{jm}+x_j\sum\limits_{m=1}^N\left(\frac{1-\delta_{jm} y_{\phi_1}}{x_j-x_m y_{\phi_1}}-\frac{\delta_{jm}-y_{\phi_1}}{x_m-x_j y_{\phi_1}}\right) \right] \nonumber\\
&&-\delta_{jl}\left[\frac{y_{\phi_1}+1}{y_{\phi_1}-1}\right]+x_l\left[\frac{1}{x_l-x_j y_{\phi_1}^{-1}}-\frac{1}{x_l-x_j y_{\phi_1}}\right],
\eea

\bea
\tilde{\mathbb{B}}_{jl}&=&\delta_{jl}\left[k+\sum\limits_{m=1}^N G_{mj}+\tilde{x}_j\sum\limits_{m=1}^N\left(\frac{\delta_{jm}-y_{\phi_2}}{\tilde{x}_m-\tilde{x}_j y_{\phi_2}}-\frac{1-\delta_{jm} y_{\phi_2}}{\tilde{x}_j-\tilde{x}_m y_{\phi_2}}\right) \right] \nonumber\\
&&-\delta_{jl}\left[\frac{1+y_{\phi_2}}{1-y_{\phi_2}}\right]+\tilde{x}_l\left[\frac{1}{\tilde{x}_l-\tilde{x}_j y_{\phi_2}}-\frac{1}{\tilde{x}_l-\tilde{x}_j y_{\phi_2}^{-1}}\right].
\eea

The Bethe potential is
\bea
\label{Eq:N010BethePotential}
\mathcal{V}&=&\sum\limits_{i=1}^N\left[\frac{k}{2}\left(\tilde{u}_i^2-u_i^2\right)-2\pi \left(\tilde{n}_i \tilde{u}_i-n_iu_i\right)\right]\nonumber \\
&&+ \sum\limits_{i,j=1}^N\left[\sum\limits_{a=3,4}{\rm Li}_2\left(e^{i\left(\tilde{u}_j-u_i+\Delta_a\right)}\right)-
\sum\limits_{a=1,2}{\rm Li}_2\left(e^{i\left(\tilde{u}_j-u_i-\Delta_a\right)}\right) - \frac{1}{2}\left(\sum\limits_{a=1}^4\Delta_a-2\pi\right)\left(\tilde{u}_j-u_i\right)\right]\nonumber \\
&&+ \sum\limits_{i,j=1}^N\frac{1}{2}\left[{\rm Li}_2\left(e^{i\left(u_j-u_i+\Delta_{\phi_1}\right)}\right)-
{\rm Li}_2\left(e^{i\left(u_j-u_i-\Delta_{\phi_1}\right)}\right)\right]\nonumber \\
&&+ \sum\limits_{i,j=1}^N\frac{1}{2}\left[{\rm Li}_2\left(e^{i\left(\tilde{u}_j-\tilde{u}_i+\Delta_{\phi_2}\right)}\right)-
{\rm Li}_2\left(e^{i\left(\tilde{u}_j-\tilde{u}_i-\Delta_{\phi_2}\right)}\right)\right],
\eea
where
\bea
\sum\limits_{i=1}^N\left[-2\pi \left(\tilde{n}_i \tilde{u}_i-n_iu_i\right)\right]&=&\left(6\pi-\sum\limits_{a=1}^4\Delta_a-2\Delta_{\phi_2}\right)\sum\limits_{i>j}^N\tilde{u}_j-\left(6\pi-\sum\limits_{a=1}^4\Delta_a-2\Delta_{\phi_1}\right)\sum\limits_{i>j}^N u_i\nonumber \\
&=&2\pi\sum\limits_{i>j}^N\left(\tilde{u}_j-u_i\right).
\eea

\begin{figure}[H]
    \centering
    \begin{subfigure}[b]{0.75\textwidth}
	    \begin{subfigure}[b]{0.5\textwidth}
	        \includegraphics[width=\textwidth]{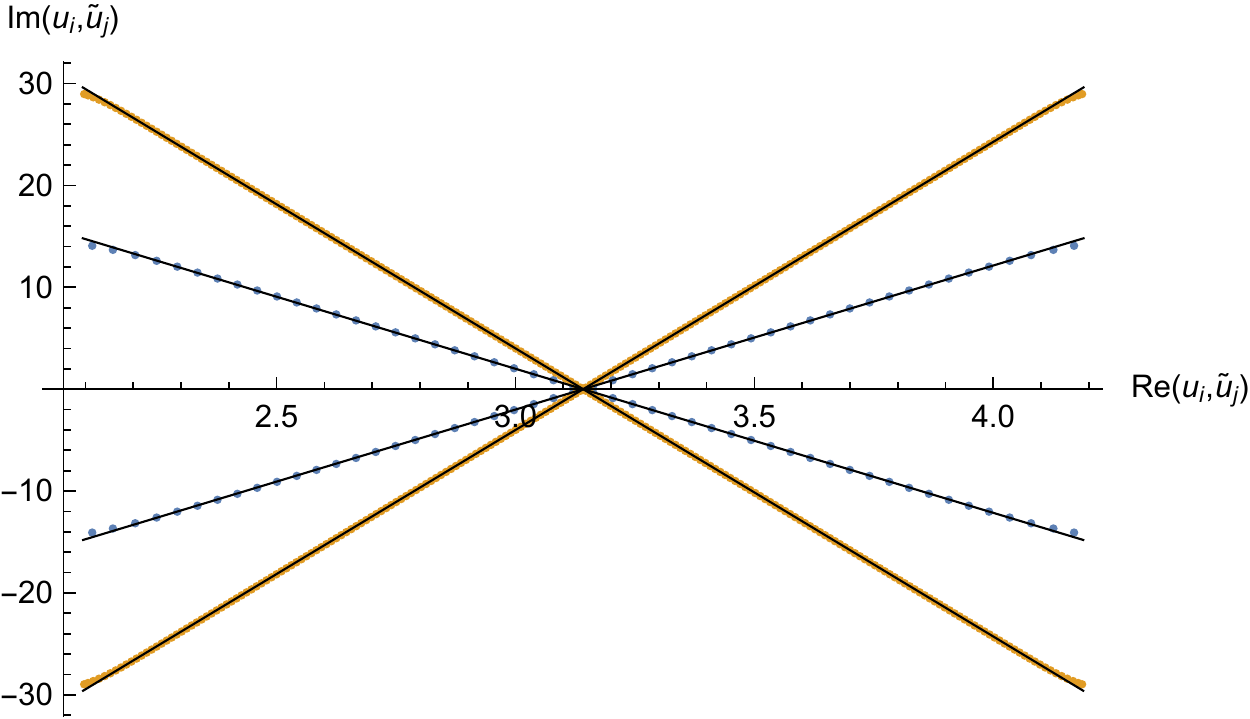}
	        \caption{Eigenvalue distribution}
	    \end{subfigure}
	    \quad %add desired spacing between images, e. g. ~, \quad, \qquad, \hfill etc. 
	    \begin{subfigure}[b]{0.5\textwidth}
	        \includegraphics[width=\textwidth]{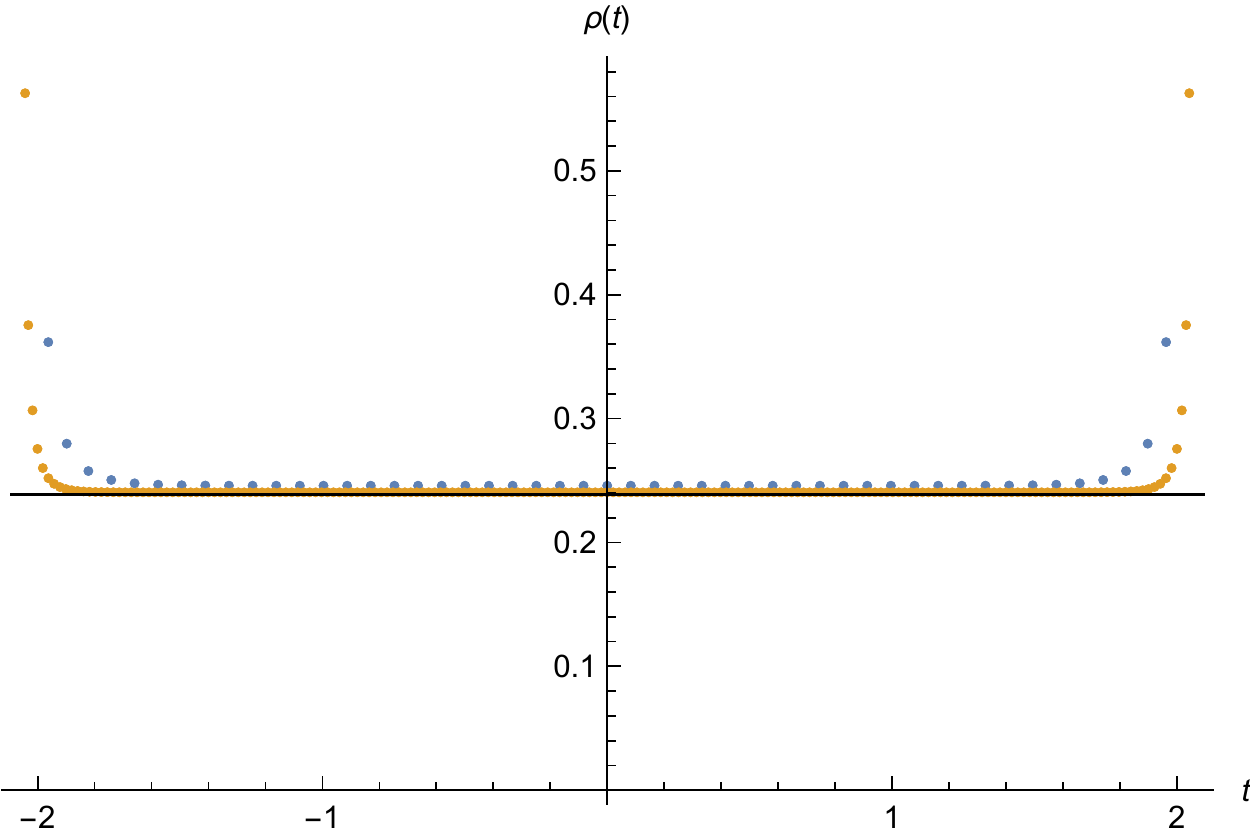}
	        \caption{Eigenvalue density $\rho(t)$}
	    \end{subfigure}\\
	    ~ %add desired spacing between images, e. g. ~, \quad, \qquad, \hfill etc. 
	    \begin{subfigure}[b]{0.5\textwidth}
	        \includegraphics[width=\textwidth]{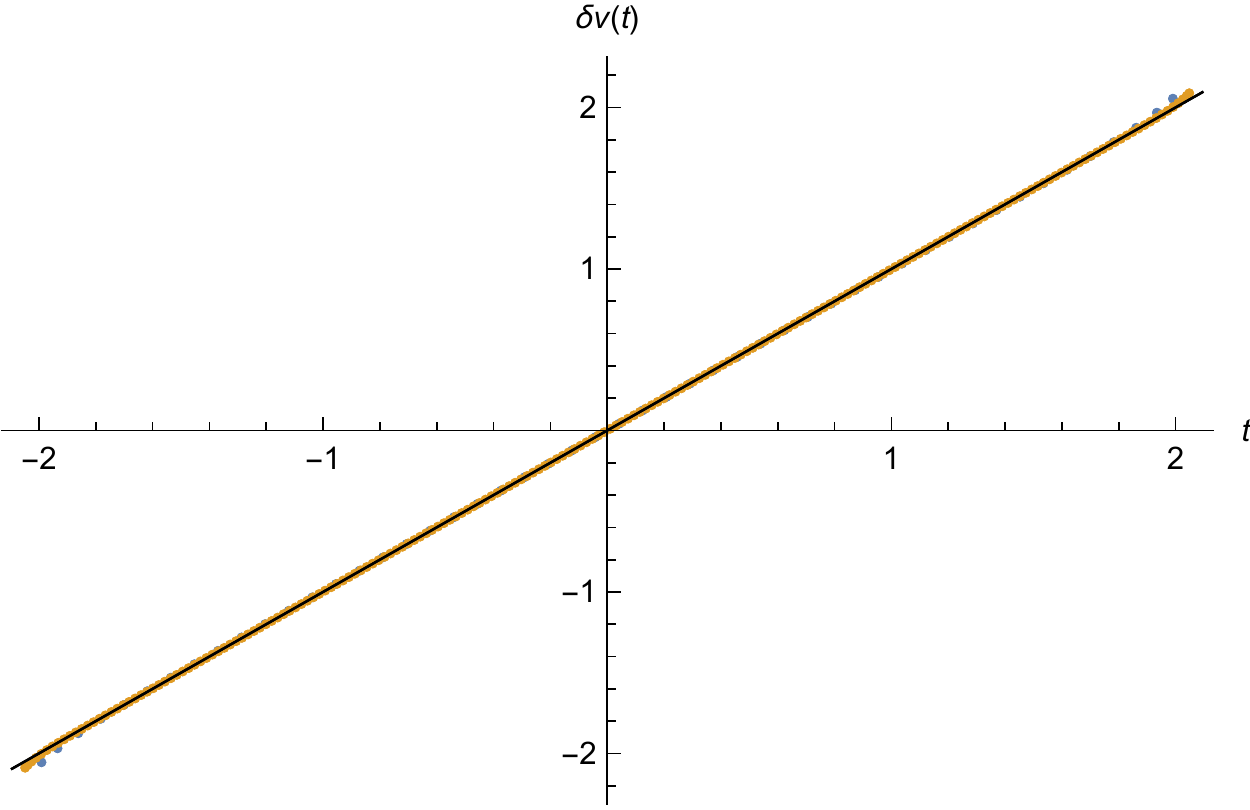}
	        \caption{Real part difference $\delta v(t)$}
	    \end{subfigure}
	    \quad %add desired spacing between images, e. g. ~, \quad, \qquad, \hfill etc. 
	    \begin{subfigure}[b]{0.5\textwidth}
	        \includegraphics[width=\textwidth]{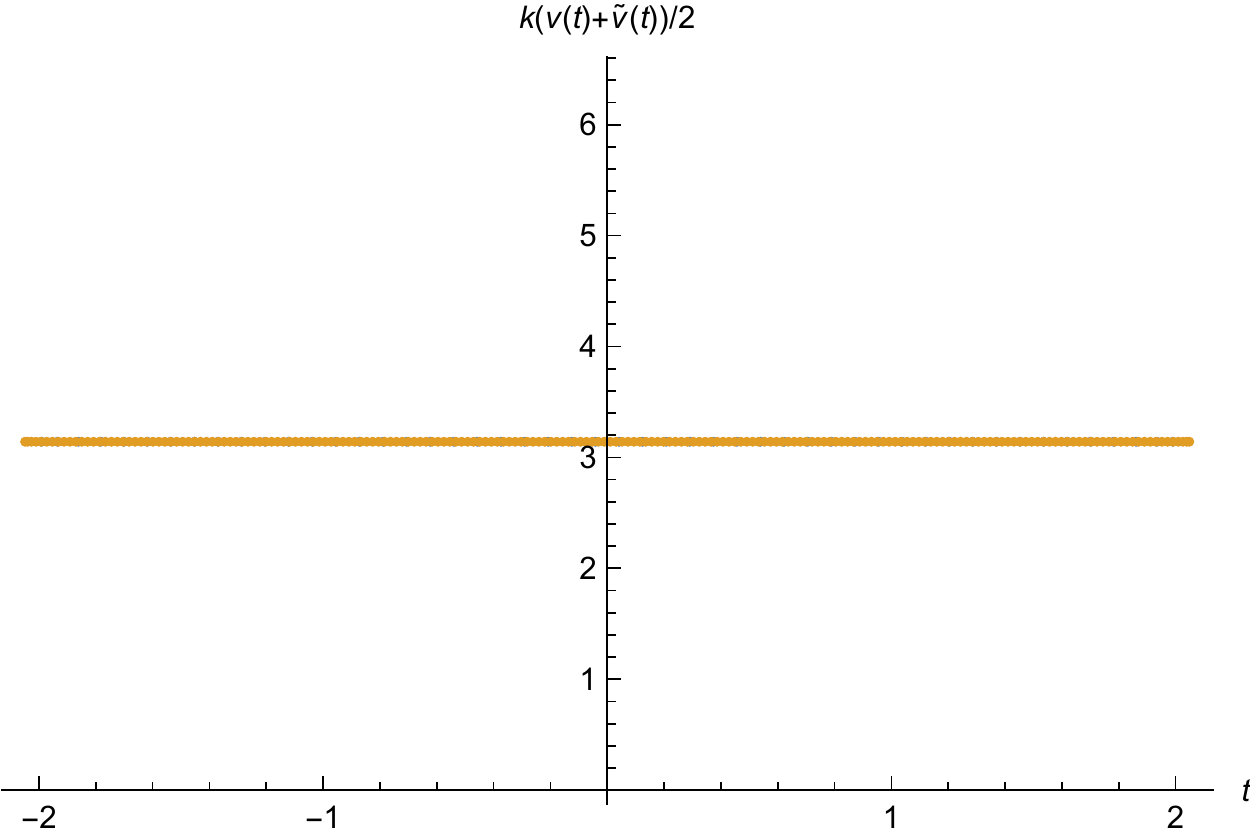}
	        \caption{Real part axis $k(v(t)+\tilde{v}(t))/2$}
	    \end{subfigure}
    \end{subfigure}
    \quad %add desired spacing between images, e. g. ~, \quad, \qquad, \hfill etc. 
    \begin{subfigure}[b]{0.2\textwidth}
        \includegraphics[width=\textwidth]{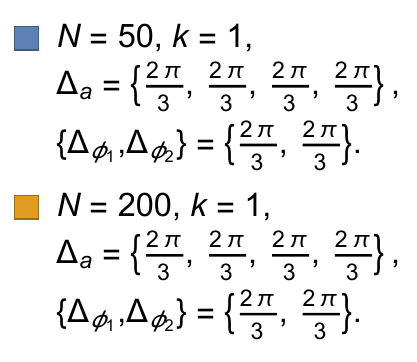}
    \end{subfigure}
    \caption{Eigenvalues for the special case $\Delta_a=\{\frac{2\pi}{3}, \frac{2\pi}{3}, \frac{2\pi}{3}, \frac{2\pi}{3}\}$ for $N=50$ (blue) and $200$ (orange) with the same other parameters.}
    \label{fig:V52_Model1_Plot_N}
\end{figure}

\begin{figure}[H]
    \centering
    \begin{subfigure}[b]{0.75\textwidth}
	    \begin{subfigure}[b]{0.5\textwidth}
	        \includegraphics[width=\textwidth]{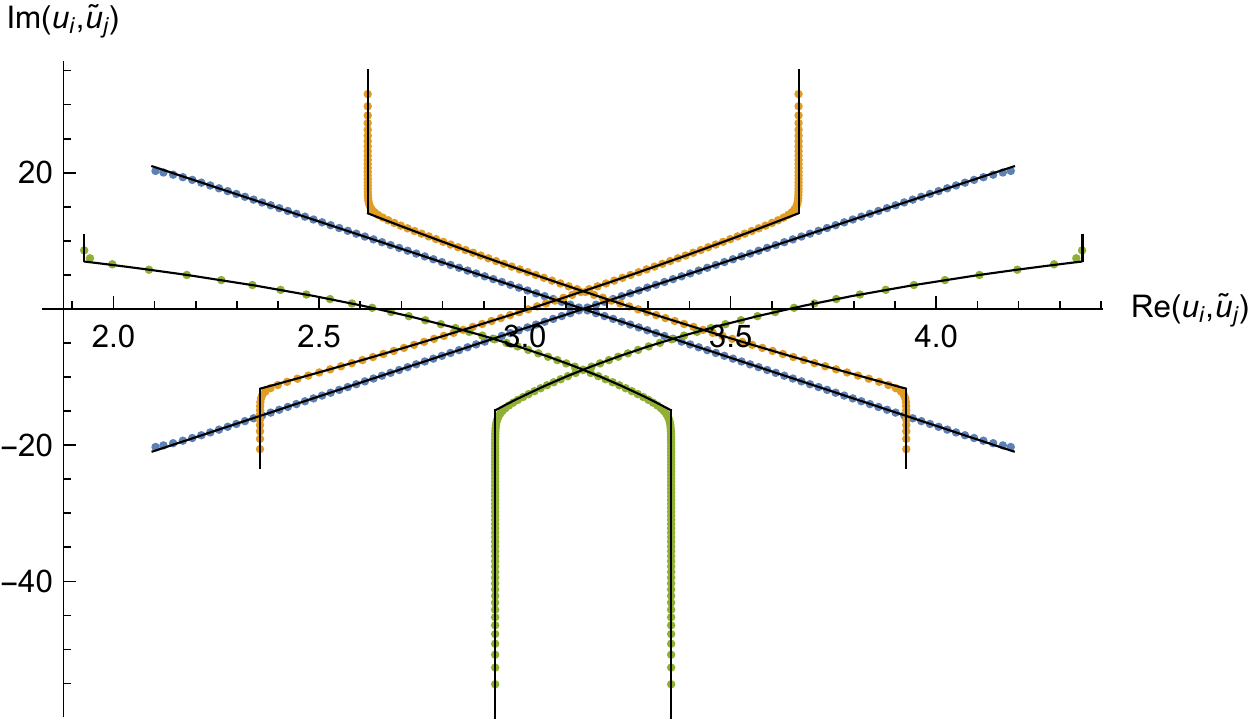}
	        \caption{Eigenvalue distribution}
	    \end{subfigure}
	    \quad %add desired spacing between images, e. g. ~, \quad, \qquad, \hfill etc. 
	    \begin{subfigure}[b]{0.5\textwidth}
	        \includegraphics[width=\textwidth]{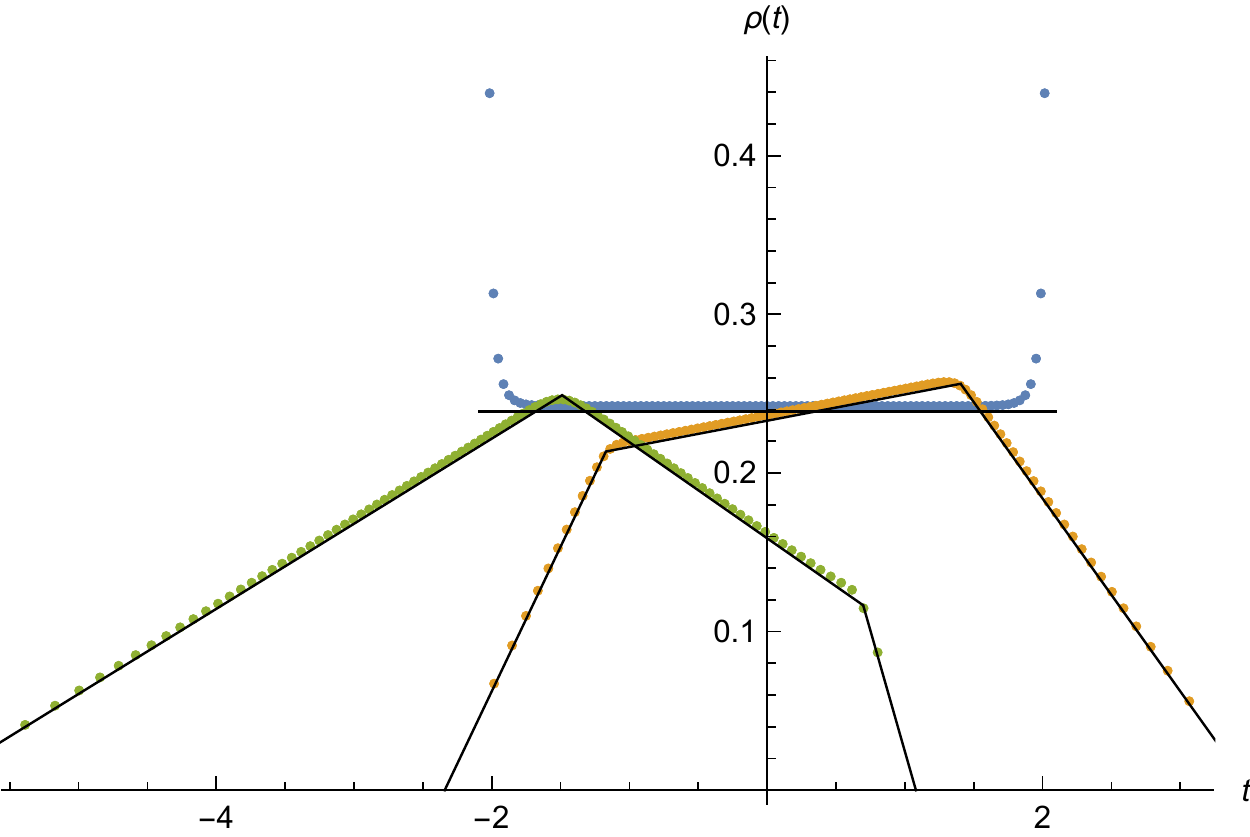}
	        \caption{Eigenvalue density $\rho(t)$}
	    \end{subfigure}\\
	    ~ %add desired spacing between images, e. g. ~, \quad, \qquad, \hfill etc. 
	    \begin{subfigure}[b]{0.5\textwidth}
	        \includegraphics[width=\textwidth]{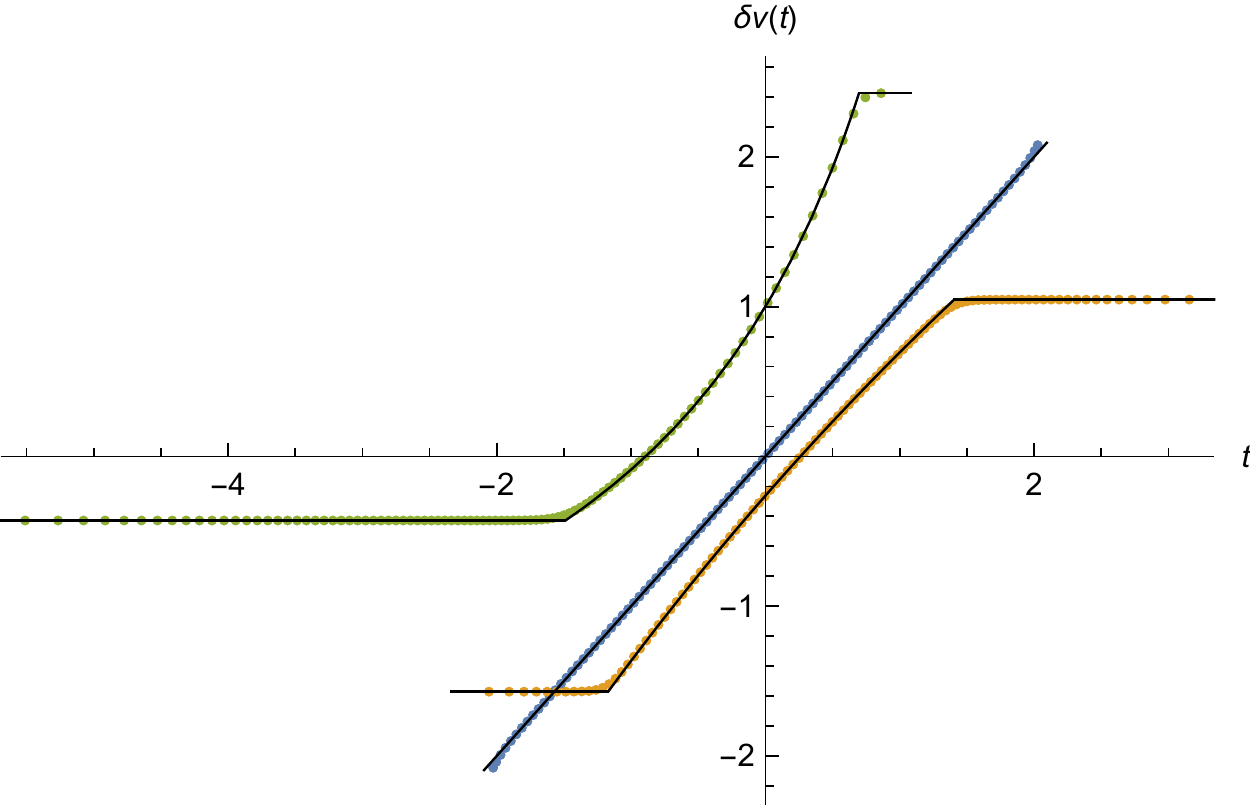}
	        \caption{Real part difference $\delta v(t)$}
	    \end{subfigure}
	    \quad %add desired spacing between images, e. g. ~, \quad, \qquad, \hfill etc. 
	    \begin{subfigure}[b]{0.5\textwidth}
	        \includegraphics[width=\textwidth]{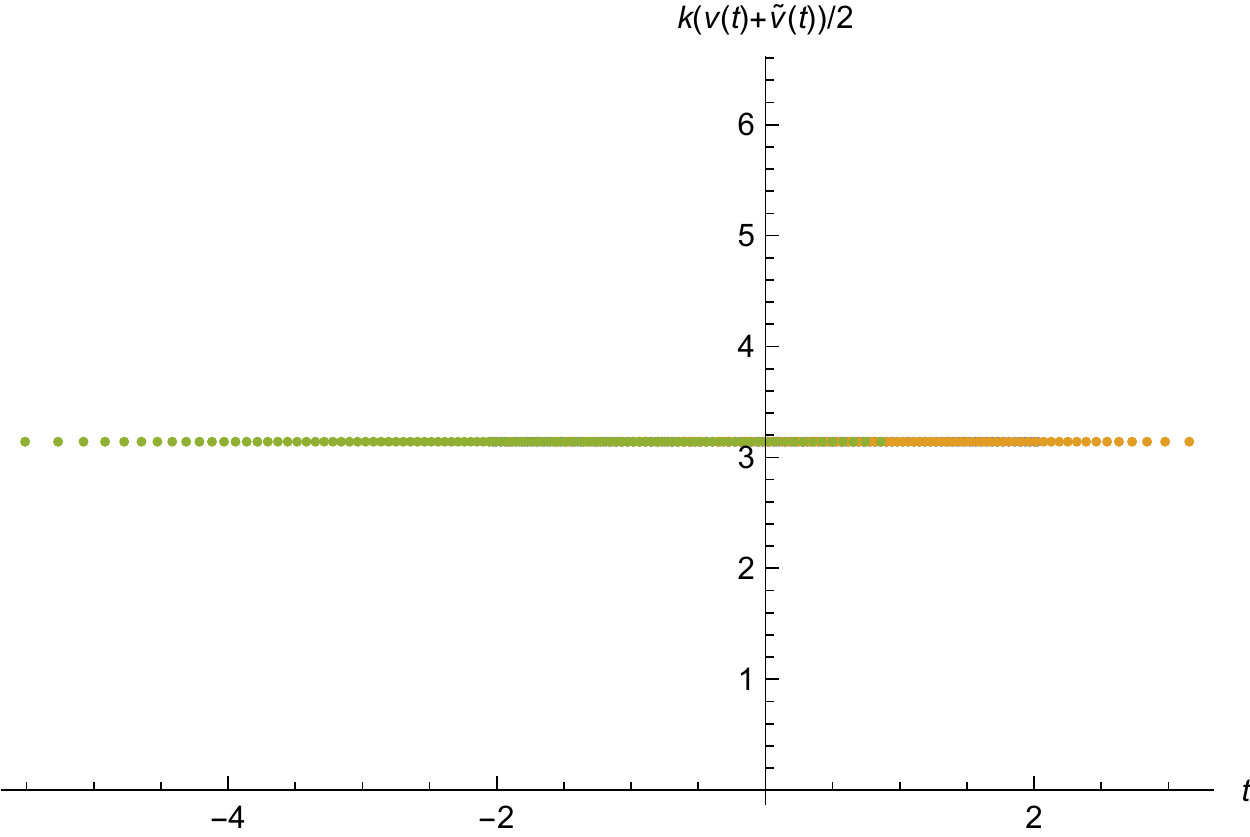}
	        \caption{Real part axis $k(v(t)+\tilde{v}(t))/2$}
	    \end{subfigure}
    \end{subfigure}
    \quad %add desired spacing between images, e. g. ~, \quad, \qquad, \hfill etc. 
    \begin{subfigure}[b]{0.2\textwidth}
        \includegraphics[width=\textwidth]{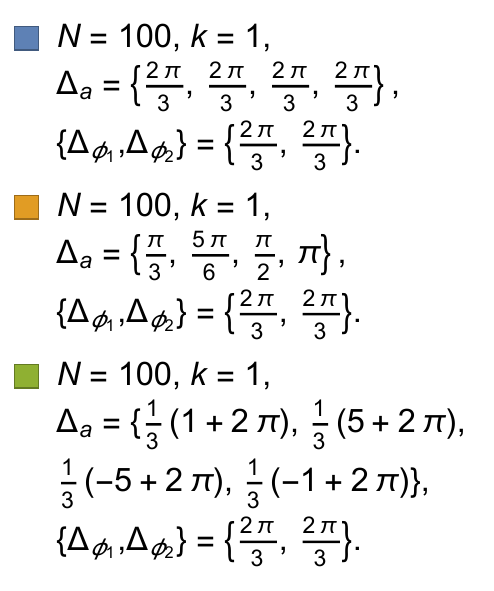}
    \end{subfigure}
    \caption{Eigenvalues for the special case $\Delta_a=\{\frac{2\pi}{3}, \frac{2\pi}{3}, \frac{2\pi}{3}, \frac{2\pi}{3}\}$ (blue), and the general case $\Delta_a=\{\frac{\pi}{3}, \frac{5\pi}{6}, \frac{\pi}{2}, \pi\}$ (orange) and $\Delta_a=\{\frac{1}{3}+\frac{2\pi}{3}, \frac{5}{3}+\frac{2\pi}{3}, -\frac{5}{3}+\frac{2\pi}{3}, -\frac{1}{3}+\frac{2\pi}{3}\}$ (green) with the same other parameters.}
    \label{fig:V52_Model1_Plot_Dela}
\end{figure}

We focus on the case $k=1$ and set the initial real part axis to be $\pi$. The numerical solutions for different values of $N$ and $\Delta_a=\{\Delta_1, \Delta_2, \Delta_3, \Delta_4\}$ are shown in Figures~\ref{fig:V52_Model1_Plot_N} and \ref{fig:V52_Model1_Plot_Dela}. The main new ingredients, as compared to the ABJM case, are the fugacities $\{\Delta_{\phi_1}, \Delta_{\phi_2}\} = \{\frac{2\pi}{3}, \frac{2\pi}{3}\}$. 

Figure~\ref{fig:V52_Model1_Plot_N} displays the eigenvalues and their imaginary and real part densities as a function of $N$. The choice of $N=50$ and $N=200$ is meant to indicate clearly that the imaginary part of the eigenvalues scales as $N^{1/2}$ in the large $N$ limit; this can be easily noted by glancing at the axes in panel $(a)$ of Figure~\ref{fig:V52_Model1_Plot_N}.

Figure~\ref{fig:V52_Model1_Plot_Dela} explores the dependence of the eigenvalues on the choice of fugacities $\{\Delta_a\}$. As expected from the ABJM  construction, and more general configuration of tails in the distribution of the eigenvalues emerges. The agreement with the leading order large $N$ distribution is maintained with deviations registered mostly along the piece-wise discontinuous slopes.

\subsubsection{The subleading term of the index at large $N$}

The index should take the form
\be
{\rm Re}\log Z=f_1(k,\Delta,\mathfrak{n})N^{3/2}+f_2(k,\Delta,\mathfrak{n})N^{1/2}+f_3(k,\Delta,\mathfrak{n})\log{N}+f_4(k,\Delta,\mathfrak{n})+\mathcal O(N^{-1/2}),
\ee
where the functions $f_1$, $f_2$, $f_3$ and $f_4$ are linear in the magnetic fluxes $\mathfrak{n}$.

The index Eq.~(\ref{Eq:V52_Model1_Index}) and ${\rm Re}\log Z$ can be computed using the numerical solutions. Under the similar decomposition
\be
{\rm Re}\log Z=A+B_1\mathfrak{n}_3+B_2\mathfrak{n}_4,
\ee
where we have used the marginality condition on the superpotential $\mathfrak{n}_1+\mathfrak{n}_4=4/3$, $\mathfrak{n}_2+\mathfrak{n}_3=4/3$ and $\mathfrak{n}_{\phi_1}=\mathfrak{n}_{\phi_2}=2/3$. Then we perform a linear least-square fit for $A$ and $B_a$ to the function
\be
f(N)=f_1N^{3/2}+f_2N^{1/2}+f_3\log{N}+f_4+\sum_{p=1}^{p_c} f_{p+4}\,\,N^{\left(1-2p\right)/2}.
\label{Eq:N010FitlogZN}
\ee

The results of the numerical fit for ${\rm Re}\log Z$ with $N$ are presented in Table~\ref{tbl:V52_Model1_FitRelogZ}. The analytical leading term computed by the index theorem in \cite{Hosseini:2016ume} and the numerical leading term $f_1N^{3/2}$ match to number of significant digits present in the table. The numerical results indicate that the coefficient $f_3$ of the $\log N$ term is precisely $-1/2$.
%%%%
\begin{table}[H]
\renewcommand{\arraystretch}{1.1}
\setlength{\tabcolsep}{5pt}
\begin{subtable}[H]{1\textwidth}
\centering
\caption{$k=1$;$\,\{\Delta_1, \Delta_2, \Delta_3, \Delta_4\}=\{\frac{2\pi}{3}, \frac{2\pi}{3}, \frac{2\pi}{3}, \frac{2\pi}{3}\}$;$\,\{\Delta_{\phi_1}, \Delta_{\phi_2}\}=\{\frac{2\pi}{3}, \frac{2\pi}{3}\}$.}
\begin{tabular}{|c||l|l|l|l|}
\hline
$N(s)$&$f_1$&$f_2$&$f_3$&$f_4$\\
\hline
$100\sim300(10)$&$-1.86168$&$+3.02526$&$-0.50066$&$-2.75740$ \\
\hline
\end{tabular}
\end{subtable}
\vspace{3mm}

\begin{subtable}[H]{1\textwidth}
\centering
\caption{$k=1$;$\,\{\Delta_1, \Delta_2, \Delta_3, \Delta_4\}=\{\frac{\pi}{3}, \frac{5\pi}{6}, \frac{\pi}{2}, \pi\}$;$\,\{\Delta_{\phi_1}, \Delta_{\phi_2}\}=\{\frac{2\pi}{3}, \frac{2\pi}{3}\}$.}
\begin{tabular}{|c||l|l|l|l|}
\hline
$N(s)$&$f_1$&$f_2$&$f_3$&$f_4$\\
\hline
$100\sim300(10)$&\tabincell{l}{$-2.18550$\\$-0.31221\mathfrak{n}_3$\\$+0.78053\mathfrak{n}_4$}&\tabincell{l}{$+3.56708$\\$+0.00781\mathfrak{n}_3$\\$-0.17562\mathfrak{n}_4$}&$-0.50082$&\tabincell{l}{$-3.39532$\\$-0.02821\mathfrak{n}_3$\\$+0.19180\mathfrak{n}_4$} \\
\hline
\end{tabular}
\end{subtable}
\vspace{3mm}

\begin{subtable}[H]{1\textwidth}
\centering
\caption{$k=1$;$\,\{\Delta_1, \Delta_2, \Delta_3, \Delta_4\}=\{\frac{2\pi}{3}+\frac{1}{3}, \frac{2\pi}{3}+\frac{5}{3}, \frac{2\pi}{3}-\frac{5}{3}, \frac{2\pi}{3}-\frac{1}{3}\}$;$\,\{\Delta_{\phi_1}, \Delta_{\phi_2}\}=\{\frac{2\pi}{3}, \frac{2\pi}{3}\}$.}
\begin{tabular}{|c||l|l|l|l|}
\hline
$N(s)$&$f_1$&$f_2$&$f_3$&$f_4$\\
\hline
$100\sim200(5)$&\tabincell{l}{$-0.79002$\\$-1.81136\mathfrak{n}_3$\\$-0.13631\mathfrak{n}_4$}&\tabincell{l}{$+4.27652$\\$+1.38166\mathfrak{n}_3$\\$-0.06990\mathfrak{n}_4$}&\tabincell{l}{$-0.50167$\\$-0.00005\mathfrak{n}_3$\\$-0.00001\mathfrak{n}_4$}&\tabincell{l}{$-4.51009$\\$-1.91182\mathfrak{n}_3$\\$+0.01935\mathfrak{n}_4$} \\
\hline
\end{tabular}
\end{subtable}

\caption{($V^{5,2}$ Model I) Numerical fit for ${\rm Re}\log Z=f_1N^{3/2}+f_2N^{1/2}+f_3\log{N}+f_4+\sum_{p=1}^{p_c=5}f_{p+4}\, N^{\left(1-2p\right)/2}$.}
\label{tbl:V52_Model1_FitRelogZ}
\end{table}
%%%%

\subsection{Model II}
Model II was originally proposed in \cite{Jafferis:2009th} with the following quiver diagram:
\bea
\begin{tikzpicture}[font=\footnotesize, scale=0.8]
\begin{scope}[auto,%
  every node/.style={draw, minimum size=0.5cm}, node distance=2cm];
\node[circle]  (UN)  at (0.3,1.7) {$N$};
\node[rectangle, right=of UN] (Ur) {$k$};
\end{scope}
\draw[decoration={markings, mark=at position 0.45 with {\arrow[scale=2.5]{>}}, mark=at position 0.5 with {\arrow[scale=2.5]{>}}, mark=at position 0.55 with {\arrow[scale=2.5]{>}}}, postaction={decorate}, shorten >=0.7pt] (-0,2) arc (30:335:0.75cm);
\draw[draw=black,solid,line width=0.2mm,->]  (UN) to[bend right=30] node[midway,below] {$Q$}node[midway,above] {}  (Ur) ; 
\draw[draw=black,solid,line width=0.2mm,<-]  (UN) to[bend left=30] node[midway,above] {$\tQ$} node[midway,above] {} (Ur) ;    
\node at (-2.2,1.7) {$\varphi_{1,2,3}$}; \label{Diagram:ModelII}
\end{tikzpicture}
\eea
The superpotential is taken to be
\begin{equation}\label{superpotential fundamental}
 W = \Tr \left\{ \varphi_3 \left[ \varphi_1, \varphi_2 \right] + \sum_{j=1}^{k} q_j \left( \varphi_1^2 + \varphi_2^2 + \varphi_3^2 \right) \tilde q^j \right\} \, .
\end{equation}
The $SO(5)$ symmetry of $V^{5,2}$ can be made manifest by using the variables
\be
X_1=\frac{1}{\sqrt{2}}\left(\varphi_1+i\varphi_2\right),\qquad X_2=\frac{1}{\sqrt{2}}\left(\varphi_1-i\varphi_2\right),\qquad X_3=i\varphi_3.
\ee
In terms of these new variables, the superpotential can be rewritten as
\be
\label{superpotential fundamental re}
 W = \Tr \left\{ X_3 \left[ X_1, X_2 \right] + \sum_{j=1}^{k} q_j \left( X_1 X_2 + X_2 X_1 - X_3^2 \right) \tilde q^j \right\} \, .
\ee

\subsubsection{Numerical solutions to the system of BAEs}
The above matter content implies that the topologically twisted index takes the general form 
\bea
Z&=&\frac{1}{N!}\sum\limits_{\mathfrak{m}\in \mathbb{Z}^N}\int_{\cal C}\prod\limits_{i=1}^N
\frac{dx_i}{2\pi i x_i} x_i^{\mathfrak{t}} \xi^{\mathfrak{m}_i} \times \prod\limits_{i\neq j}^N\left(1-\frac{x_i}{x_j}\right) \nonumber\\
&&\times \prod\limits_{i,j=1}^N\prod\limits_{a=1}^3\left(\frac{\sqrt{\frac{x_i}{x_j}y_{X_a}}}{1-\frac{x_i}{x_j}y_{X_a}}\right)^{\mathfrak{m}_i-\frac{1}{2}\mathfrak{n}_{X_a}+\frac{1}{2}}
\left(\frac{\sqrt{\frac{x_j}{x_i}y_{X_a}}}{1-\frac{x_j}{x_i}y_{X_a}}\right)^{-\mathfrak{m}_i-\frac{1}{2}\mathfrak{n}_{X_a}+\frac{1}{2}} \nonumber\\
&&\times \prod\limits_{i=1}^N\prod\limits_{j=1}^k\Bigg(\frac{\sqrt{x_i y_{q_j}}}{1-x_i y_{q_j}}\Bigg)^{\mathfrak{m}_i-\mathfrak{n}_{q_j}+1}\Bigg(\frac{\sqrt{\frac{1}{x_i} \tilde{y}_{q_j}}}{1-\frac{1}{x_i} \tilde{y}_{q_j}}\Bigg)^{-\mathfrak{m}_i-\tilde{\mathfrak{n}}_{q_j}+1}.
\eea

Performing the summation over magnetic fluxes by introducing a large cut-off $M$ we get
\bea
Z&=&\frac{1}{N!}\sum\limits_{\mathfrak{m}\in \mathbb{Z}^N}\int_{\cal C}\prod\limits_{i=1}^N
\frac{dx_i}{2\pi i x_i} x_i^{\mathfrak{t}} \times \prod\limits_{i\neq j}^N\left(1-\frac{x_i}{x_j}\right) \times \prod\limits_{i,j=1}^N\prod\limits_{a=1}^3
\left(\frac{\sqrt{y_{X_a}}}{1-\frac{x_j}{x_i}y_{X_a}}\right)^{1-\mathfrak{n}_{X_a}} \nonumber\\
&&\times \prod\limits_{i=1}^N\prod\limits_{j=1}^k\Bigg(\frac{\sqrt{x_i y_{q_j}}}{1-x_i y_{q_j}}\Bigg)^{1-\mathfrak{n}_{q_j}}\Bigg(\frac{\sqrt{\frac{1}{x_i} \tilde{y}_{q_j}}}{1-\frac{1}{x_i} \tilde{y}_{q_j}}\Bigg)^{1-\tilde{\mathfrak{n}}_{q_j}} \times \prod\limits_{i=1}^N\frac{\big(e^{iB_i}\big)^M}{e^{iB_i}-1},
\eea
where the Bethe Ansatz equations are
\bea
1&=&e^{iB_i}=\xi \prod\limits_{a=1}^3\prod\limits_{j=1}^N\left(\frac{x_i-x_j y_{X_a}}{x_j-x_i y_{X_a}}\right)\times \prod\limits_{j=1}^k\Bigg(\frac{\sqrt{x_i y_{q_j}}}{1-x_i y_{q_j}}\Bigg) \Bigg(\frac{\sqrt{\frac{1}{x_i} \tilde{y}_{q_j}}}{1-\frac{1}{x_i} \tilde{y}_{q_j}}\Bigg)^{-1}.
\eea

The compact expression for the index once the BA solutions are known is
\bea
\label{Eq:V52_Model2_Index}
Z\left(y_a,\mathfrak{n}_a\right)&=&e^{i\frac{\pi}{3}N k}\prod_{a=1}^3 y_{X_a}^{-\frac{1}{2}N^2 \mathfrak{n}_{X_a}} \prod\limits_{j=1}^k y_{q_j}^{-\frac{1}{2}N\mathfrak{n}_{q_j}}\tilde{y}_{q_j}^{-\frac{1}{2}N\tilde{\mathfrak{n}}_{q_j}}\nonumber\\
&&\times\sum_{I\in BAE}\left[\frac{1}{\det\mathbb{B}}
\frac{\prod_{i=1}^N x_i^{N+\mathfrak{t}} \prod_{i\neq j}\left(1-\frac{x_i}{x_j}\right)}{\prod_{i,j=1}^N\prod_{a=1}^3\left(x_i-x_j y_{X_a}\right)^{1-\mathfrak{n}_{X_a}}} \right.\nonumber\\
&&\left.\times\prod_{i=1}^N \frac{x_i^{\frac{2}{3}k}}{\prod_{j=1}^k \left(1-x_i y_{q_j}\right)^{\left(1-\mathfrak{n}_{q_j}\right)}\left(x_i-\tilde{y}_{q_j}\right)^{\left(1-\tilde{\mathfrak{n}}_{q_j}\right)}}\right]. \eea

The matrix $\mathbb{B}$ is
\be
\mathbb{B}\Big|_{\rm BAEs}=
\left(
\begin{array}{cc}
\mathbb{B}_{jl}
\end{array}
\right),
\ee
where 
\bea
\mathbb{B}_{jl}&=&\delta_{jl}\left[x_j\sum\limits_{a=1}^3\sum\limits_{m=1}^N\left(\frac{1-\delta_{jm} y_{X_a}}{x_j-x_m y_{X_a}}-\frac{\delta_{jm}-y_{X_a}}{x_m-x_j y_{X_a}}\right)+x_j\sum\limits_{i=1}^k\left(\frac{1}{x_j-\tilde{y}_{q_i}}-\frac{1}{x_j-y_{q_i}^{-1}}\right) \right] \nonumber\\
&&-\delta_{jl}\left[\sum\limits_{a=1}^3\frac{y_{X_a}+1}{y_{X_a}-1}\right]+x_l\sum\limits_{a=1}^3\left[\frac{1}{x_l-x_j y_{X_a}^{-1}}-\frac{1}{x_l-x_j y_{X_a}}\right].
\eea

The Bethe potential is
\bea
\label{Eq:N010BethePotential}
\mathcal{V}&=&\sum\limits_{i=1}^N\left[-\Delta_m u_i+2\pi n_i u_i\right]+\frac{1}{2}\sum\limits_{a=1}^3\sum\limits_{i,j=1}^N\left[{\rm Li}_2\left(e^{i\left(u_j-u_i+\Delta_{X_a}\right)}\right)-
{\rm Li}_2\left(e^{i\left(u_j-u_i-\Delta_{X_a}\right)}\right)\right]\nonumber \\
&&+\sum\limits_{i=1}^N\sum\limits_{j=1}^k\left[{\rm Li}_2\left(e^{i\left(-u_i+\tilde{\Delta}_{q_j}\right)}\right)-{\rm Li}_2\left(e^{i\left(-u_i-\Delta_{q_j}\right)}\right)\right]\nonumber \\
&&+ \frac{1}{2}\sum\limits_{i=1}^N\sum\limits_{j=1}^k\left[\left(\tilde{\Delta}_{q_j}+\Delta_{q_j}-2\pi\right)u_i\right],
\eea
where
\be
\sum\limits_{i=1}^N 2\pi n_i u_i=\left(3\pi-\sum\limits_{a=1}^3\Delta_{X_a}\right)\sum\limits_{i>j}^N\left(u_j-u_i\right)=\pi\sum\limits_{i>j}^N\left(u_j-u_i\right).
\ee
Notice that for even $N$ we are left with a common factor $\pi\sum_{i=1}^N u_i$ which can be reabsorbed in the definition of the topological fugacity $\xi$ \cite{Hosseini:2016tor}.

For simplicity we set $\Delta_{q_j}=\Delta_q$ and $\tilde{\Delta}_{q_j}=\tilde{\Delta}_q$ for all $j$. Since there is only one gauge group we directly set the initial real part to be
\be
v(t)=\frac{\tilde{\Delta}_q-\Delta_q}{2}.
\ee

The quiver described in quiver diagram \ref{Diagram:ModelII} is quite different from the ABJM quiver, in particular, it  has only one node. We have only one set of eigenvalues to consider. The set of fugacities involved $y_{X_a}$, $y_{q_j}$ and $\tilde{y}_{q_j}$ are quite different  as well, in the sense that there is no particular subset that one would naturally identify with the $y_a$ of the ABJM model. 

\begin{figure}[H]
    \centering
    \begin{subfigure}[b]{0.75\textwidth}
	    \begin{subfigure}[b]{0.5\textwidth}
	        \includegraphics[width=\textwidth]{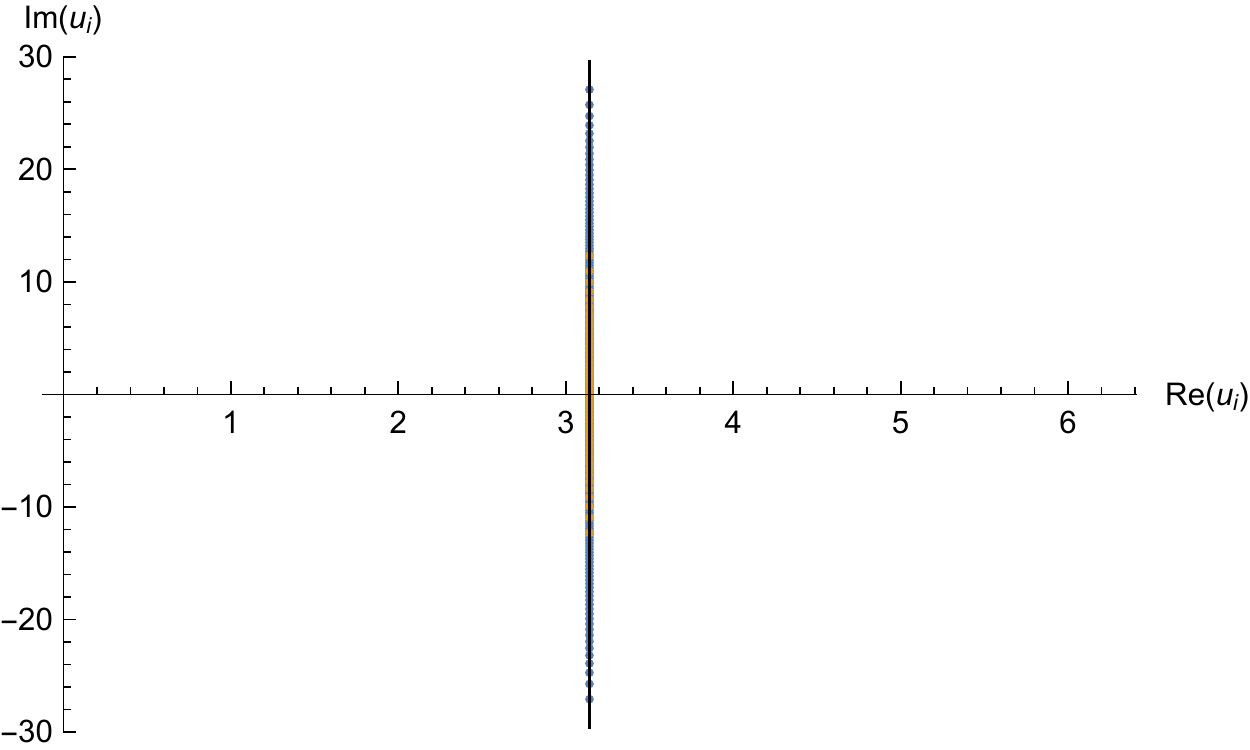}
	        \caption{Eigenvalue distribution}
	    \end{subfigure}
	    \quad %add desired spacing between images, e. g. ~, \quad, \qquad, \hfill etc. 
	    \begin{subfigure}[b]{0.5\textwidth}
	        \includegraphics[width=\textwidth]{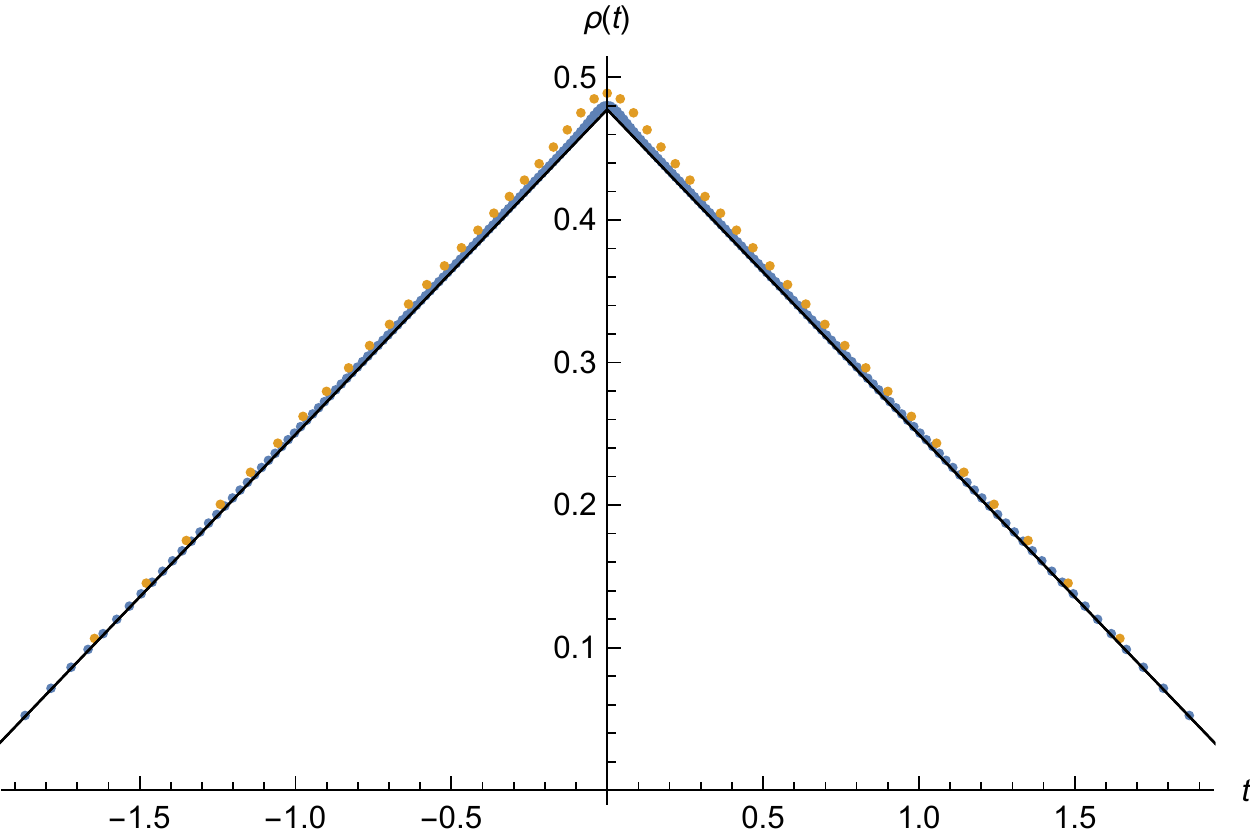}
	        \caption{Eigenvalue density $\rho(t)$}
	    \end{subfigure}
    \end{subfigure}
    \quad %add desired spacing between images, e. g. ~, \quad, \qquad, \hfill etc. 
    \begin{subfigure}[b]{0.17\textwidth}
        \includegraphics[width=\textwidth]{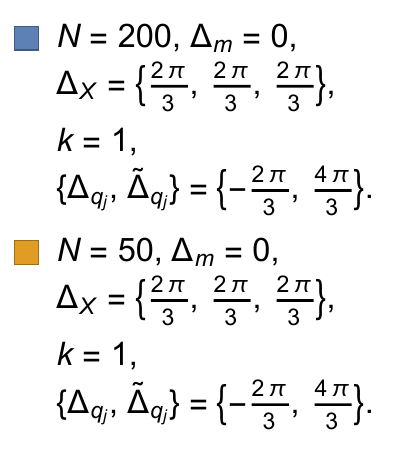}
    \end{subfigure}
    \caption{Eigenvalues for the special case $\Delta_X=\{\frac{2\pi}{3}, \frac{2\pi}{3}, \frac{2\pi}{3}\}$ for $N=50$ (orange) and $200$ (blue) with the same other parameters.}
    \label{fig:V52_Model2_Plot_N}
\end{figure}

\begin{figure}[H]
    \centering
    \begin{subfigure}[b]{0.75\textwidth}
	    \begin{subfigure}[b]{0.5\textwidth}
	        \includegraphics[width=\textwidth]{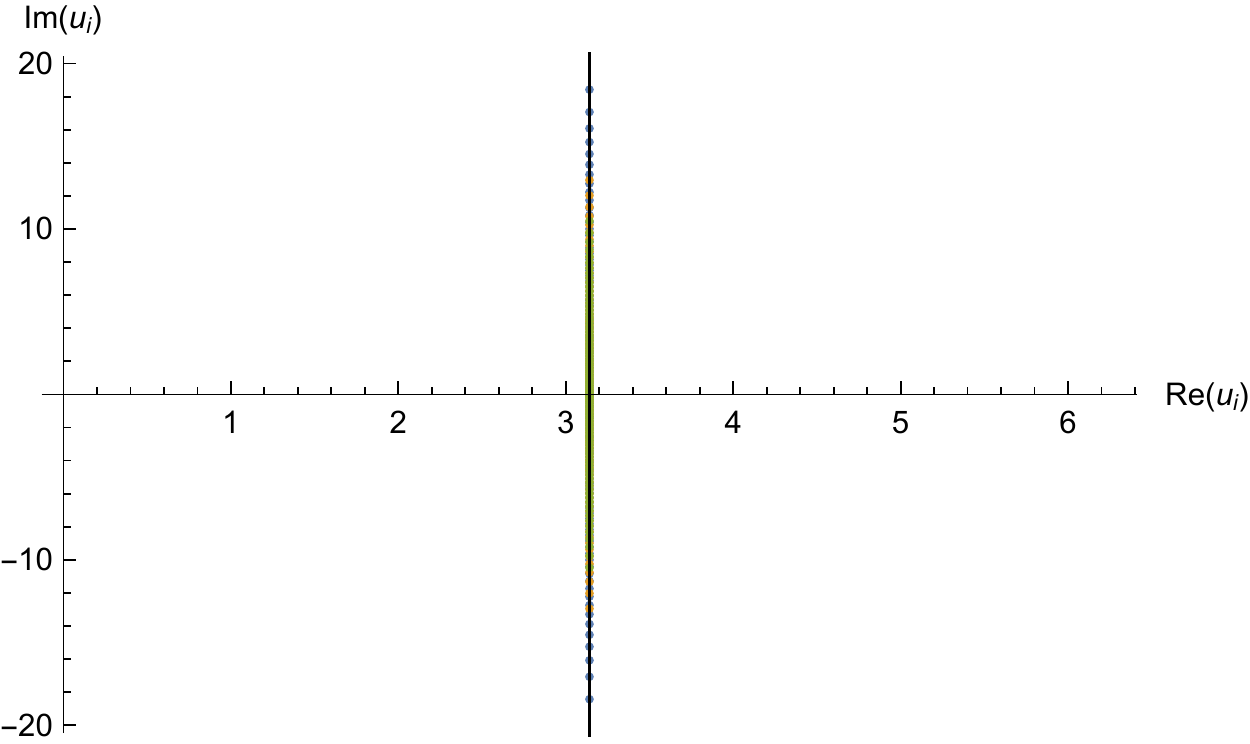}
	        \caption{Eigenvalue distribution}
	    \end{subfigure}
	    \quad %add desired spacing between images, e. g. ~, \quad, \qquad, \hfill etc. 
	    \begin{subfigure}[b]{0.5\textwidth}
	        \includegraphics[width=\textwidth]{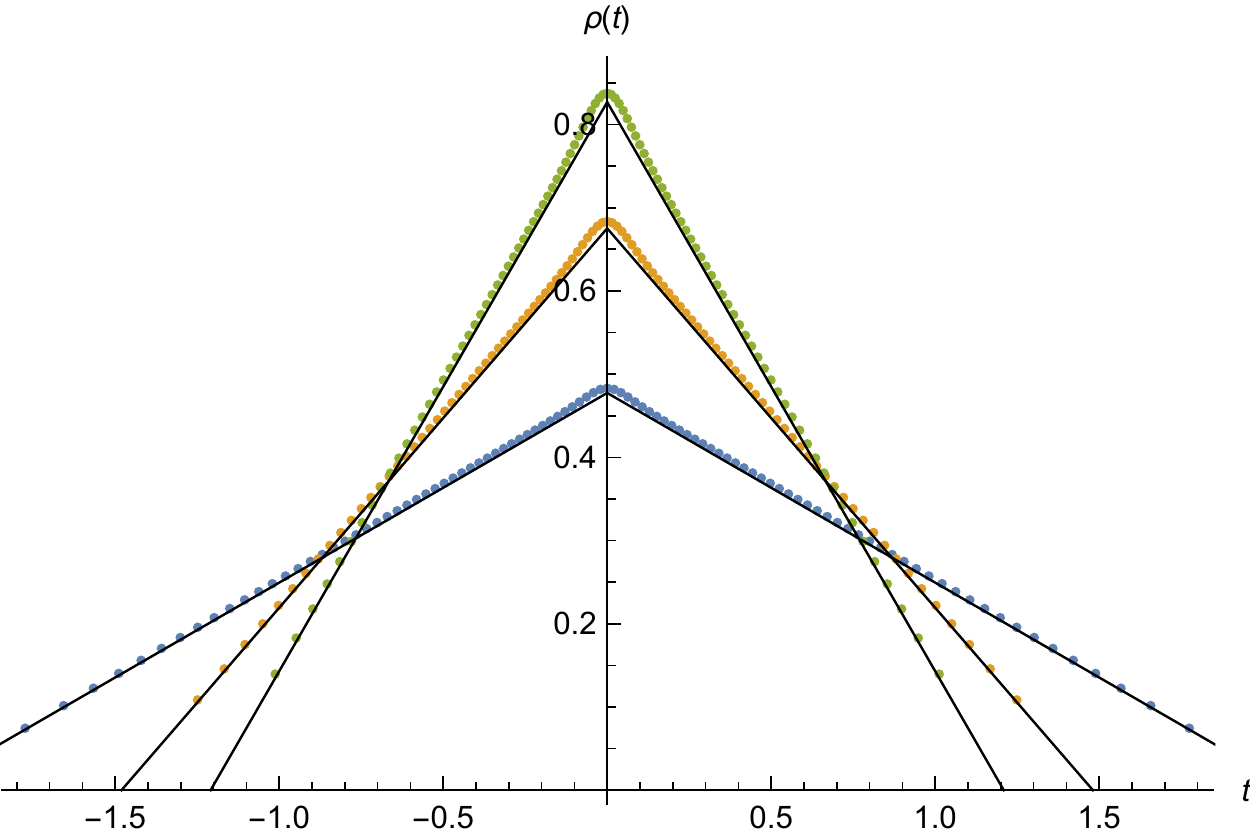}
	        \caption{Eigenvalue density $\rho(t)$}
	    \end{subfigure}
    \end{subfigure}
    \quad %add desired spacing between images, e. g. ~, \quad, \qquad, \hfill etc. 
    \begin{subfigure}[b]{0.17\textwidth}
        \includegraphics[width=\textwidth]{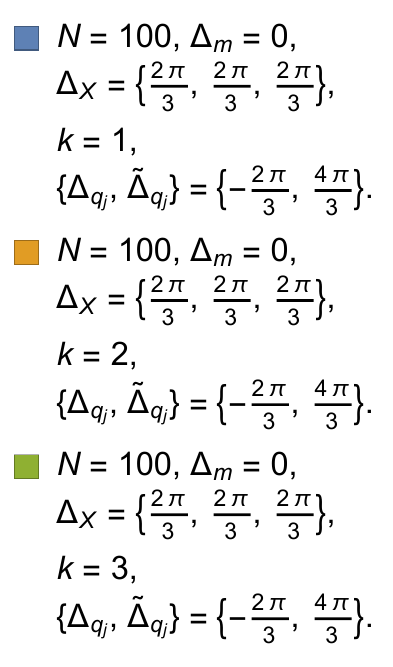}
    \end{subfigure}
    \caption{Eigenvalues for the special case $\Delta_X=\{\frac{2\pi}{3}, \frac{2\pi}{3}, \frac{2\pi}{3}\}$ for $k=1$ (blue), $2$ (orange) and $3$ (green) with the same other parameters.}
    \label{fig:V52_Model2_Plot_k}
\end{figure}

\begin{figure}[H]
    \centering
    \begin{subfigure}[b]{0.75\textwidth}
	    \begin{subfigure}[b]{0.5\textwidth}
	        \includegraphics[width=\textwidth]{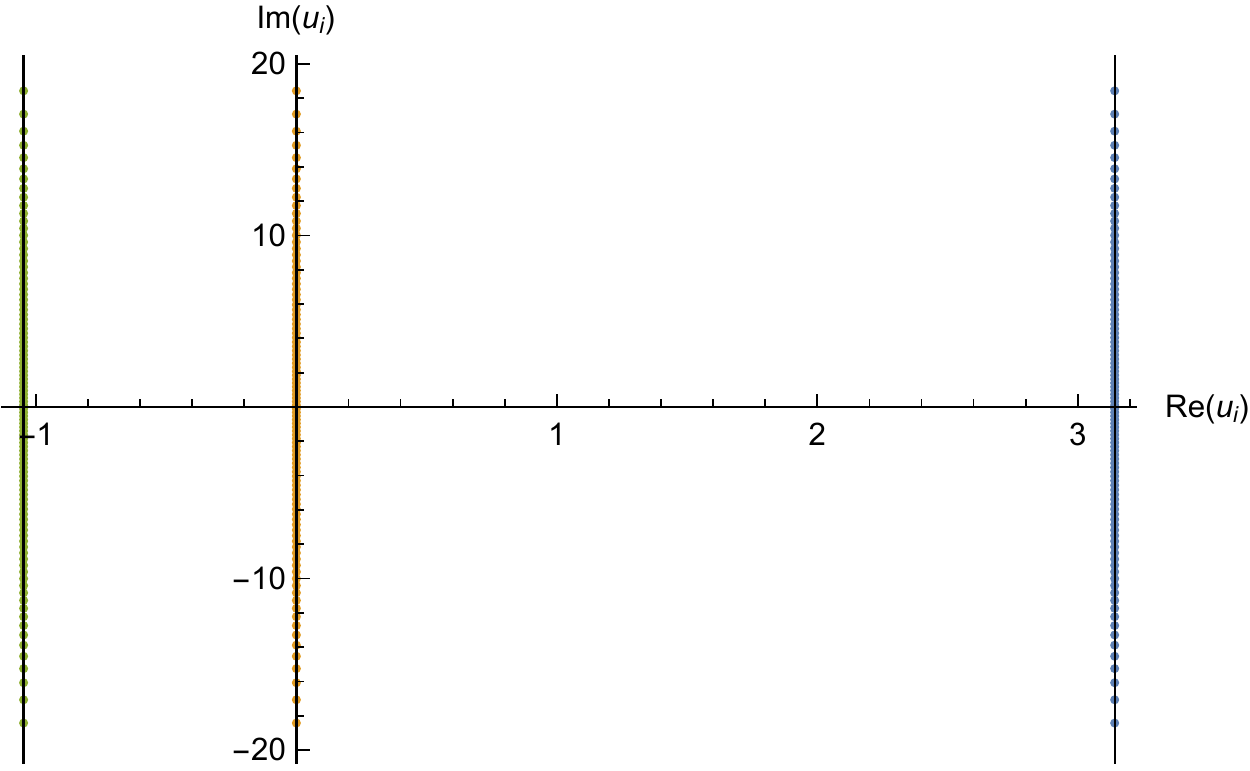}
	        \caption{Eigenvalue distribution}
	    \end{subfigure}
	    \quad %add desired spacing between images, e. g. ~, \quad, \qquad, \hfill etc. 
	    \begin{subfigure}[b]{0.5\textwidth}
	        \includegraphics[width=\textwidth]{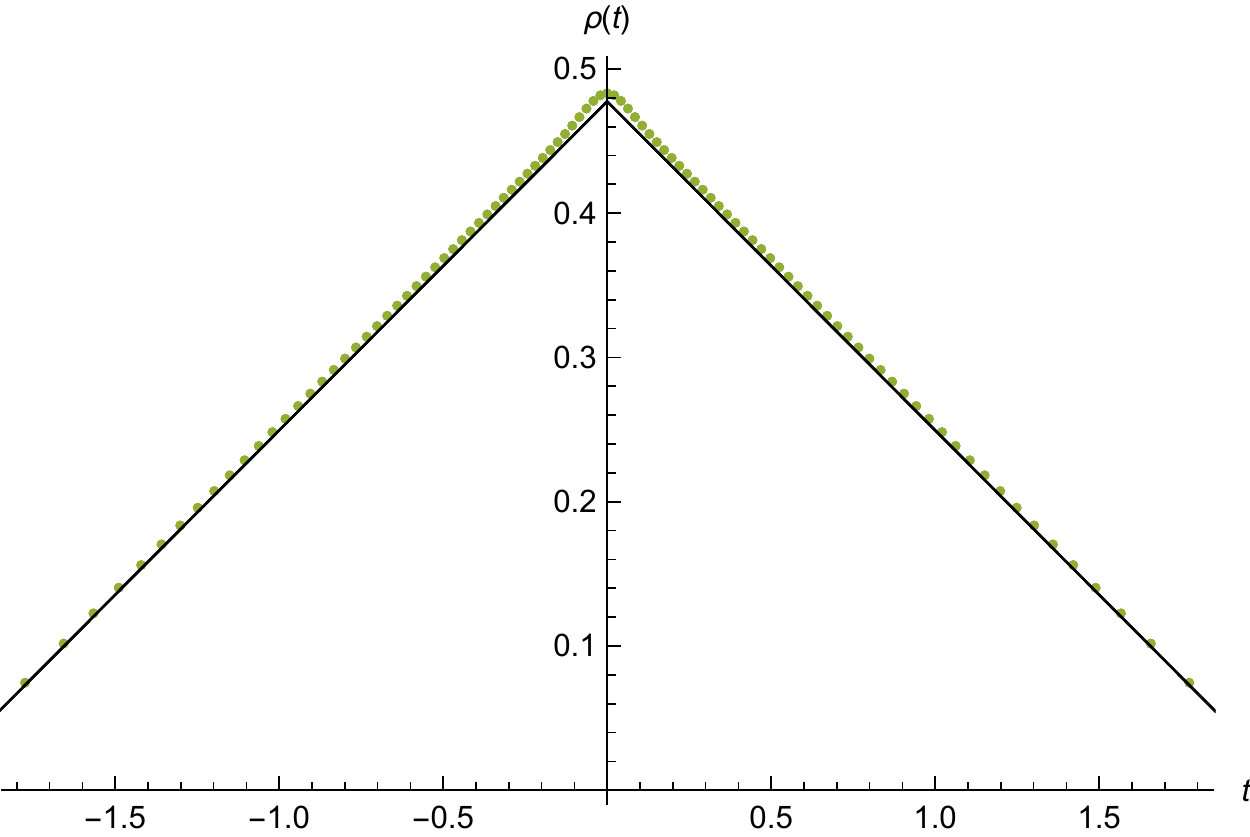}
	        \caption{Eigenvalue density $\rho(t)$}
	    \end{subfigure}
    \end{subfigure}
    \quad %add desired spacing between images, e. g. ~, \quad, \qquad, \hfill etc. 
    \begin{subfigure}[b]{0.17\textwidth}
        \includegraphics[width=\textwidth]{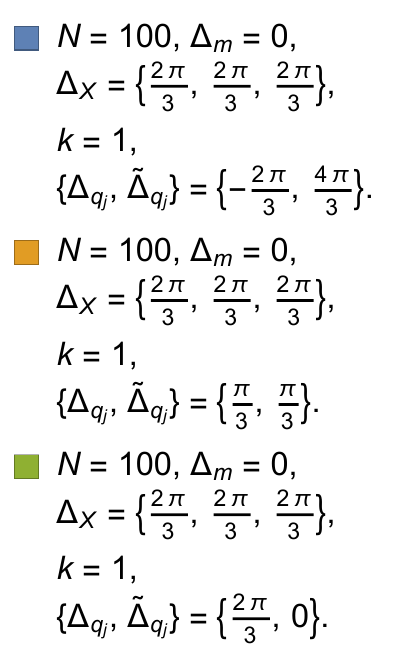}
    \end{subfigure}
    \caption{Eigenvalues for the special case $\Delta_X=\{\frac{2\pi}{3}, \frac{2\pi}{3}, \frac{2\pi}{3}\}$ for $\{\Delta_q, \tilde{\Delta}_q\}=\{-\frac{2\pi}{3}, \frac{4\pi}{3}\}$ (blue), $\{\frac{\pi}{3}, \frac{\pi}{3}\}$ (orange) and $\{\frac{2\pi}{3}, 0\}$ (green) with the same other parameters.}
    \label{fig:V52_Model2_Plot_Delq}
\end{figure}

\begin{figure}[H]
    \centering
    \begin{subfigure}[b]{0.75\textwidth}
	    \begin{subfigure}[b]{0.5\textwidth}
	        \includegraphics[width=\textwidth]{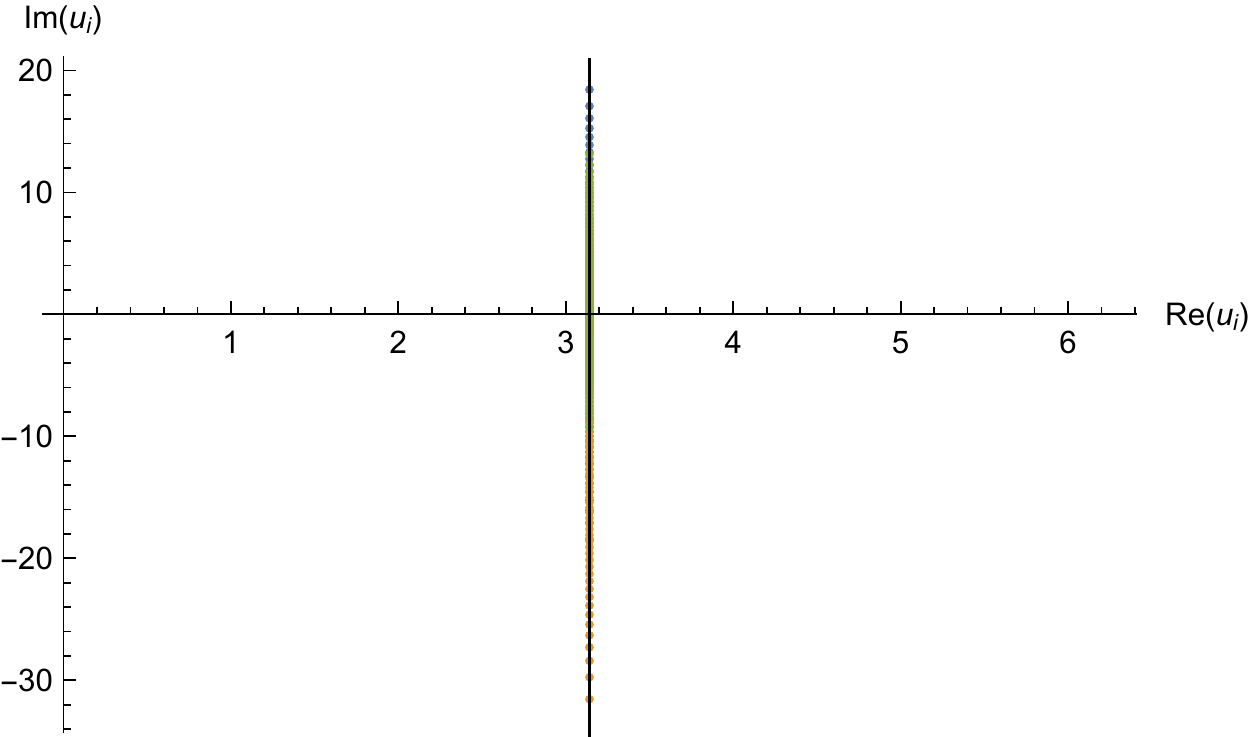}
	        \caption{Eigenvalue distribution}
	    \end{subfigure}
	    \quad %add desired spacing between images, e. g. ~, \quad, \qquad, \hfill etc. 
	    \begin{subfigure}[b]{0.5\textwidth}
	        \includegraphics[width=\textwidth]{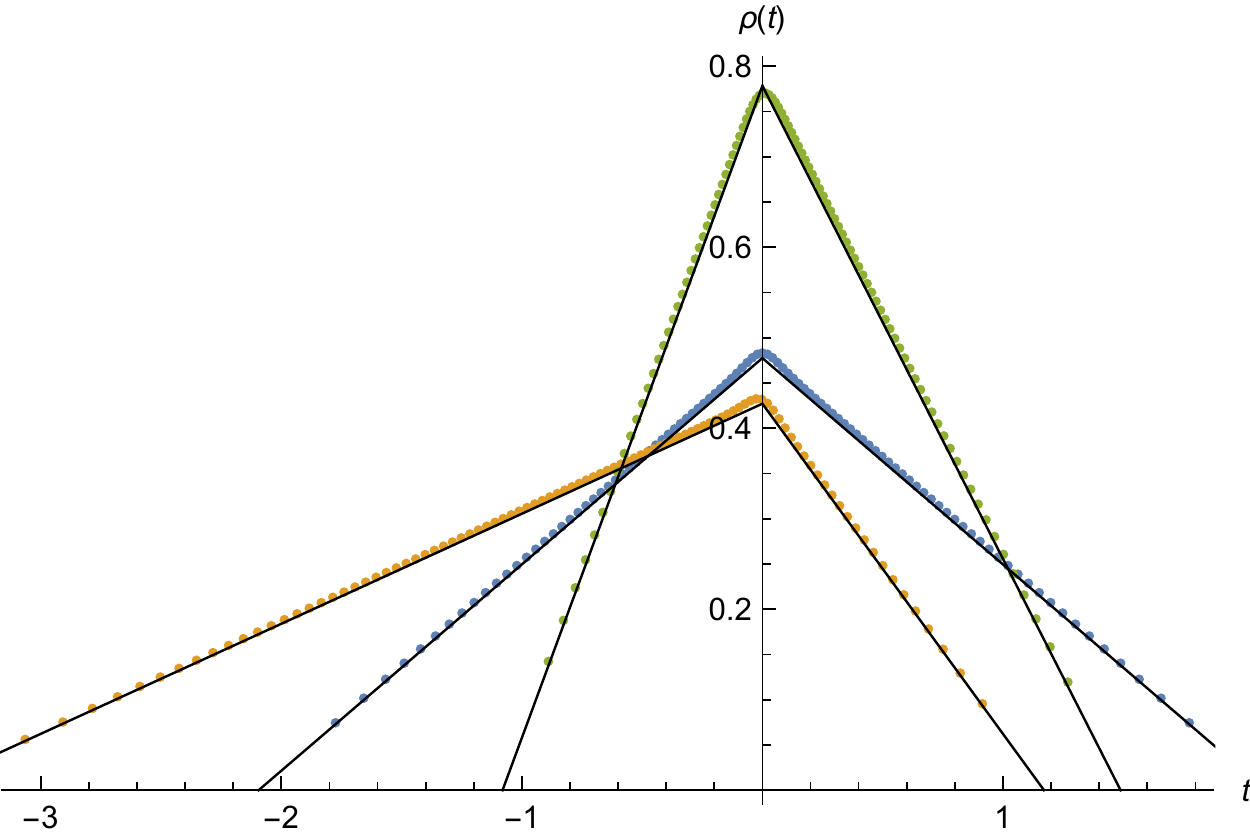}
	        \caption{Eigenvalue density $\rho(t)$}
	    \end{subfigure}
    \end{subfigure}
    \quad %add desired spacing between images, e. g. ~, \quad, \qquad, \hfill etc. 
    \begin{subfigure}[b]{0.17\textwidth}
        \includegraphics[width=\textwidth]{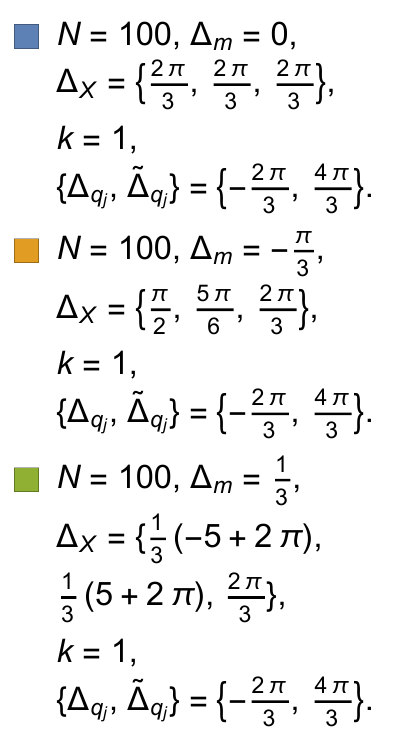}
    \end{subfigure}
    \caption{Eigenvalues for the special case $\Delta_m=0, \Delta_X=\{\frac{2\pi}{3}, \frac{2\pi}{3}, \frac{2\pi}{3}\}$ (blue), and for the general cases $\Delta_m=-\frac{\pi}{3}, \Delta_X=\{\frac{\pi}{2}, \frac{5\pi}{6}, \frac{2\pi}{3}\}$ (orange) and $\Delta_m=\frac{1}{3}, \Delta_X=\{-\frac{5}{3}+\frac{2\pi}{3}, \frac{5}{3}+\frac{2\pi}{3}, \frac{2\pi}{3}\}$ (green) with the same other parameters.}
    \label{fig:V52_Model2_Plot_Delm_DelX}
\end{figure}

The numerical solutions for different values of $N$, $k$, $\{\Delta_q, \tilde{\Delta}_q\}$ and $\Delta_m,\,\Delta_X=\{\Delta_{X_1},\Delta_{X_2},\Delta_{X_3}\}$ are shown in Figures~\ref{fig:V52_Model2_Plot_N} - \ref{fig:V52_Model2_Plot_Delm_DelX}. We find the value of the real parts of the exact eigenvalues is the same as our initial value.

Figure~\ref{fig:V52_Model2_Plot_N} shows that the real part of the eigenvalues is a constant. It is harder to spot the scaling from panel (a) but it is still $N^{1/2}$. 

In Figure~\ref{fig:V52_Model2_Plot_k} we consider the effect of changing the number of flavors $k$. As expected from the leading order solution \cite{Hosseini:2016ume},  changing $k$ mostly affects the slope in the eigenvalue distribution $\rho(t)$ which increases proportional to $k$.

Figure~\ref{fig:V52_Model2_Plot_Delq} demonstrates that the role of the chemical potentials $\{\Delta_{q}, \tilde{\Delta}_{q}\}$ display the real part of the eigenvalues $u_i$  and have, otherwise no effect on the imaginary eigenvalue density $\rho(t)$. 

Finally, we display the dependence of the eigenvalues $u_i$ on the choice of the chemical potential $\Delta_X$ in Figure~\ref{fig:V52_Model2_Plot_Delm_DelX} keeping the same real parts.

%%%%%%%%%%%%%%%%%%%%%%%%%%%%%%%%%%%%
\subsubsection{The subleading term of the index at large $N$}

The index should take the form
\be
{\rm Re}\log Z=f_1(k,\Delta,\mathfrak{n},\mathfrak{t})N^{3/2}+f_2(k,\Delta,\mathfrak{n},\mathfrak{t})N^{1/2}+f_3(k,\Delta,\mathfrak{n},\mathfrak{t})\log{N}+f_4(k,\Delta,\mathfrak{n},\mathfrak{t})+\mathcal O(N^{-1/2}),
\ee
where the functions $f_1$, $f_2$, $f_3$ and $f_4$ are linear in the magnetic fluxes $\mathfrak{n}$ and $\mathfrak{t}$.

The index Eq.~(\ref{Eq:V52_Model2_Index}) and ${\rm Re}\log Z$ can be computed using the numerical solutions. For simplicity we set $\mathfrak{n}_{q_j}=\mathfrak{n}_q$ and $\tilde{\mathfrak{n}}_{q_j}=\tilde{\mathfrak{n}}_q$ for all $j$. Under the similar decomposition
\be
{\rm Re}\log Z=A+B_1\mathfrak{n}_{X_1}+B_2\mathfrak{n}_{q}+B_3\mathfrak{t},
\ee
where we have used the marginality condition on the superpotential $\mathfrak{n}_{X_1}+\mathfrak{n}_{X_2}=4/3$, $\mathfrak{n}_{X_3}=2/3$ and $\mathfrak{n}_q+\tilde{\mathfrak{n}}_q=2/3$. Then we perform a linear least-squares fit for $A$ and $B_a$ to the function
\be
f(N)=f_1N^{3/2}+f_2N^{1/2}+f_3\log{N}+f_4+\sum_{p=1}^{p_c}f_{p+4}\,\, N^{\left(1-2p\right)/2}.
\label{Eq:N010FitlogZN}
\ee

The results of the numerical fit for ${\rm Re}\log Z$ with $N$ are presented in Table~\ref{tbl:V52_Model2_FitRelogZ}. The analytical leading term computed by the index theorem in \cite{Hosseini:2016ume} and the numerical leading term $f_1N^{3/2}$ match to number of significant digits present in the table. And the numerical results are also independent of $\{\Delta_q, \tilde{\Delta}_q\}$. The numerical results indicate that the coefficient $f_3$ of the $\log N$ term is precisely $-1/2$. 

Furthermore, from the numerical results it is shown that by taking
\be
\Delta_{X_1}=\Delta_3,\quad\Delta_m=k\left(\frac{2\pi}{3}-\Delta_4\right),\quad\mathfrak{n}_{X_1}=\mathfrak{n}_3,\quad\mathfrak{t}=-k\left[\frac{2}{3}\left(1-g\right)-\mathfrak{n}_4\right],
\ee
where $g$ is the genus and $g=0$ for this case, Model I matches exactly with Model II for every term, including  to logarithmic terms which are equal as they are independent of the chemical potentials of the magnetic charges.

%%%%
\begin{table}[htp]
\renewcommand{\arraystretch}{1.2}
\setlength{\tabcolsep}{5pt}
\begin{subtable}[H]{1\textwidth}
\centering
\caption{$\Delta_m=0$;$\,\{\Delta_{X_1}, \Delta_{X_2}, \Delta_{X_3}\}=\{\frac{2\pi}{3}, \frac{2\pi}{3}, \frac{2\pi}{3}\}$.}
\begin{tabular}{|c|c c|c||l|l|l|l|}
\hline
$k$&$\Delta_q$&$\tilde{\Delta}_q$&$N(s)$&$f_1$&$f_2$&$f_3$&$f_4$\\
\hline
\multirow{3}{*}{$1$}&$-\frac{2\pi}{3}$&$\frac{4\pi}{3}$&$100\sim300(10)$&$-1.86168$&$+3.02526$&$-0.50066$&$-2.75740$ \\
\cline{2-8}
&$\frac{\pi}{3}$&$\frac{\pi}{3}$&$100\sim300(10)$&$-1.86168$&$+3.02526$&$-0.50066$&$-2.75740$ \\
\cline{2-8}
&$\frac{2\pi}{3}$&$0$&$100\sim300(10)$&$-1.86168$&$+3.02526$&$-0.50066$&$-2.75740$ \\
\hline
$2$&$-\frac{2\pi}{3}$&$\frac{4\pi}{3}$&$100\sim200(5)$&$-2.63282$&$+3.37334$&$-0.50090$&$-3.26096$ \\
\hline
$3$&$-\frac{2\pi}{3}$&$\frac{4\pi}{3}$&$100\sim200(5)$&$-3.22453$&$+4.43378$&$-0.50115$&$-4.75453$ \\
\hline
\end{tabular}
\end{subtable}
\vspace{3mm}

\begin{subtable}[H]{1\textwidth}
\centering
\caption{$\Delta_m=-\frac{\pi}{3}$;$\,\{\Delta_{X_1}, \Delta_{X_2}, \Delta_{X_3}\}=\{\frac{\pi}{2}, \frac{5\pi}{6}, \frac{2\pi}{3}\}$.}
\begin{tabular}{|c|c c|c||l|l|l|l|}
\hline
$k$&$\Delta_q$&$\tilde{\Delta}_q$&$N(s)$&$f_1$&$f_2$&$f_3$&$f_4$\\
\hline
\multirow{7}{*}{$1$}&$-\frac{2\pi}{3}$&$\frac{4\pi}{3}$&$100\sim300(10)$&\tabincell{l}{$-1.66514$\\$-0.31221\mathfrak{n}_{X_1}$\\$+0.78053\mathfrak{t}$}&\tabincell{l}{$+3.44999$\\$+0.00781\mathfrak{n}_{X_1}$\\$-0.17562\mathfrak{t}$}&\tabincell{l}{$-0.50082$}&\tabincell{l}{$-3.26746$\\$-0.02821\mathfrak{n}_{X_1}$\\$+0.19180\mathfrak{t}$}\\
\cline{2-8}
&$\frac{\pi}{3}$&$\frac{\pi}{3}$&$100\sim300(10)$&\tabincell{l}{$-1.66514$\\$-0.31221\mathfrak{n}_{X_1}$\\$+0.78053\mathfrak{t}$}&\tabincell{l}{$+3.44999$\\$+0.00781\mathfrak{n}_{X_1}$\\$-0.17562\mathfrak{t}$}&\tabincell{l}{$-0.50082$}&\tabincell{l}{$-3.26746$\\$-0.02821\mathfrak{n}_{X_1}$\\$+0.19180\mathfrak{t}$}\\
\cline{2-8}
&$\frac{2\pi}{3}$&$0$&$100\sim300(10)$&\tabincell{l}{$-1.66514$\\$-0.31221\mathfrak{n}_{X_1}$\\$+0.78053\mathfrak{t}$}&\tabincell{l}{$+3.44999$\\$+0.00781\mathfrak{n}_{X_1}$\\$-0.17562\mathfrak{t}$}&\tabincell{l}{$-0.50082$}&\tabincell{l}{$-3.26746$\\$-0.02821\mathfrak{n}_{X_1}$\\$+0.19180\mathfrak{t}$}\\
\hline
$2$&$-\frac{2\pi}{3}$&$\frac{4\pi}{3}$&$100\sim200(5)$&\tabincell{l}{$-2.30372$\\$-0.49365\mathfrak{n}_{X_1}$\\$+0.24683\mathfrak{t}$}&\tabincell{l}{$+3.19233$\\$+0.41961\mathfrak{n}_{X_1}$\\$+0.08639\mathfrak{t}$}&\tabincell{l}{$-0.50098$}&\tabincell{l}{$-3.22095$\\$-0.34133\mathfrak{n}_{X_1}$\\$-0.06902\mathfrak{t}$}\\
\hline
$3$&$-\frac{2\pi}{3}$&$\frac{4\pi}{3}$&$100\sim200(5)$&\tabincell{l}{$-2.81460$\\$-0.61569\mathfrak{n}_{X_1}$\\$+0.13193\mathfrak{t}$}&\tabincell{l}{$+3.99013$\\$+0.89056\mathfrak{n}_{X_1}$\\$+0.09188\mathfrak{t}$}&\tabincell{l}{$-0.50123$\\$-0.00006\mathfrak{n}_{X_1}$\\$+0.00003\mathfrak{t}$}&\tabincell{l}{$-4.52640$\\$-0.83862\mathfrak{n}_{X_1}$\\$-0.08196\mathfrak{t}$}\\
\hline
\end{tabular}
\end{subtable}
\vspace{3mm}

\begin{subtable}[H]{1\textwidth}
\centering
\caption{$\Delta_m=\frac{1}{3}$;$\,\{\Delta_{X_1}, \Delta_{X_2}, \Delta_{X_3}\}=\{\frac{2\pi}{3}-\frac{5}{3}, \frac{2\pi}{3}+\frac{5}{3}, \frac{2\pi}{3}\}$.}
\begin{tabular}{|c|c c|c||l|l|l|l|}
\hline
$k$&$\Delta_q$&$\tilde{\Delta}_q$&$N(s)$&$f_1$&$f_2$&$f_3$&$f_4$\\
\hline
$1$&$-\frac{2\pi}{3}$&$\frac{4\pi}{3}$&$100\sim200(5)$&\tabincell{l}{$-0.88090$\\$-1.81136\mathfrak{n}_{X_1}$\\$-0.13631\mathfrak{t}$}&\tabincell{l}{$+4.22992$\\$+1.38166\mathfrak{n}_{X_1}$\\$-0.06990\mathfrak{t}$}&\tabincell{l}{$-0.50168$\\$-0.00005\mathfrak{n}_{X_1}$\\$-0.00001\mathfrak{t}$}&\tabincell{l}{$-4.49720$\\$-1.91182\mathfrak{n}_{X_1}$\\$+0.01935\mathfrak{t}$}\\
\hline
$2$&$-\frac{2\pi}{3}$&$\frac{4\pi}{3}$&$100\sim200(5)$&\tabincell{l}{$-1.24227$\\$-2.58650\mathfrak{n}_{X_1}$\\$-0.04773\mathfrak{t}$}&\tabincell{l}{$+3.87599$\\$+5.44770\mathfrak{n}_{X_1}$\\$-0.07670\mathfrak{t}$}&\tabincell{l}{$-0.50409$\\$+0.00620\mathfrak{n}_{X_1}$\\$+0.00063\mathfrak{t}$}&\tabincell{l}{$-5.28616$\\$-8.41742\mathfrak{n}_{X_1}$\\$+0.05162\mathfrak{t}$}\\
\hline
$3$&$-\frac{2\pi}{3}$&$\frac{4\pi}{3}$&$100\sim200(5)$&\tabincell{l}{$-1.52070$\\$-3.17340\mathfrak{n}_{X_1}$\\$-0.02594\mathfrak{t}$}&\tabincell{l}{$+4.40337$\\$+10.53032\mathfrak{n}_{X_1}$\\$-0.06618\mathfrak{t}$}&\tabincell{l}{$-0.49827$\\$-0.05617\mathfrak{n}_{X_1}$\\$-0.00672\mathfrak{t}$}&\tabincell{l}{$-7.73447$\\$-18.79411\mathfrak{n}_{X_1}$\\$+0.09834\mathfrak{t}$}\\
\hline
\end{tabular}
\end{subtable}

\caption{($V^{5,2}$ Model II) Numerical fit for ${\rm Re}\log Z=f_1N^{3/2}+f_2N^{1/2}+f_3\log{N}+f_4+\sum_{g=1}^{g_c=5} N^{\left(1-2g\right)/2}$.}
\label{tbl:V52_Model2_FitRelogZ}
\end{table}
%%%%

%%%%%%%%%%%%%%%%%%%%%%%%%%%%%%%%%%%%%%%%%%%%%%%%%%%%%%%
\section{The topologically twisted index of $Q^{1,1,1}$}\label{Sec:Q111}

The field theory dual to M-theory of AdS$_4\times Q^{1,1,1}/\mathbb{Z}_k$  was originally discussed in \cite{Benini:2009qs,Jafferis:2009th}, see also \cite{Cremonesi:2010ae}. The quiver diagram is a particular flavored type of the ABJM quiver: 
\bea
\begin{tikzpicture}[baseline, font=\scriptsize, scale=0.8]
\begin{scope}[auto,%
  every node/.style={draw, minimum size=0.5cm}, node distance=4cm];
  % the vertices
\node[circle] (UN1) at (0, 0) {$N_{+k}$};
\node[circle, right=of UN1] (UN2)  {$N_{-k}$};
\node[rectangle] at (3.2,2.2) (UNa1)  {$n_{a1}$};
\node[rectangle] at (3.2,3.5) (UNa2)  {$n_{a2}$};
\node[rectangle] at (3.2,-2.2) (UNb1)  {$n_{b1}$};
\node[rectangle] at (3.2,-3.5) (UNb2)  {$n_{b2}$};
\end{scope}
  % the edges
\draw[solid,line width=0.2mm,<-]  (UN1) to[bend right=30] node[midway,above] {$B_2 $}node[midway,above] {}  (UN2) ;
\draw[solid,line width=0.2mm,->]  (UN1) to[bend right=-10] node[midway,above] {$A_1$}node[midway,above] {}  (UN2) ; 
\draw[solid,line width=0.2mm,<-]  (UN1) to[bend left=-10] node[midway,above] {$B_1$} node[midway,above] {} (UN2) ;  
\draw[solid,line width=0.2mm,->]  (UN1) to[bend left=30] node[midway,above] {$A_2$} node[midway,above] {} (UN2) ;   
\draw[solid,line width=0.2mm,->]  (UN1)  to[bend right=30] node[midway,right] {}   (UNb1);
\draw[solid,line width=0.2mm,->]  (UNb1) to[bend right=30] node[midway,left] {} (UN2) ; 
\draw[solid,line width=0.2mm,->]  (UN1)  to[bend right=30]  node[midway,right] {} (UNb2);
\draw[solid,line width=0.2mm,->]  (UNb2) to[bend right=30] node[midway,left] {} (UN2); 
\draw[solid,line width=0.2mm,->]  (UN2)  to[bend right=30] node[pos=0.9,right] {}   (UNa1);
\draw[solid,line width=0.2mm,->]  (UNa1) to[bend right=30] node[pos=0.1,left] {} (UN1) ; 
\draw[solid,line width=0.2mm,->]  (UN2)  to[bend right=30]  node[pos=0.9,right] {} (UNa2);
\draw[solid,line width=0.2mm,->]  (UNa2) to[bend right=30] node[pos=0.1,left] {}  (UN1); 
\node at (4.5,-2.4) {$\tQ^{(1)}$};
\node at (2.,-2.4) {$Q^{(1)}$};
\node at (5.5,-2.8) {$\tQ^{(2)}$};
\node at (0.9,-2.9) {$Q^{(2)}$};
\node at (4.5,2.4) {$q^{(1)}$};
\node at (2.,2.4) {$\tilde{q}^{(1)}$};
\node at (5.5,2.8) {$q^{(2)}$};
\node at (1.,2.8) {$\tilde{q}^{(2)}$};
\end{tikzpicture}\label{Quiver:Q111}
\eea
with the superpotential
\bea\label{supflvABJM}
W &= &\Tr \left( A_1 B_1 A_2 B_2 - A_1 B_2 A_2 B_1  \right) \nn \\
& &+\Tr\left[\sum_{j=1}^{n_{a1}} q_{j}^{(1)} A_1 \tilde q_{j}^{(1)}
 +\sum_{j=1}^{n_{a2}} q_{j}^{(2)} A_2 \tilde q_{j}^{(2)}
 +\sum_{j=1}^{n_{b1}} Q_{j}^{(1)} B_1 \tilde Q_{j}^{(1)}
 +\sum_{j=1}^{n_{b2}} Q_{j}^{(2)} B_2 \tilde Q_{j}^{(2)}\right].
\eea

The free energy of the field theory on $S^3$ was shown to match the gravity side in \cite{Cheon:2011vi,Jafferis:2011zi}. Aspects of the superconformal index have been discussed in  \cite{Imamura:2011uj,Cheon:2011th}. The large $N$ analysis of the topologically twisted index was presented in \cite{Hosseini:2016ume}  whose notation and leading order analysis we follow very closely. 

\subsection{Numerical solutions to the system of BAEs}

From the ingredients of the quiver diagram \ref{Quiver:Q111} one constructs the topologically twisted index as 
\bea
Z&=&\frac{1}{(N!)^2}\sum\limits_{\mathfrak{m},\tilde{\mathfrak{m}}\in \mathbb{Z}^N}\int_{\cal C}\prod\limits_{i=1}^N
\frac{dx_i}{2\pi i x_i}\frac{d\tx_i}{2\pi i \tx_i} x_i^{\mathfrak{t}}\tx_i^{\tilde{\mathfrak{t}}} \xi^{\mathfrak{m}_i} \tilde{\xi}^{-\tilde{\mathfrak{m}}_i} \times \prod\limits_{i\neq j}^N\left(1-\frac{x_i}{x_j}\right)\left(1-\frac{\tx_i}{\tx_j}\right) \nonumber\\
&&\times \prod\limits_{i,j=1}^N\prod\limits_{a=1,2}
\left(\frac{\sqrt{\frac{x_i}{\tx_j}y_a}}{1-\frac{x_i}{\tx_j}y_a}\right)^{\mathfrak{m}_i-\tilde{\mathfrak{m}}_j-\mathfrak{n}_a+1}
\prod\limits_{b=3,4}
\left(\frac{\sqrt{\frac{\tx_j}{x_i}y_b}}{1-\frac{\tx_j}{x_i}y_b}\right)^{\tilde{\mathfrak{m}}_j-\mathfrak{m}_i-\mathfrak{n}_b+1} \nonumber\\
&&\times \prod\limits_{i=1}^N\prod\limits_{k=1,2}\Bigg(\frac{\sqrt{\frac{1}{x_i}\tilde{y}_{ak}}}{1-\frac{1}{x_i} \tilde{y}_{ak}}\Bigg)^{n(-\mathfrak{m}_i-\tilde{\mathfrak{n}}_{ak}+1)} \times \prod\limits_{j=1}^N\prod\limits_{k=1,2}\Bigg(\frac{\sqrt{\tilde{x}_j y_{ak}}}{1-\tilde{x}_j y_{ak}}\Bigg)^{n(\tilde{\mathfrak{m}}_j-\mathfrak{n}_{ak}+1)}.
\eea

Permorming the summation over magnetic fluxes introducing a large cut-off $M$.
\bea
Z&=&\frac{1}{(N!)^2}\sum\limits_{\mathfrak{m},\tilde{\mathfrak{m}}\in \mathbb{Z}^N}\int_{\cal C}\prod\limits_{i=1}^N
\frac{dx_i}{2\pi i x_i}\frac{d\tx_i}{2\pi i \tx_i} x_i^{\mathfrak{t}}\tx_i^{\tilde{\mathfrak{t}}} \times \prod\limits_{i\neq j}^N\left(1-\frac{x_i}{x_j}\right)\left(1-\frac{\tx_i}{\tx_j}\right) \nonumber\\
&&\times \prod\limits_{i,j=1}^N\prod\limits_{a=1,2}
\left(\frac{\sqrt{\frac{x_i}{\tx_j}y_a}}{1-\frac{x_i}{\tx_j}y_a}\right)^{1-\mathfrak{n}_a}
\prod\limits_{b=3,4}
\left(\frac{\sqrt{\frac{\tx_j}{x_i}y_b}}{1-\frac{\tx_j}{x_i}y_b}\right)^{1-\mathfrak{n}_b} \nonumber\\
&&\times \prod\limits_{i=1}^N\prod\limits_{k=1,2}\Bigg(\frac{\sqrt{\frac{1}{x_i}\tilde{y}_{ak}}}{1-\frac{1}{x_i} \tilde{y}_{ak}}\Bigg)^{n(1-\tilde{\mathfrak{n}}_{ak})} \times \prod\limits_{j=1}^N\prod\limits_{k=1,2}\Bigg(\frac{\sqrt{\tilde{x}_j y_{ak}}}{1-\tilde{x}_j y_{ak}}\Bigg)^{n(1-\mathfrak{n}_{ak})} \nonumber\\
&&\times \prod\limits_{i=1}^N\frac{\big(e^{iB_i}\big)^M}{e^{iB_i}-1} \prod\limits_{j=1}^N\frac{\big(e^{i\tilde{B}_j}\big)^M}{e^{i\tilde{B}_j}-1},
\eea
where the Bethe Ansatz equations are
\bea
1&=&e^{iB_i}=\xi\prod\limits_{j=1}^N\frac{\left(1-y_3 \frac{\tilde{x}_j}{x_i}\right)\left(1-y_4 \frac{\tilde{x}_j}{x_i}\right)}{\left(1-y_1^{-1} \frac{\tilde{x}_j}{x_i}\right)\left(1-y_2^{-1} \frac{\tilde{x}_j}{x_i}\right)} \times \prod\limits_{k=1,2}\Bigg(\frac{\sqrt{\frac{1}{x_i}\tilde{y}_{ak}}}{1-\frac{1}{x_i} \tilde{y}_{ak}}\Bigg)^{-n}, \nonumber\\
1&=&e^{i\tilde{B}_j}=\tilde{\xi}\prod\limits_{i=1}^N\frac{\left(1-y_3 \frac{\tilde{x}_j}{x_i}\right)\left(1-y_4 \frac{\tilde{x}_j}{x_i}\right)}{\left(1-y_1^{-1} \frac{\tilde{x}_j}{x_i}\right)\left(1-y_2^{-1} \frac{\tilde{x}_j}{x_i}\right)} \times \prod\limits_{k=1,2}\Bigg(\frac{\sqrt{\tilde{x}_j y_{ak}}}{1-\tilde{x}_j y_{ak}}\Bigg)^{-n}.
\eea

The compact expression is
\bea
\label{Eq:Q111_Index}
Z\left(y_a,\mathfrak{n}_a\right)&=&\prod\limits_{k=1,2}\left[{\tilde{y}_{ak}}^{\frac{1}{2}Nn\left(1-\tilde{\mathfrak{n}}_{ak}\right)}y_{ak}^{\frac{1}{2}Nn\left(1-\mathfrak{n}_{ak}\right)}\right]\times\prod_{a=1}^4 y_a^{-\frac{1}{2}N^2 \mathfrak{n}_a}\nonumber\\
&&\times\sum_{I\in BAE}\left[\frac{1}{\det\mathbb{B}}
\frac{\prod_{i=1}^N x_i^{N+\mathfrak{t}} \tilde{x}_i^{N+\tilde{\mathfrak{t}}} \prod_{i\neq j}\left(1-\frac{x_i}{x_j}\right)\left(1-\frac{\tilde{x}_i}{\tilde{x}_j}\right)}{\prod_{i,j=1}^N\prod_{a=1,2}\left(\tilde{x}_j-y_ax_i\right)^{1-\mathfrak{n}_a}\prod_{a=3,4}\left(x_i-y_a\tilde{x}_j\right)^{1-\mathfrak{n}_a}} \right.\nonumber\\
&&\left.\times\prod_{i=1}^N\prod\limits_{k=1,2}\frac{x_i^{\frac{1}{2}n(1-\tilde{\mathfrak{n}}_{ak})}\tilde{x}_i^{\frac{1}{2}n(1-\mathfrak{n}_{ak})}}{\left(x_i-\tilde{y}_{ak}\right)^{n(1-\tilde{\mathfrak{n}}_{ak})}\left(1-\tilde{x}_i y_{ak}\right)^{n(1-\mathfrak{n}_{ak})}}\right].
\eea

The matrix $\mathbb{B}$ is
\be
\mathbb{B}\Big|_{\rm BAEs}=
\left(
\begin{array}{cc}
\delta_{jl}\left[-\sum\limits_{m=1}^N G_{jm}+\sum\limits_{k=1,2}\frac{nx_j}{x_j-\tilde{y}_{ak}}-n \right] & G_{jl}\\
-G_{lj} & \delta_{jl}\left[\sum\limits_{m=1}^N G_{mj}+\sum\limits_{k=1,2}\frac{n\tilde{x}_j}{\tilde{x}_j-y_{ak}^{-1}}-n  \right]
\end{array}
\right).
\ee

The Bethe potential is
\bea
\label{Eq:Q111BethePotential}
\mathcal{V}&=&\sum\limits_{i=1}^N\left[-\Delta_m^{(1)} u_i+\Delta_m^{(2)} \tilde{u}_i-2\pi \left(\tilde{n}_i \tilde{u}_i-n_iu_i\right)\right] \nonumber \\
&&+ \sum\limits_{i,j=1}^N\left[\sum\limits_{a=3,4}{\rm Li}_2\left(e^{i\left(\tilde{u}_j-u_i+\Delta_a\right)}\right)-
\sum\limits_{a=1,2}{\rm Li}_2\left(e^{i\left(\tilde{u}_j-u_i-\Delta_a\right)}\right)\right]\nonumber \\
&&+ n\sum\limits_{i=1}^N\sum\limits_{k=1,2}\left[{\rm Li}_2\left(e^{i\left(-u_i+\tilde{\Delta}_{ak}\right)}\right)-{\rm Li}_2\left(e^{i\left(-\tilde{u}_i-\Delta_{ak}\right)}\right)\right]\nonumber \\
&&+ \frac{n}{2}\sum\limits_{i=1}^N\sum\limits_{k=1,2}\left[\left(\tilde{\Delta}_{ak}-\pi\right)u_i+\left(\Delta_{ak}-\pi\right)\tilde{u}_i\right]-\frac{n}{2}\sum\limits_{i=1}^N\left[u_i^2-\tilde{u}_i^2\right],
\eea
where
\be
\sum\limits_{i=1}^N\left[-2\pi \left(\tilde{n}_i \tilde{u}_i-n_iu_i\right)\right]=\left(4\pi-\sum\limits_{a=1}^4\Delta_a\right)\sum\limits_{i>j}^N\left(\tilde{u}_j-u_i\right)=2\pi\sum\limits_{i>j}^N\left(\tilde{u}_j-u_i\right).
\ee

It is assumed that $0<-v(t)+\tilde{\Delta}_{a1}<2\pi$, $0<-v(t)+\tilde{\Delta}_{a2}<2\pi$, $0<\tilde{v}(t)+\Delta_{a1}<2\pi$ and $0<\tilde{v}(t)+\Delta_{a2}<2\pi$. Using $\Delta_1=\Delta_2=\pi-\Delta_3=\pi-\Delta_4=\Delta$ and the marginality condition from the superpotential  $\Delta_1+\Delta_{a1}+\tilde{\Delta}_{a1}=2\pi$, $\Delta_2+\Delta_{a2}+\tilde{\Delta}_{a2}=2\pi$ we get
\be
\frac{v(t)+\tilde{v}(t)}{2}\in\left\{
\begin{aligned}
\left(\frac{-\Delta-2\Delta_{a1}}{2}, \frac{+\Delta+2\tilde{\Delta}_{a2}}{2}\right),\quad\Delta_{a1}<\Delta_{a2}\\
\left(\frac{-\Delta-2\Delta_{a2}}{2}, \frac{+\Delta+2\tilde{\Delta}_{a1}}{2}\right),\quad\Delta_{a1}\geq\Delta_{a2}
\end{aligned}
\right.
\ee
Thus we set the initial real part axis to be
\be
\frac{v(t)+\tilde{v}(t)}{2}=\frac{2\pi-\Delta-\Delta_{a1}-\Delta_{a2}}{2}.
\ee

Note that we have specialized from the general flavored ABJM quiver \ref{Quiver:Q111} to the particular case corresponding to the dual of AdS$_4\times Q^{1,1,1}/\mathbb{Z}_n$ which in the notation we have introduced implies: $k=0, n_{a1}=n_{a2}=n, n_{b1}=n_{b2}=0$ where the subindex $bi$ corresponds to the fields $Q$ and $\tilde{Q}$. Our goal is thus, to explore the numerical behavior of the index as a function of $N$ and $\Delta_a$ just as in the ABJM case but also as functions of $\{\Delta_m^{(1)}, \Delta_m^{(2)}\}$, $n$ and $\{\Delta_{a1}, \Delta_{a2},\tilde{\Delta}_{a1}, \tilde{\Delta}_{a2}\}$.

\begin{figure}[H]
    \centering
    \begin{subfigure}[b]{0.75\textwidth}
	    \begin{subfigure}[b]{0.5\textwidth}
	        \includegraphics[width=\textwidth]{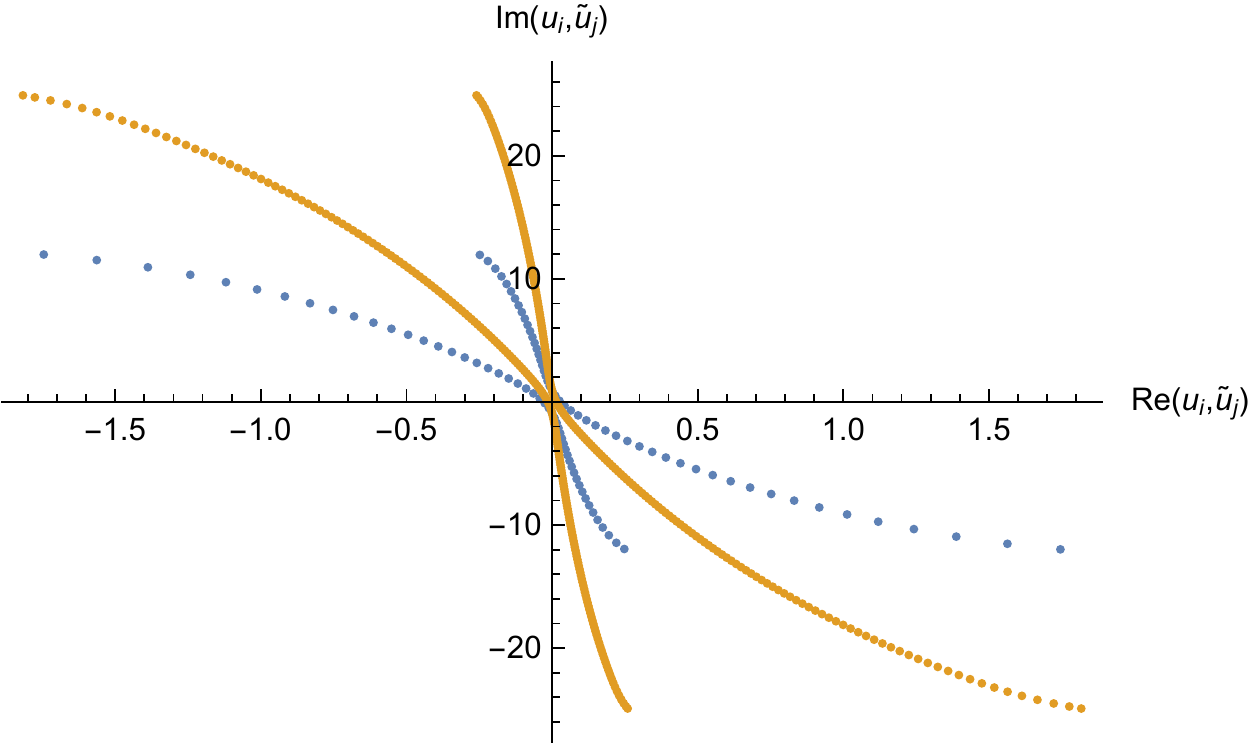}
	        \caption{Eigenvalue distribution}
	    \end{subfigure}
	    \quad %add desired spacing between images, e. g. ~, \quad, \qquad, \hfill etc. 
	    \begin{subfigure}[b]{0.5\textwidth}
	        \includegraphics[width=\textwidth]{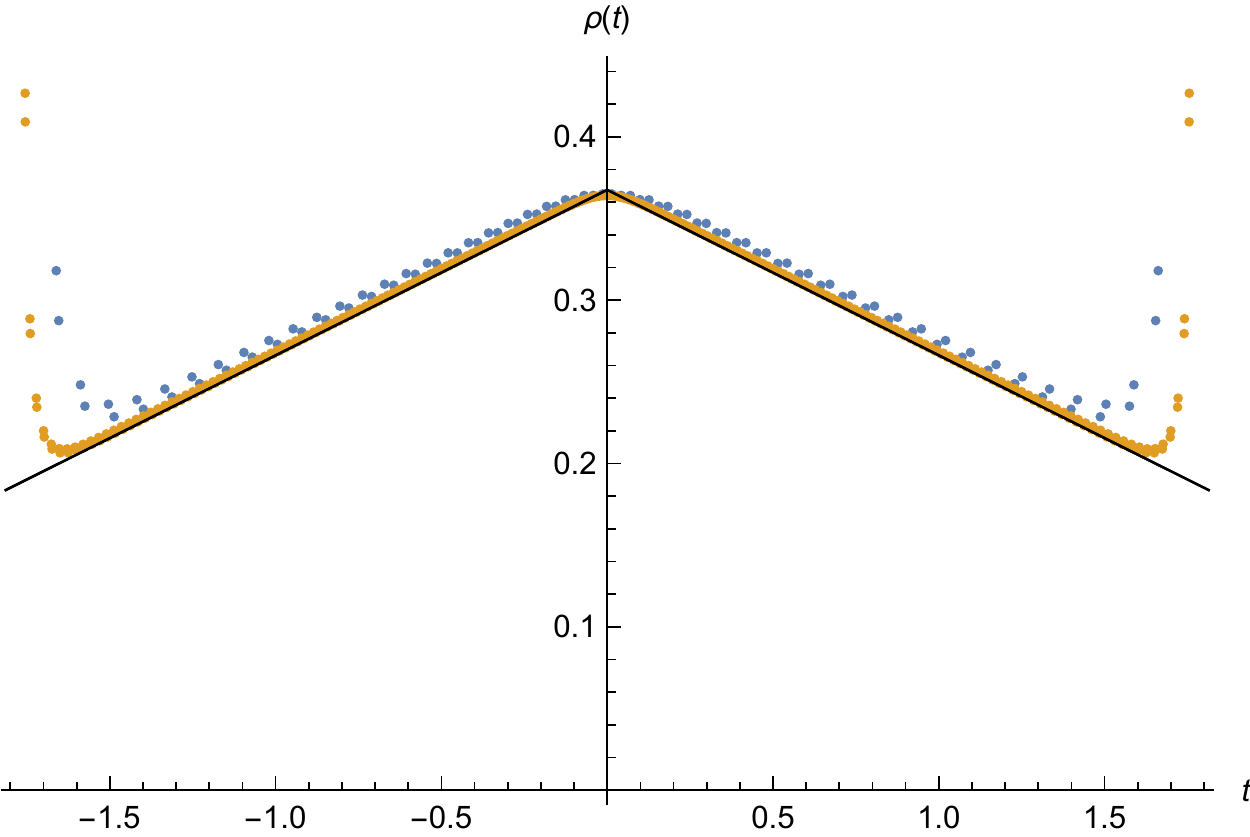}
	        \caption{Eigenvalue density $\rho(t)$}
	    \end{subfigure}\\
	    ~ %add desired spacing between images, e. g. ~, \quad, \qquad, \hfill etc. 
	    \begin{subfigure}[b]{0.5\textwidth}
	        \includegraphics[width=\textwidth]{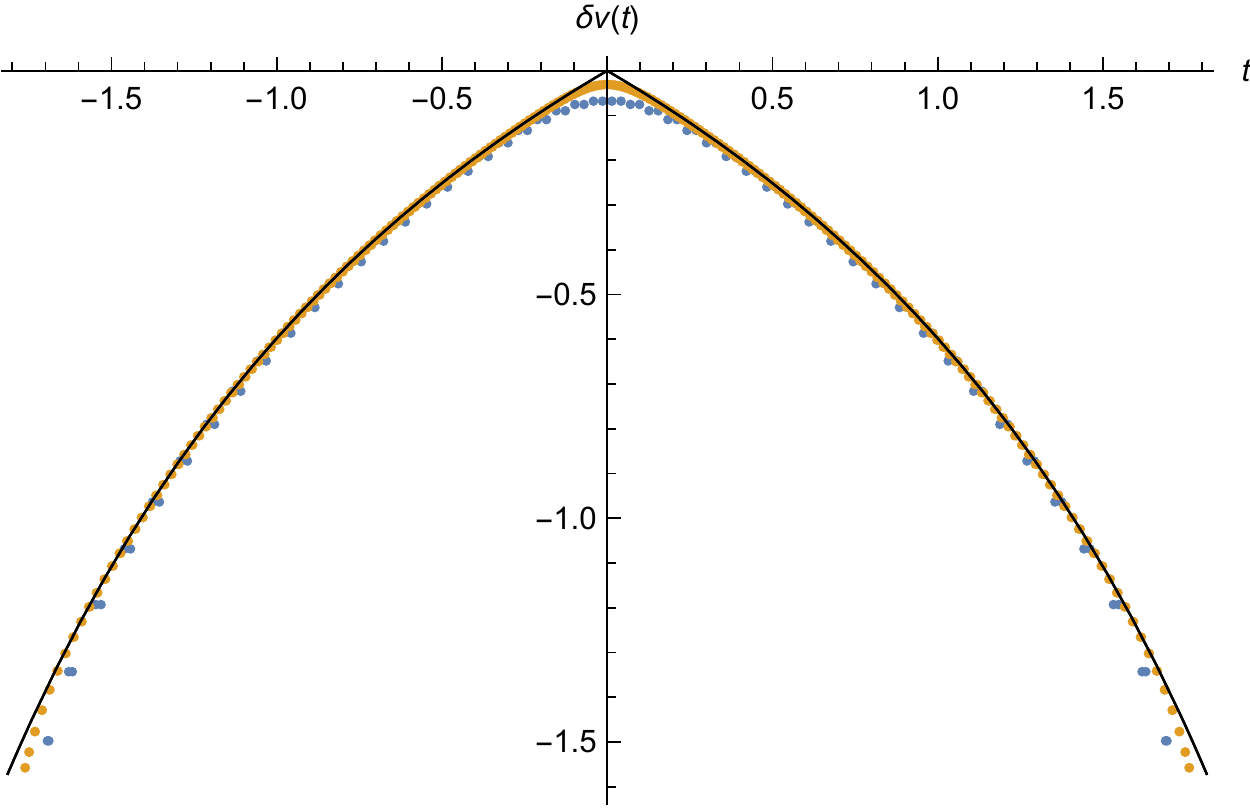}
	        \caption{Real part difference $\delta v(t)$}
	    \end{subfigure}
	    \quad %add desired spacing between images, e. g. ~, \quad, \qquad, \hfill etc. 
	    \begin{subfigure}[b]{0.5\textwidth}
	        \includegraphics[width=\textwidth]{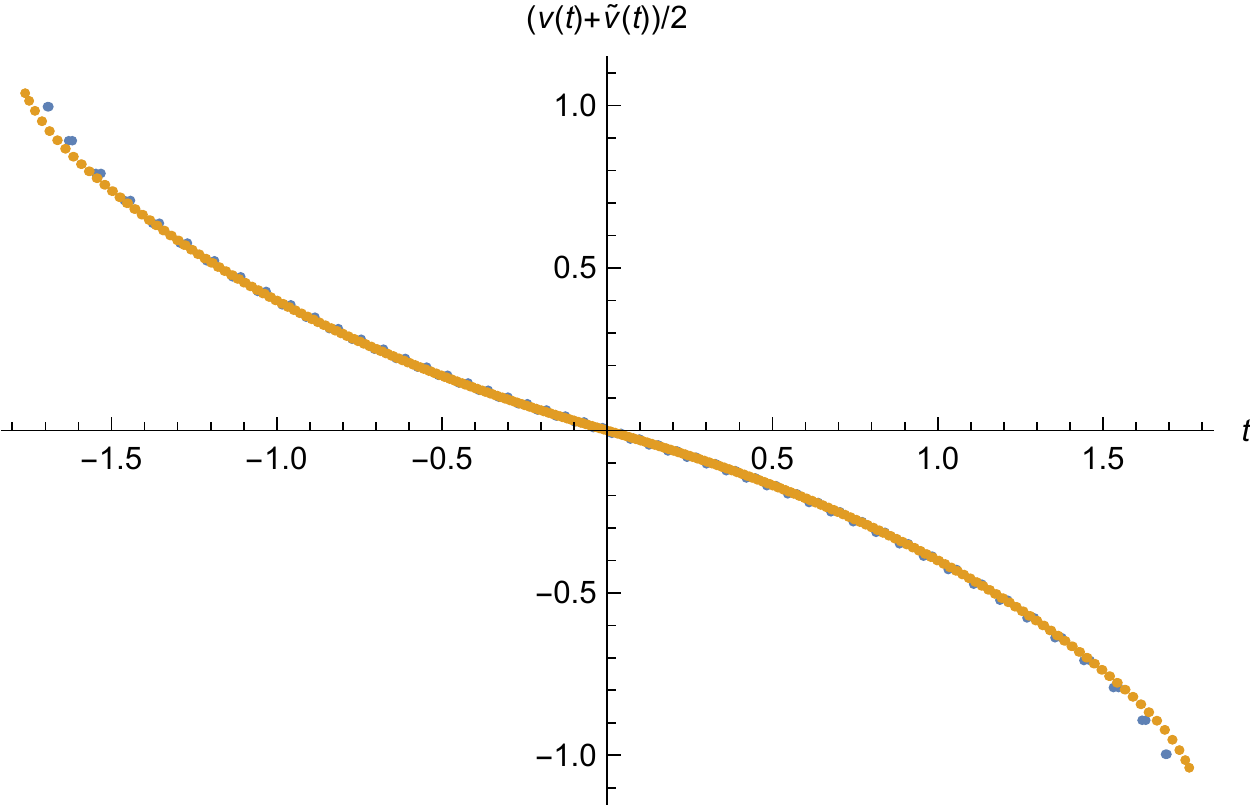}
	        \caption{Real part axis $(v(t)+\tilde{v}(t))/2$}
	    \end{subfigure}
    \end{subfigure}
    \quad %add desired spacing between images, e. g. ~, \quad, \qquad, \hfill etc. 
    \begin{subfigure}[b]{0.2\textwidth}
        \includegraphics[width=\textwidth]{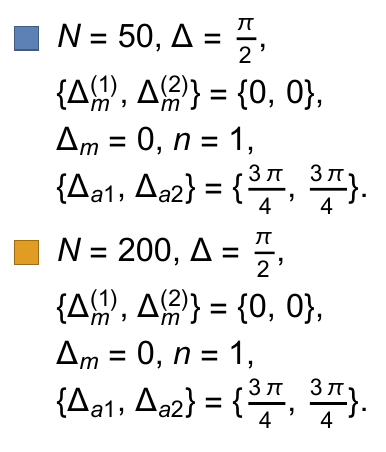}
    \end{subfigure}
    \caption{Eigenvalues for $\Delta_a=\{\frac{\pi}{2}, \frac{\pi}{2}, \frac{\pi}{2}, \frac{\pi}{2}\}$ for $N=50$ (blue) and $200$ (orange) with the same other parameters.}
    \label{fig:Q111_Plot_N}
\end{figure}

\begin{figure}[H]
    \centering
    \begin{subfigure}[b]{0.75\textwidth}
	    \begin{subfigure}[b]{0.5\textwidth}
	        \includegraphics[width=\textwidth]{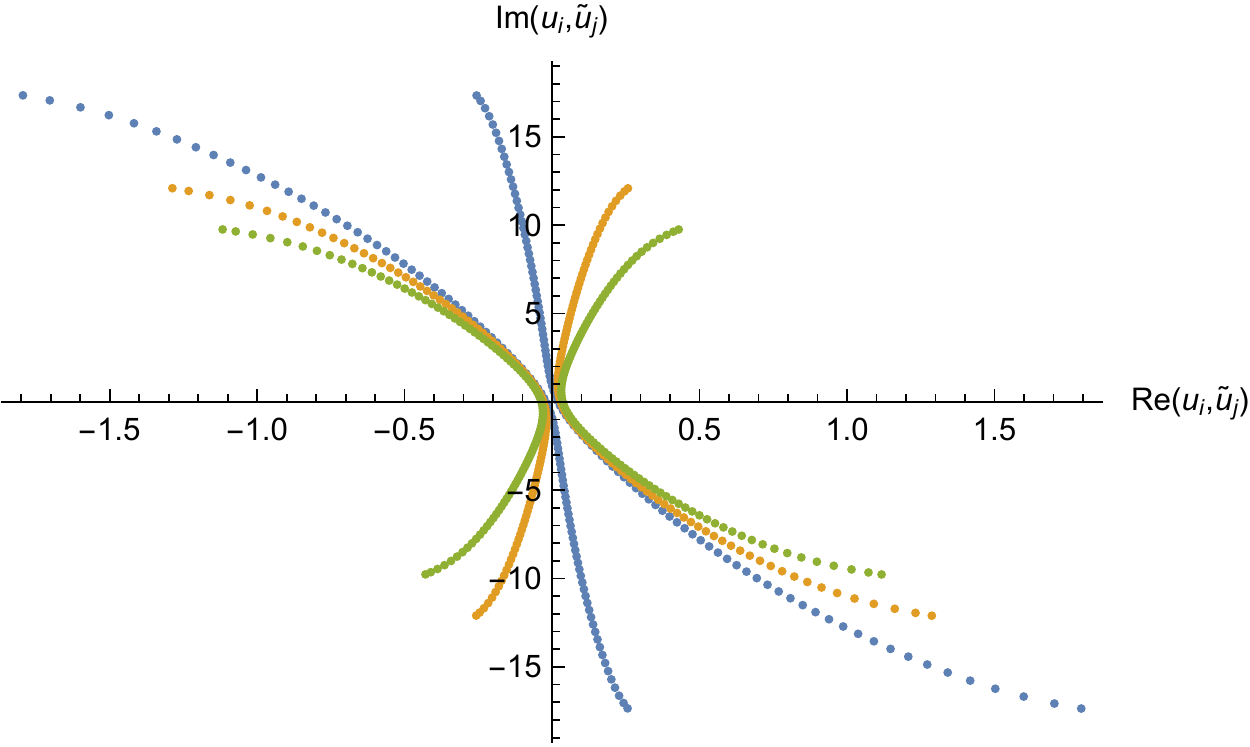}
	        \caption{Eigenvalue distribution}
	    \end{subfigure}
	    \quad %add desired spacing between images, e. g. ~, \quad, \qquad, \hfill etc. 
	    \begin{subfigure}[b]{0.5\textwidth}
	        \includegraphics[width=\textwidth]{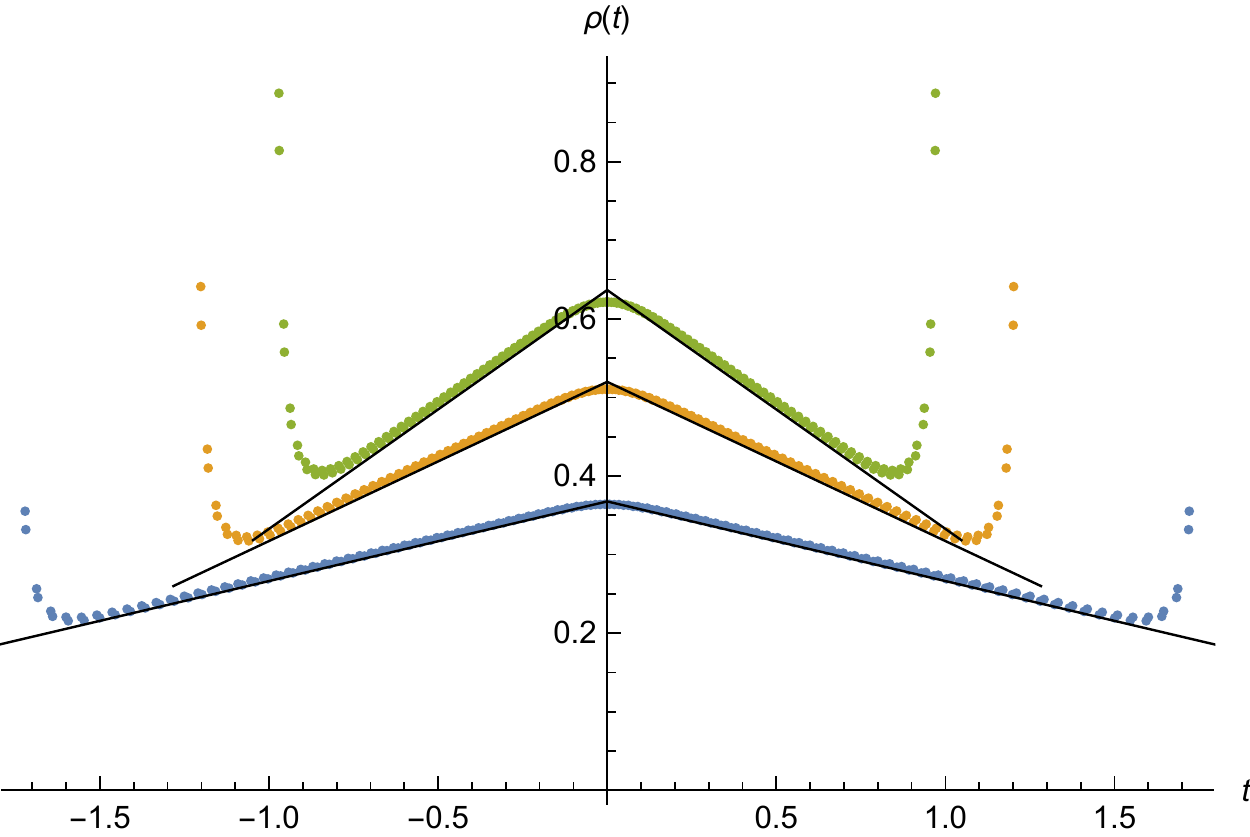}
	        \caption{Eigenvalue density $\rho(t)$}
	    \end{subfigure}\\
	    ~ %add desired spacing between images, e. g. ~, \quad, \qquad, \hfill etc. 
	    \begin{subfigure}[b]{0.5\textwidth}
	        \includegraphics[width=\textwidth]{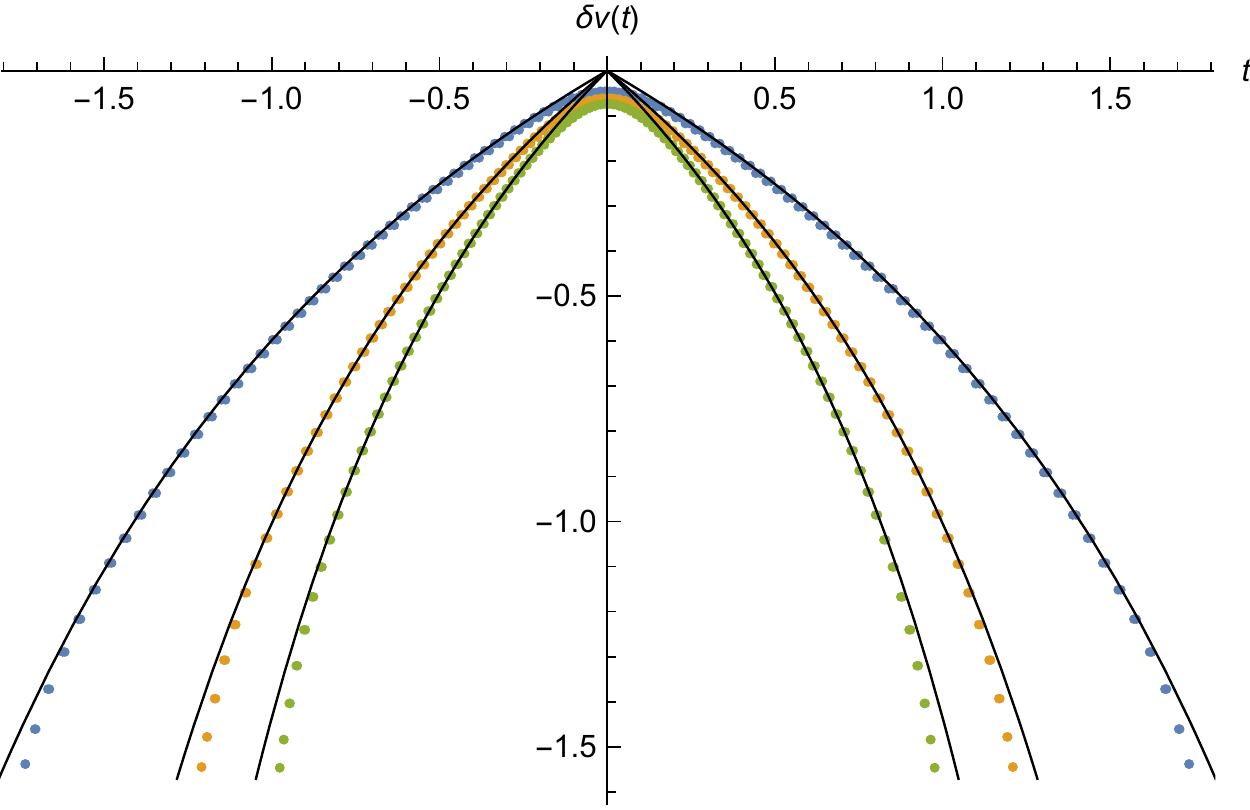}
	        \caption{Real part difference $\delta v(t)$}
	    \end{subfigure}
	    \quad %add desired spacing between images, e. g. ~, \quad, \qquad, \hfill etc. 
	    \begin{subfigure}[b]{0.5\textwidth}
	        \includegraphics[width=\textwidth]{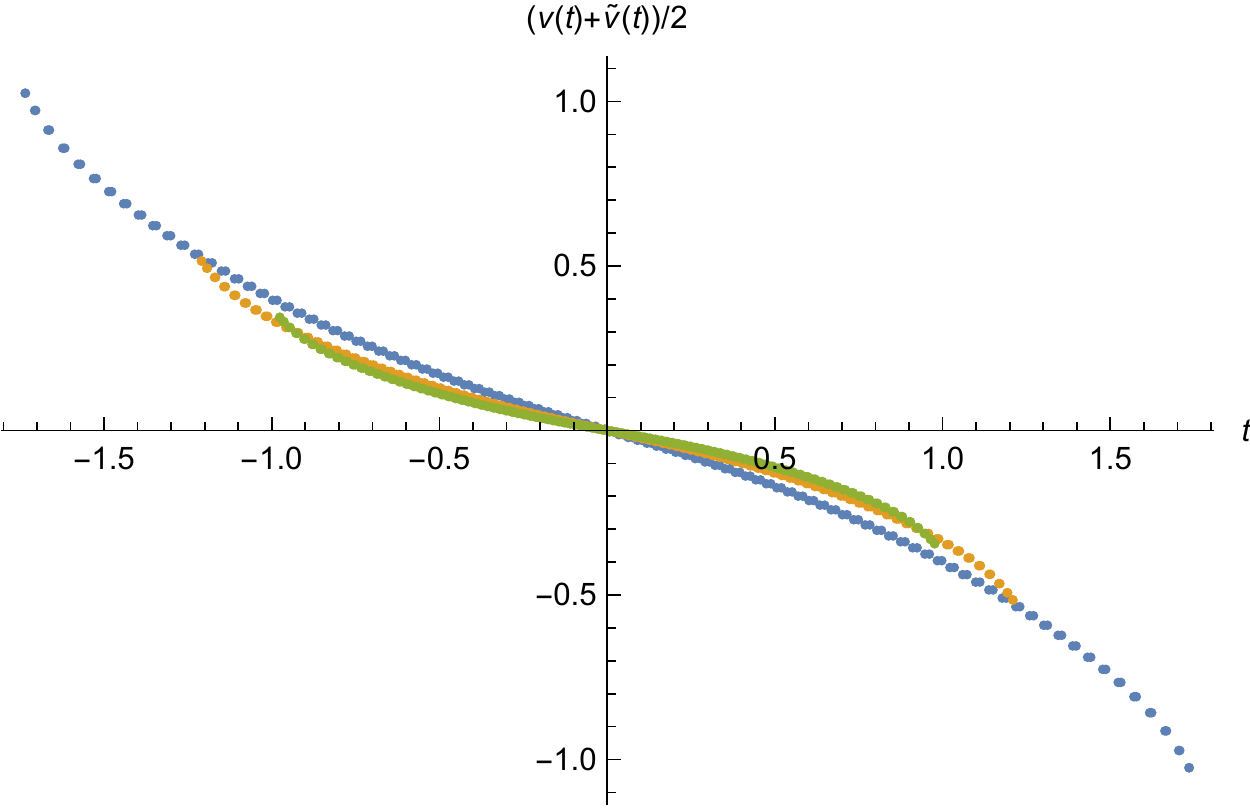}
	        \caption{Real part axis $(v(t)+\tilde{v}(t))/2$}
	    \end{subfigure}
    \end{subfigure}
    \quad %add desired spacing between images, e. g. ~, \quad, \qquad, \hfill etc. 
    \begin{subfigure}[b]{0.2\textwidth}
        \includegraphics[width=\textwidth]{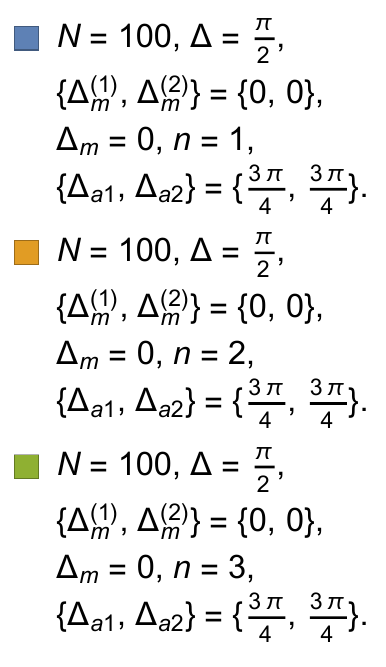}
    \end{subfigure}
    \caption{Eigenvalues for $\Delta_a=\{\frac{\pi}{2}, \frac{\pi}{2}, \frac{\pi}{2}, \frac{\pi}{2}\}$ for $n=1$ (blue), $2$ (orange) and $3$ (green) with the same other parameters.}
    \label{fig:Q111_Plot_nq}
\end{figure}

\begin{figure}[H]
    \centering
    \begin{subfigure}[b]{0.75\textwidth}
	    \begin{subfigure}[b]{0.5\textwidth}
	        \includegraphics[width=\textwidth]{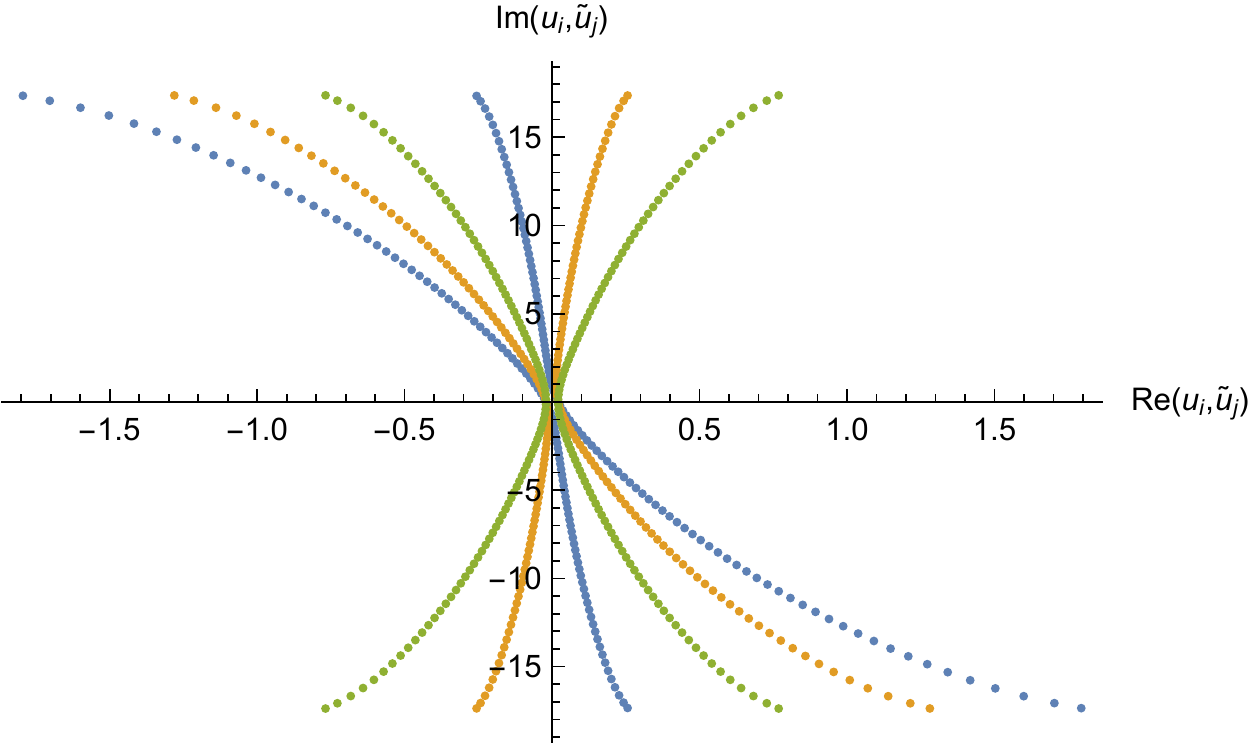}
	        \caption{Eigenvalue distribution}
	    \end{subfigure}
	    \quad %add desired spacing between images, e. g. ~, \quad, \qquad, \hfill etc. 
	    \begin{subfigure}[b]{0.5\textwidth}
	        \includegraphics[width=\textwidth]{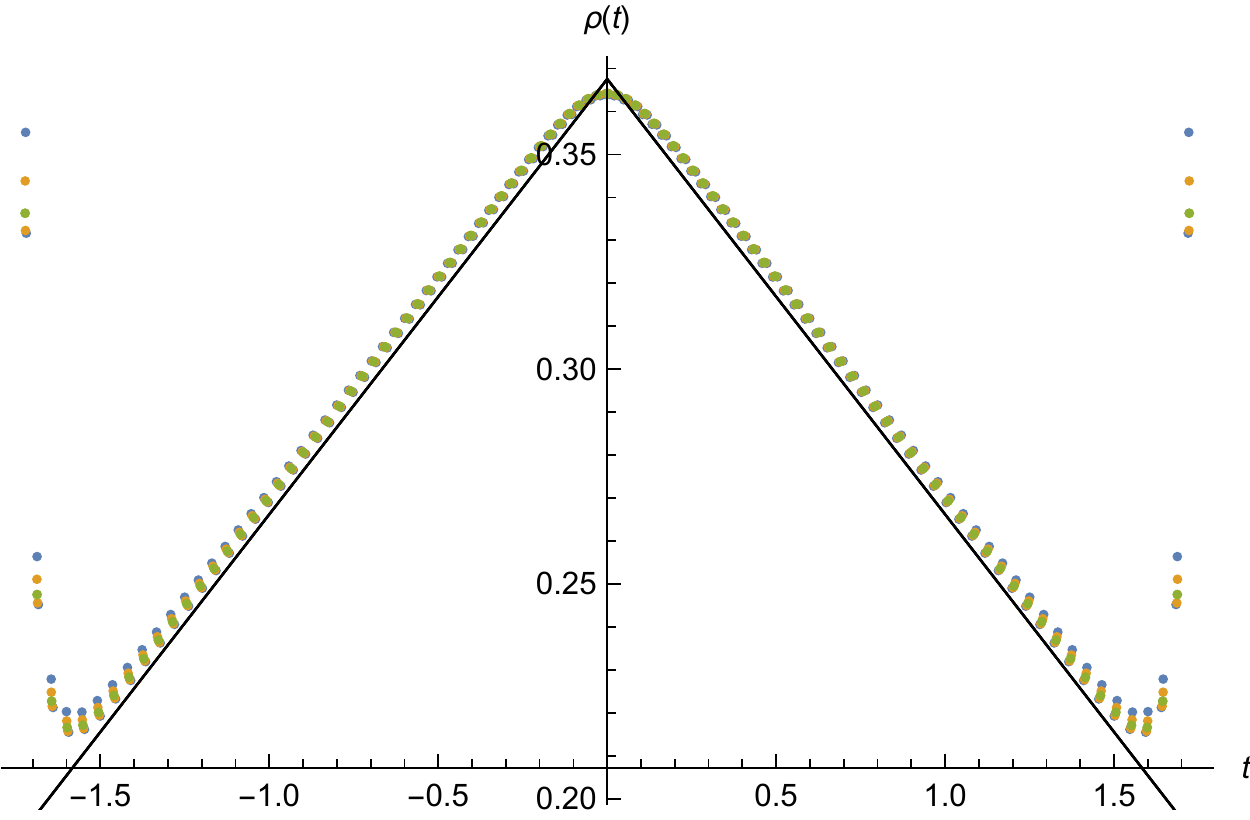}
	        \caption{Eigenvalue density $\rho(t)$}
	    \end{subfigure}\\
	    ~ %add desired spacing between images, e. g. ~, \quad, \qquad, \hfill etc. 
	    \begin{subfigure}[b]{0.5\textwidth}
	        \includegraphics[width=\textwidth]{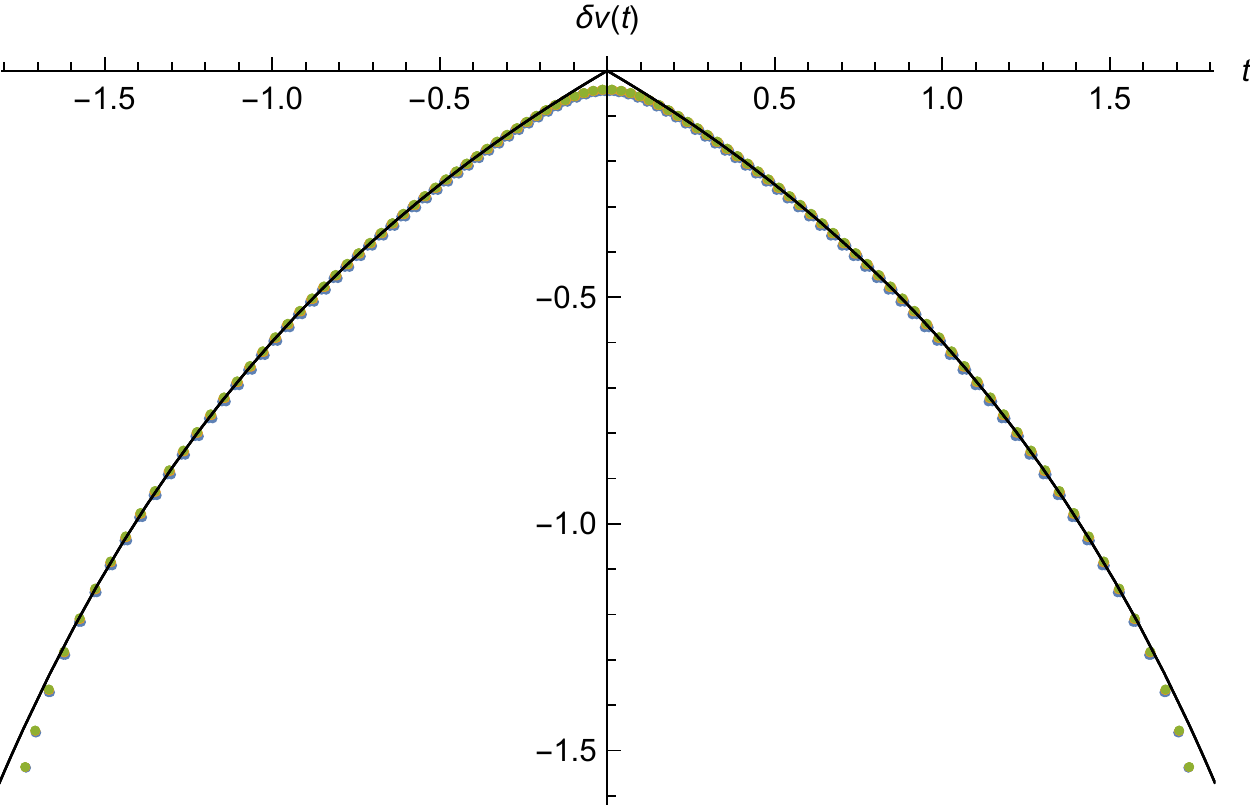}
	        \caption{Real part difference $\delta v(t)$}
	    \end{subfigure}
	    \quad %add desired spacing between images, e. g. ~, \quad, \qquad, \hfill etc. 
	    \begin{subfigure}[b]{0.5\textwidth}
	        \includegraphics[width=\textwidth]{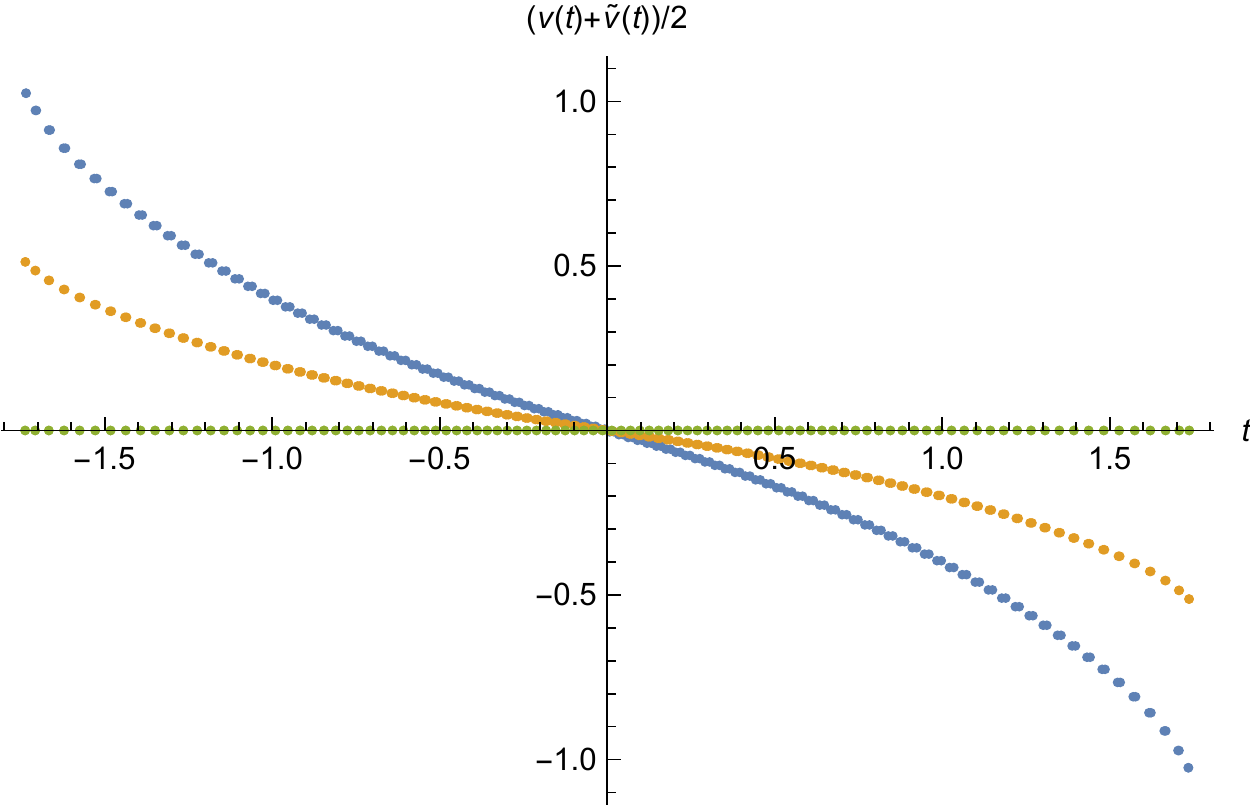}
	        \caption{Real part axis $(v(t)+\tilde{v}(t))/2$}
	    \end{subfigure}
    \end{subfigure}
    \quad %add desired spacing between images, e. g. ~, \quad, \qquad, \hfill etc. 
    \begin{subfigure}[b]{0.2\textwidth}
        \includegraphics[width=\textwidth]{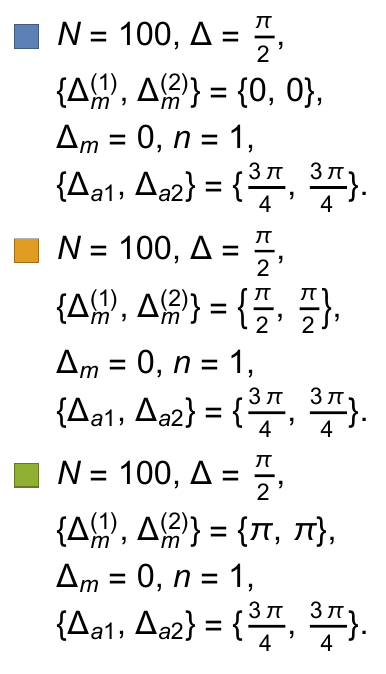}
    \end{subfigure}
    \caption{Eigenvalues for $\Delta_a=\{\frac{\pi}{2}, \frac{\pi}{2}, \frac{\pi}{2}, \frac{\pi}{2}\}$ for $\{\Delta_m^{(1)}, \Delta_m^{(2)}\}=\{0,0\}$ (blue), $\{\frac{\pi}{2},\frac{\pi}{2}\}$ (orange) and $\{\pi,\pi\}$ (green) keeping the same $\Delta_m=0$ with the same other parameters.}
    \label{fig:Q111_Plot_Delm12}
\end{figure}

\begin{figure}[H]
    \centering
    \begin{subfigure}[b]{0.75\textwidth}
	    \begin{subfigure}[b]{0.5\textwidth}
	        \includegraphics[width=\textwidth]{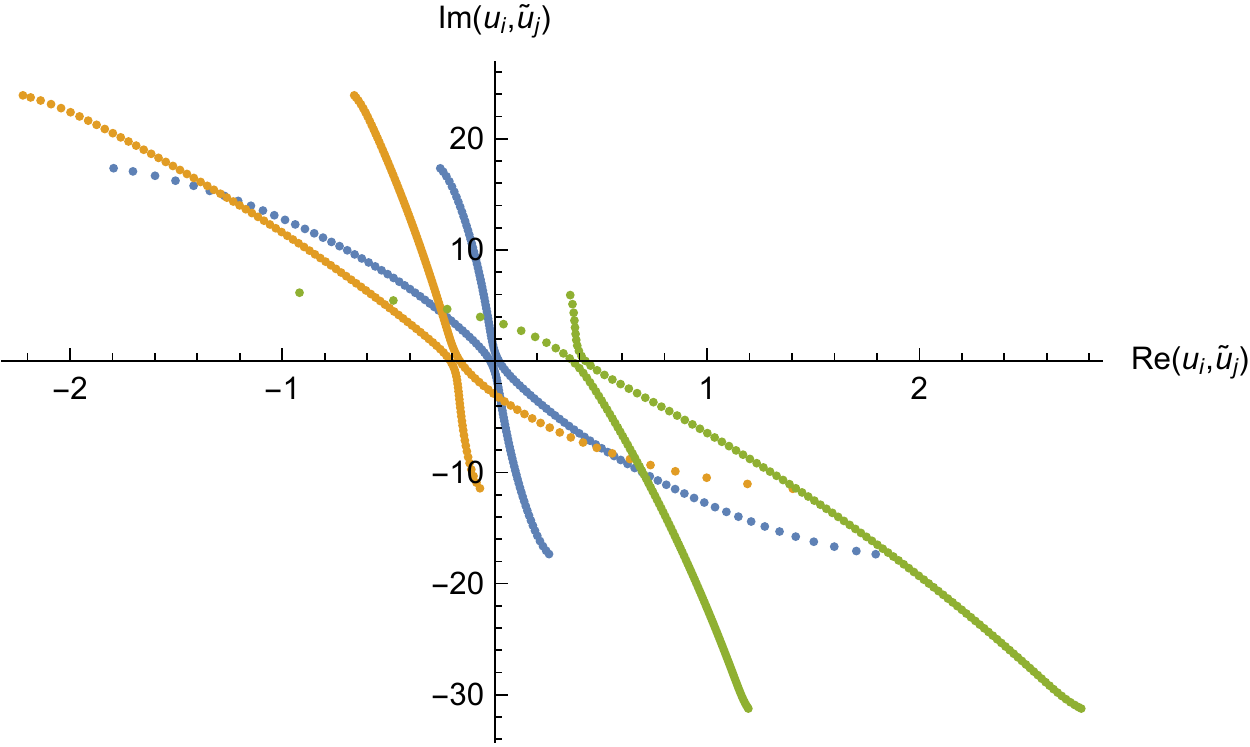}
	        \caption{Eigenvalue distribution}
	    \end{subfigure}
	    \quad %add desired spacing between images, e. g. ~, \quad, \qquad, \hfill etc. 
	    \begin{subfigure}[b]{0.5\textwidth}
	        \includegraphics[width=\textwidth]{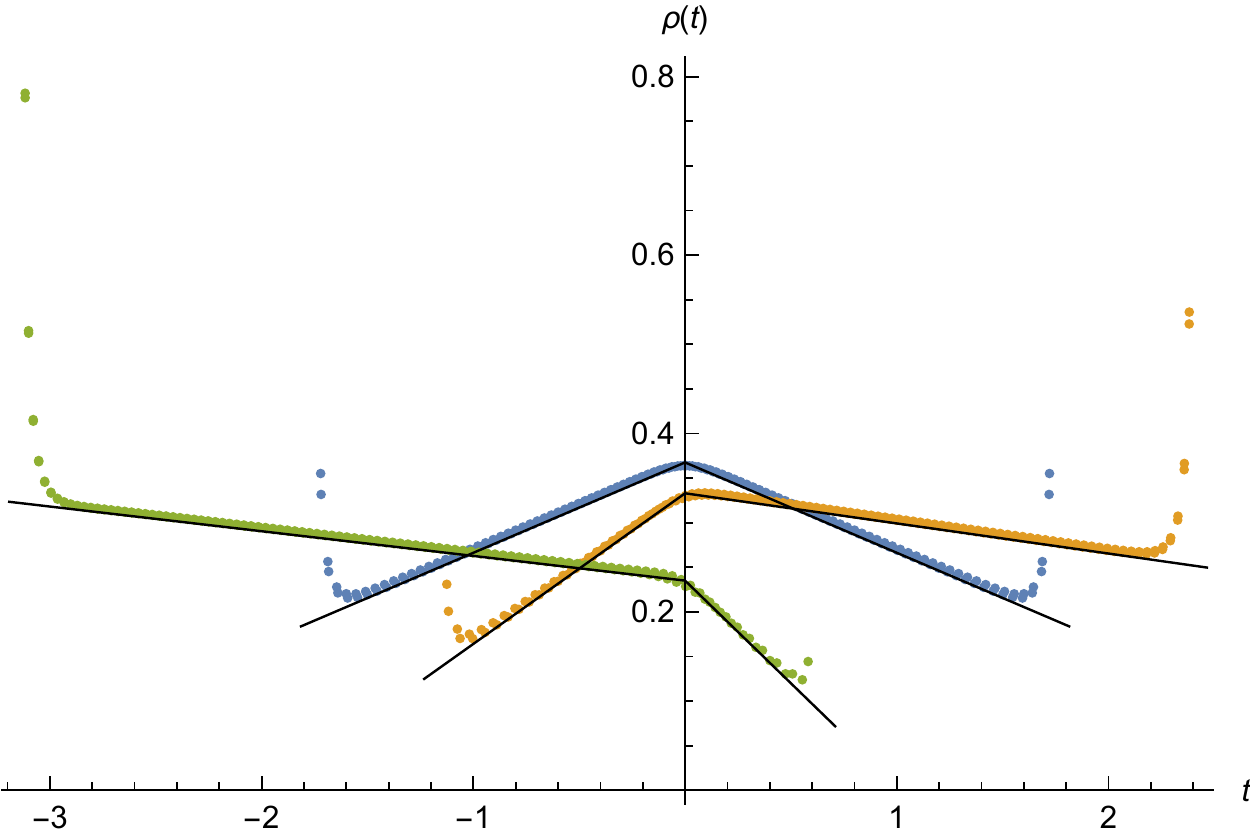}
	        \caption{Eigenvalue density $\rho(t)$}
	    \end{subfigure}\\
	    ~ %add desired spacing between images, e. g. ~, \quad, \qquad, \hfill etc. 
	    \begin{subfigure}[b]{0.5\textwidth}
	        \includegraphics[width=\textwidth]{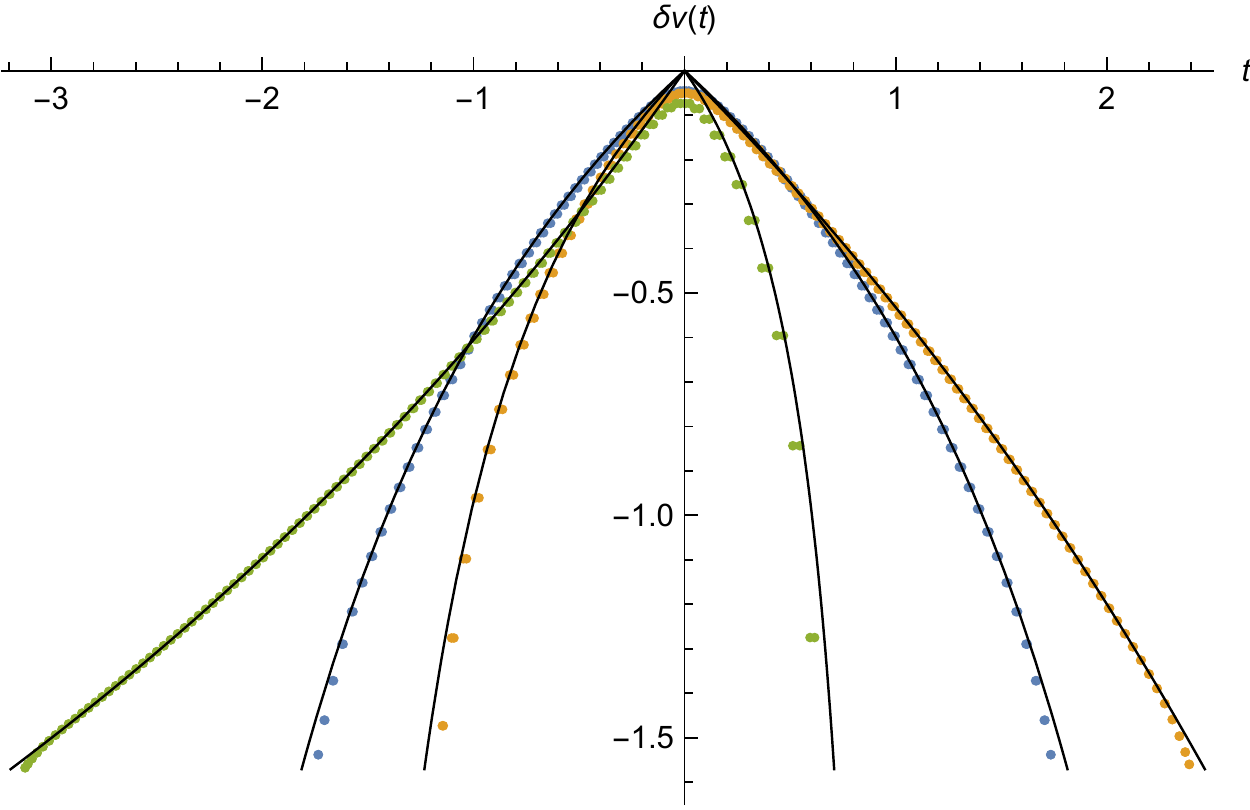}
	        \caption{Real part difference $\delta v(t)$}
	    \end{subfigure}
	    \quad %add desired spacing between images, e. g. ~, \quad, \qquad, \hfill etc. 
	    \begin{subfigure}[b]{0.5\textwidth}
	        \includegraphics[width=\textwidth]{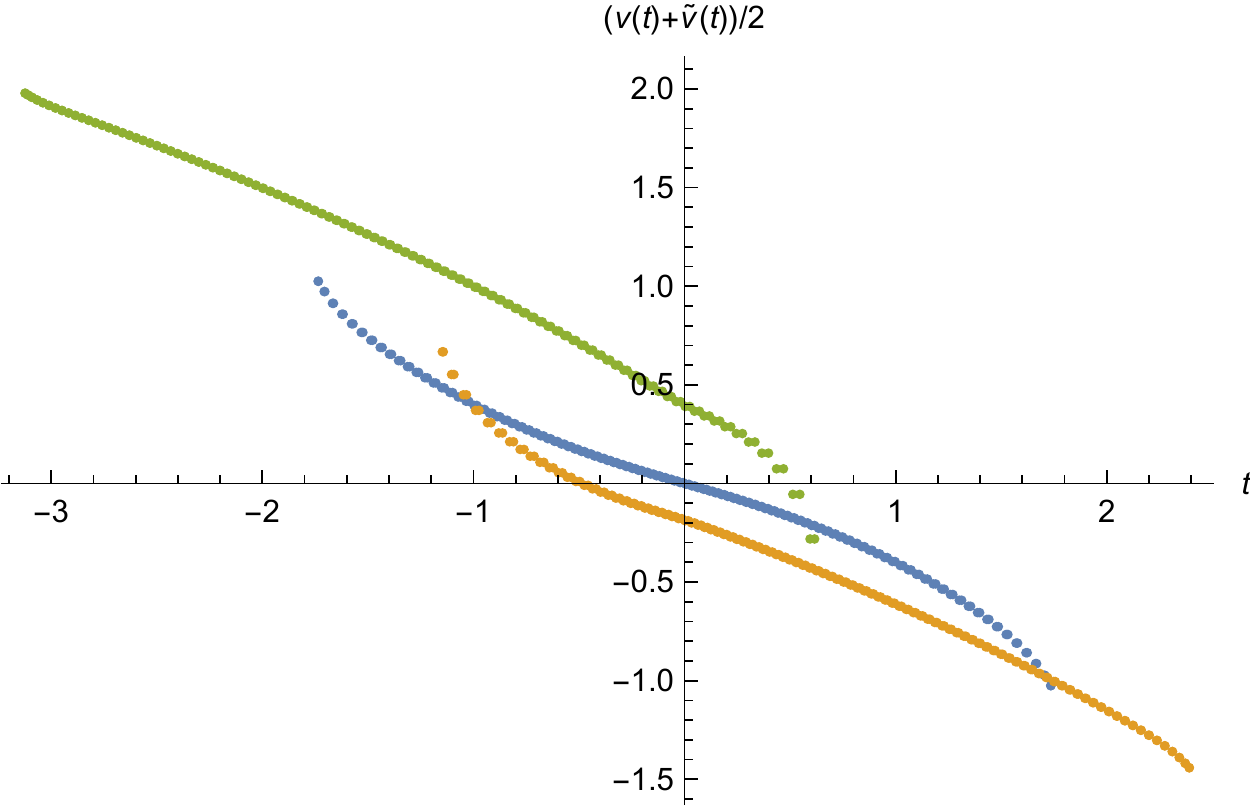}
	        \caption{Real part axis $(v(t)+\tilde{v}(t))/2$}
	    \end{subfigure}
    \end{subfigure}
    \quad %add desired spacing between images, e. g. ~, \quad, \qquad, \hfill etc. 
    \begin{subfigure}[b]{0.2\textwidth}
        \includegraphics[width=\textwidth]{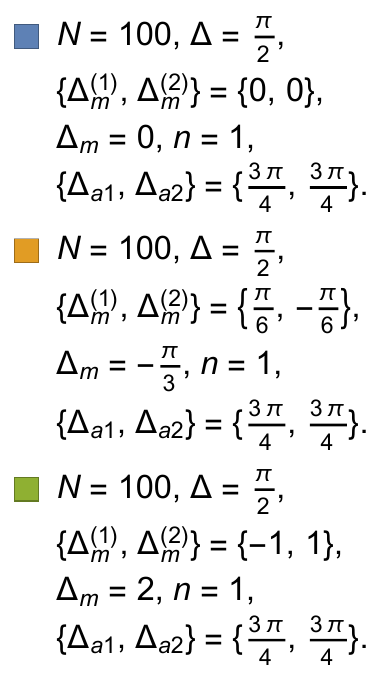}
    \end{subfigure}
    \caption{Eigenvalues for $\Delta_a=\{\frac{\pi}{2}, \frac{\pi}{2}, \frac{\pi}{2}, \frac{\pi}{2}\}$ for $\Delta_m=0, \{\Delta_m^{(1)}, \Delta_m^{(2)}\}=\{0,0\}$ (blue), $\Delta_m=-\frac{\pi}{3}, \{\Delta_m^{(1)}, \Delta_m^{(2)}\}=\{-\frac{\pi}{6},\frac{\pi}{6}\}$ (orange) and $\Delta_m=2, \{\Delta_m^{(1)}, \Delta_m^{(2)}\}=\{1,-1\}$ (green) with the same other parameters.}
    \label{fig:Q111_Plot_Delm}
\end{figure}

\begin{figure}[H]
    \centering
    \begin{subfigure}[b]{0.75\textwidth}
	    \begin{subfigure}[b]{0.5\textwidth}
	        \includegraphics[width=\textwidth]{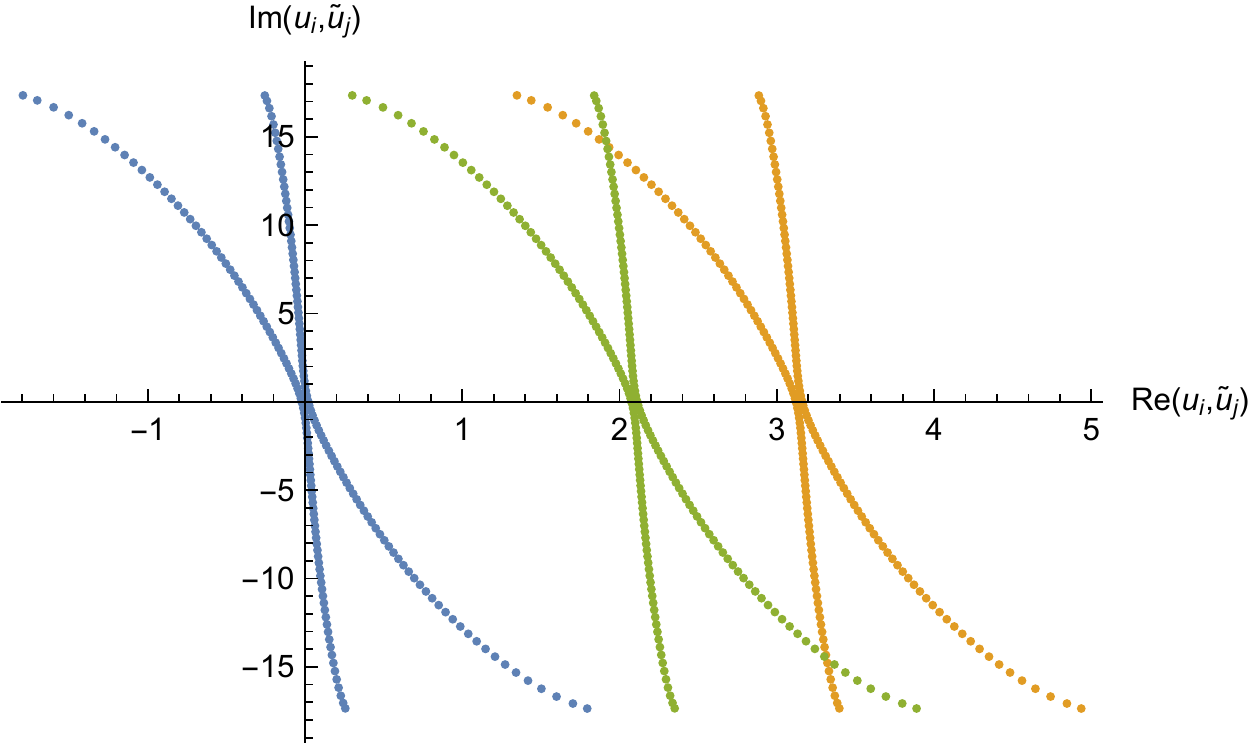}
	        \caption{Eigenvalue distribution}
	    \end{subfigure}
	    \quad %add desired spacing between images, e. g. ~, \quad, \qquad, \hfill etc. 
	    \begin{subfigure}[b]{0.5\textwidth}
	        \includegraphics[width=\textwidth]{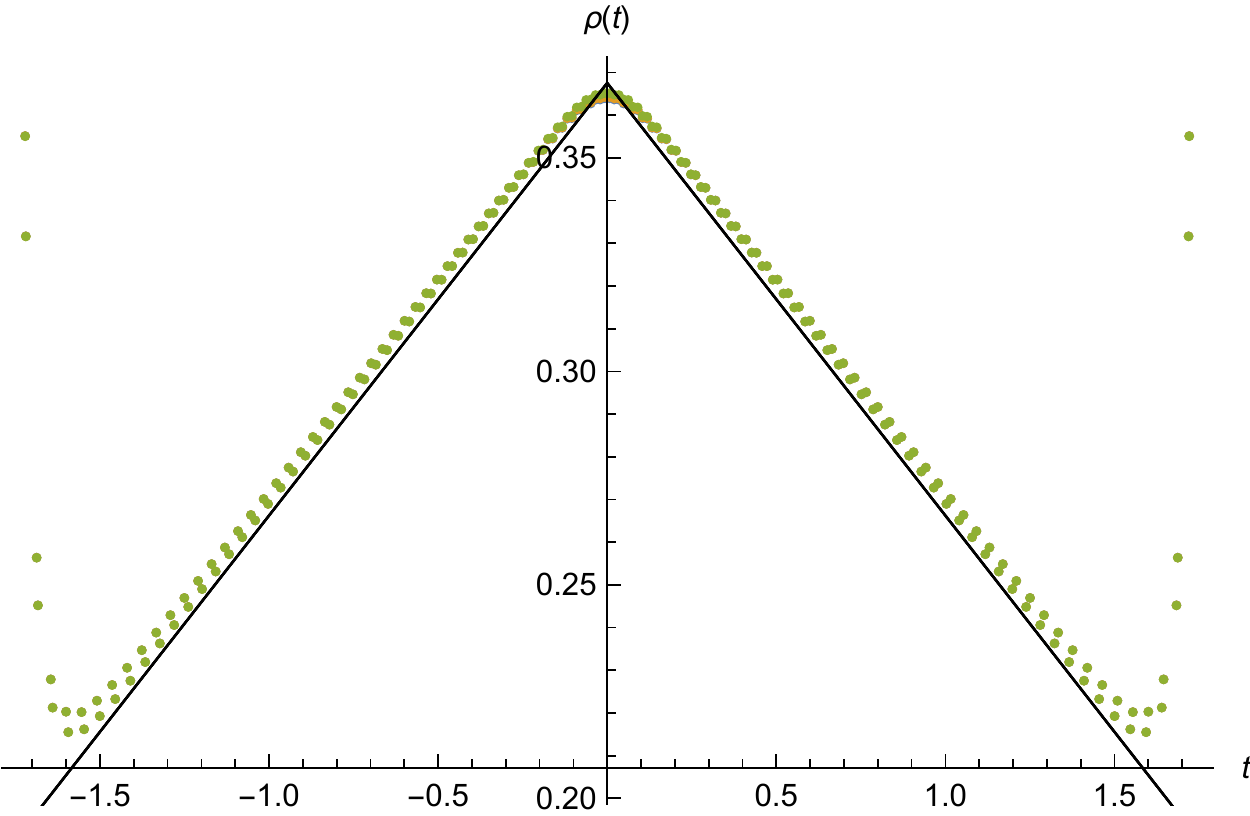}
	        \caption{Eigenvalue density $\rho(t)$}
	    \end{subfigure}\\
	    ~ %add desired spacing between images, e. g. ~, \quad, \qquad, \hfill etc. 
	    \begin{subfigure}[b]{0.5\textwidth}
	        \includegraphics[width=\textwidth]{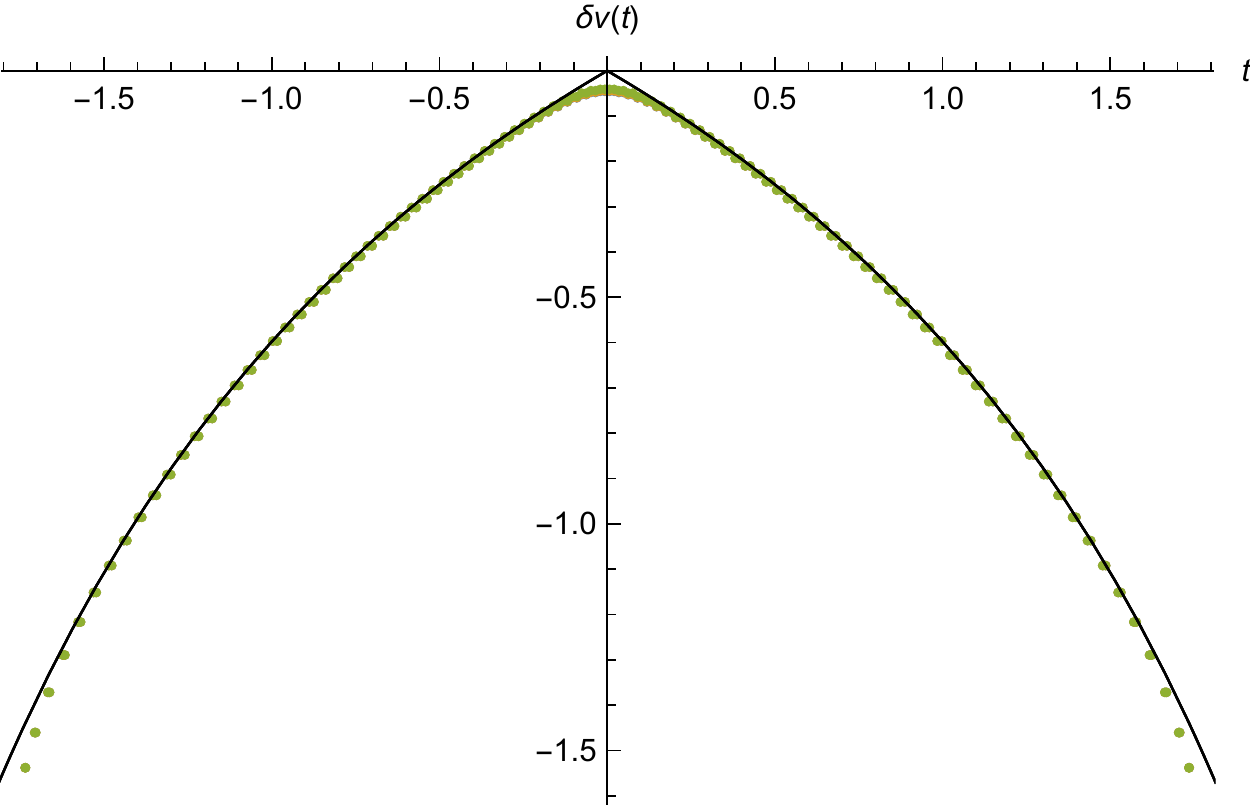}
	        \caption{Real part difference $\delta v(t)$}
	    \end{subfigure}
	    \quad %add desired spacing between images, e. g. ~, \quad, \qquad, \hfill etc. 
	    \begin{subfigure}[b]{0.5\textwidth}
	        \includegraphics[width=\textwidth]{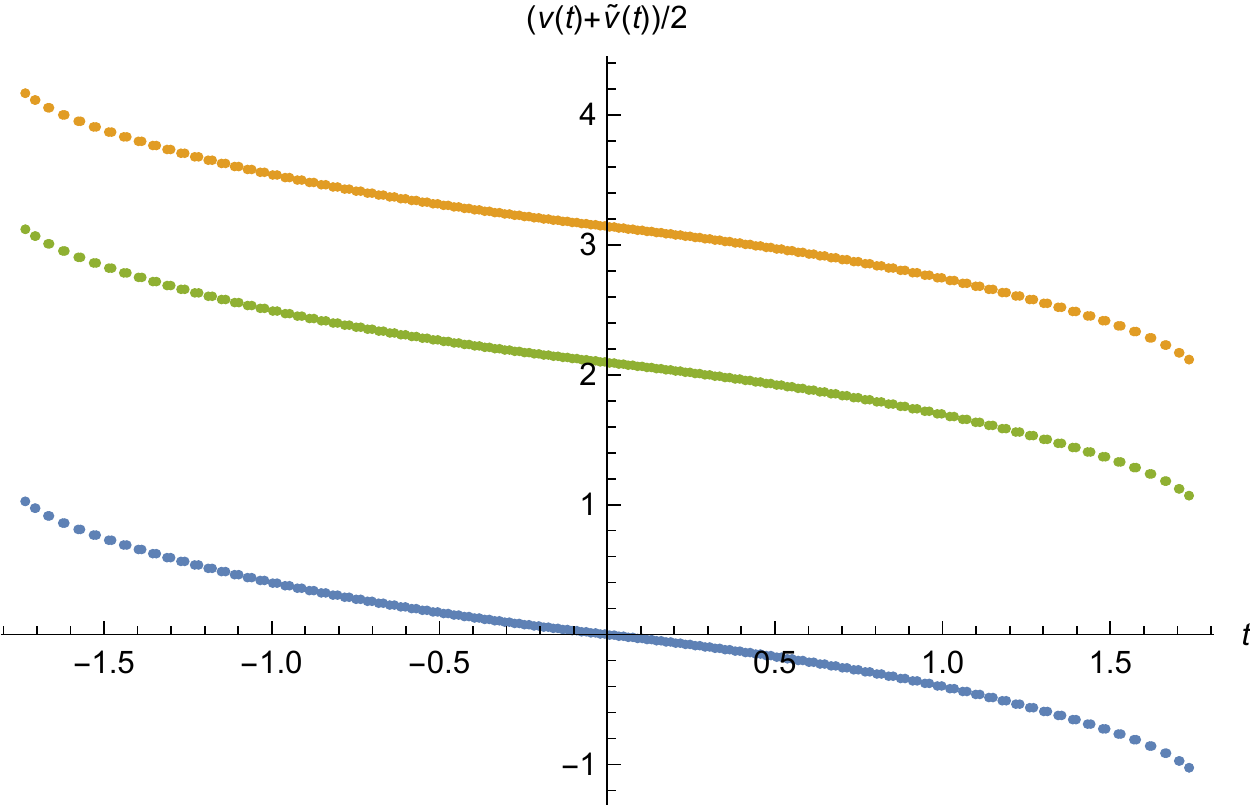}
	        \caption{Real part axis $(v(t)+\tilde{v}(t))/2$}
	    \end{subfigure}
    \end{subfigure}
    \quad %add desired spacing between images, e. g. ~, \quad, \qquad, \hfill etc. 
    \begin{subfigure}[b]{0.2\textwidth}
        \includegraphics[width=\textwidth]{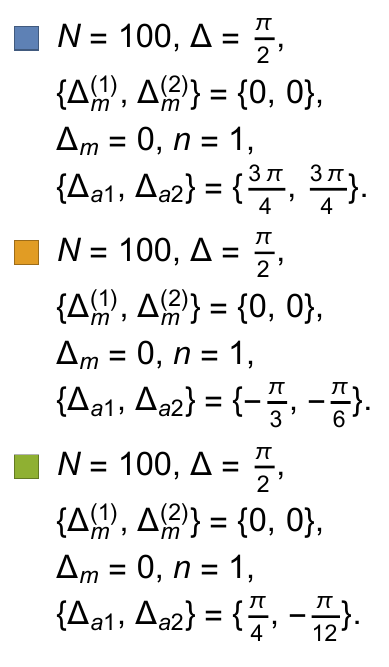}
    \end{subfigure}
    \caption{Eigenvalues for $\Delta_a=\{\frac{\pi}{2}, \frac{\pi}{2}, \frac{\pi}{2}, \frac{\pi}{2}\}$ for $\{\Delta_{a1}, \Delta_{a2}, \tilde{\Delta}_{a1}, \tilde{\Delta}_{a2}\}=\{\frac{3\pi}{4}, \frac{3\pi}{4}, \frac{3\pi}{4}, \frac{3\pi}{4}\}$ (blue), $\{-\frac{\pi}{3}, -\frac{\pi}{6}, \frac{11\pi}{6}, \frac{5\pi}{3}\}$ (orange) and $\{\frac{\pi}{4}, -\frac{\pi}{12}, \frac{5\pi}{4}, \frac{19\pi}{12}\}$ (green) with the same other parameters.}
    \label{fig:Q111_Plot_Delaq}
\end{figure}

\begin{figure}[H]
    \centering
    \begin{subfigure}[b]{0.75\textwidth}
	    \begin{subfigure}[b]{0.5\textwidth}
	        \includegraphics[width=\textwidth]{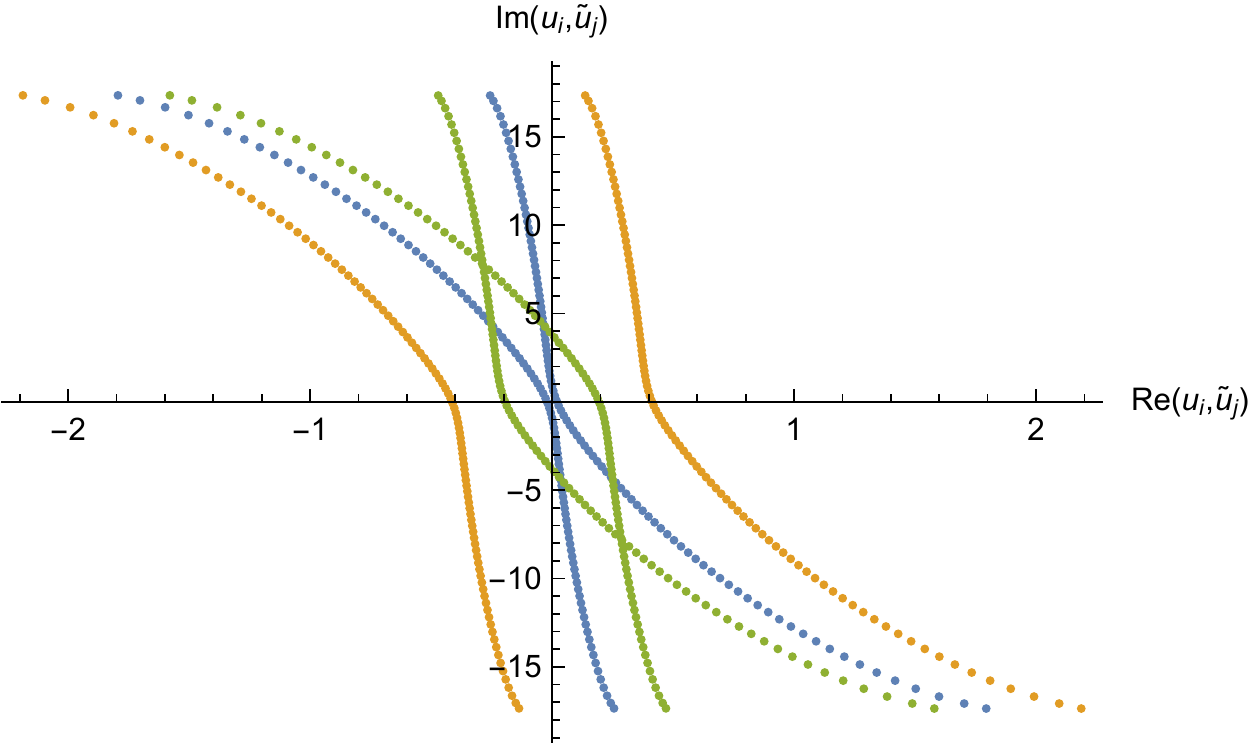}
	        \caption{Eigenvalue distribution}
	    \end{subfigure}
	    \quad %add desired spacing between images, e. g. ~, \quad, \qquad, \hfill etc. 
	    \begin{subfigure}[b]{0.5\textwidth}
	        \includegraphics[width=\textwidth]{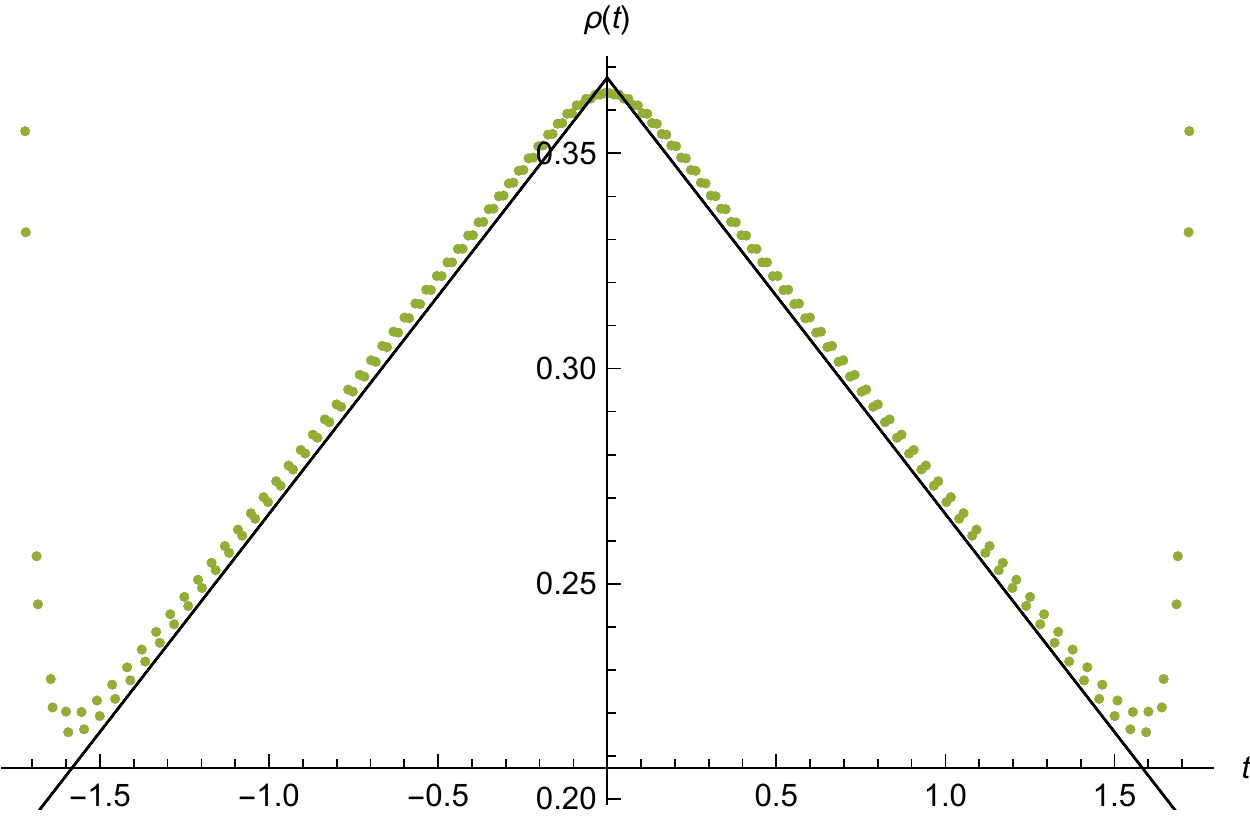}
	        \caption{Eigenvalue density $\rho(t)$}
	    \end{subfigure}\\
	    ~ %add desired spacing between images, e. g. ~, \quad, \qquad, \hfill etc. 
	    \begin{subfigure}[b]{0.5\textwidth}
	        \includegraphics[width=\textwidth]{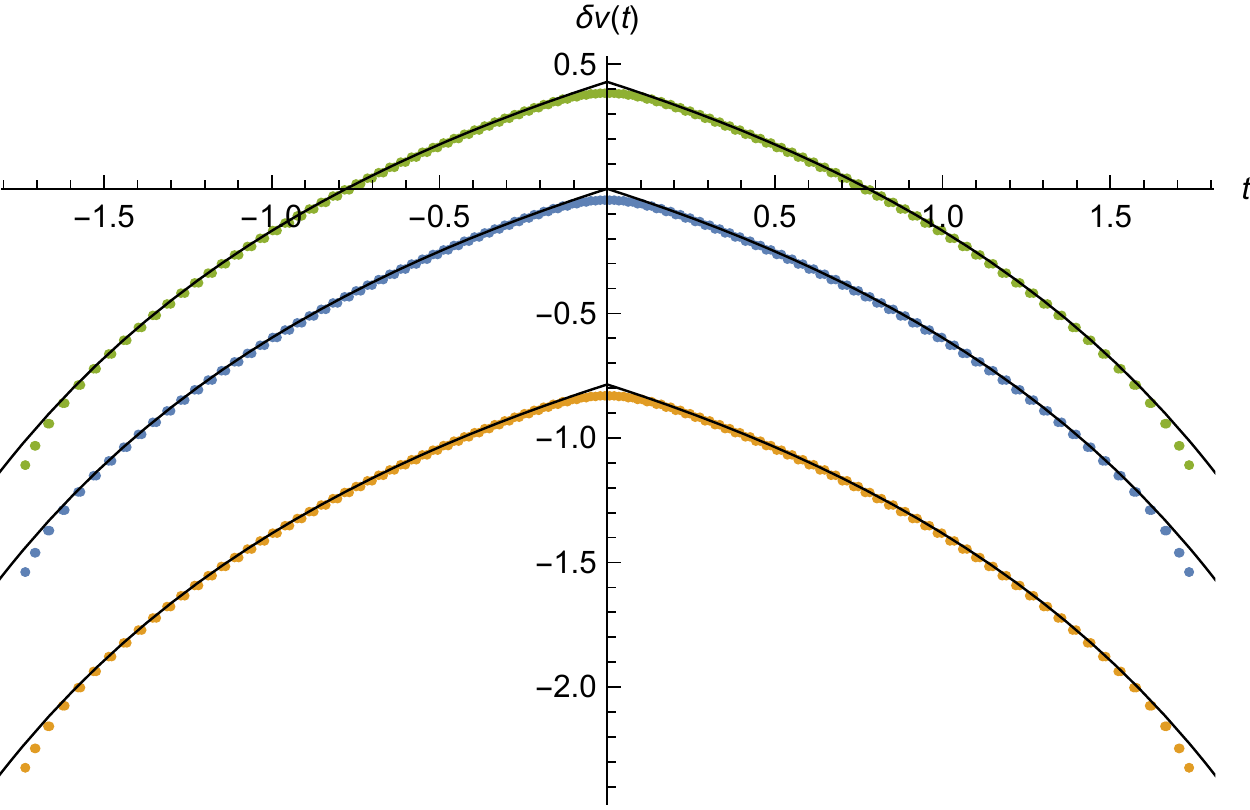}
	        \caption{Real part difference $\delta v(t)$}
	    \end{subfigure}
	    \quad %add desired spacing between images, e. g. ~, \quad, \qquad, \hfill etc. 
	    \begin{subfigure}[b]{0.5\textwidth}
	        \includegraphics[width=\textwidth]{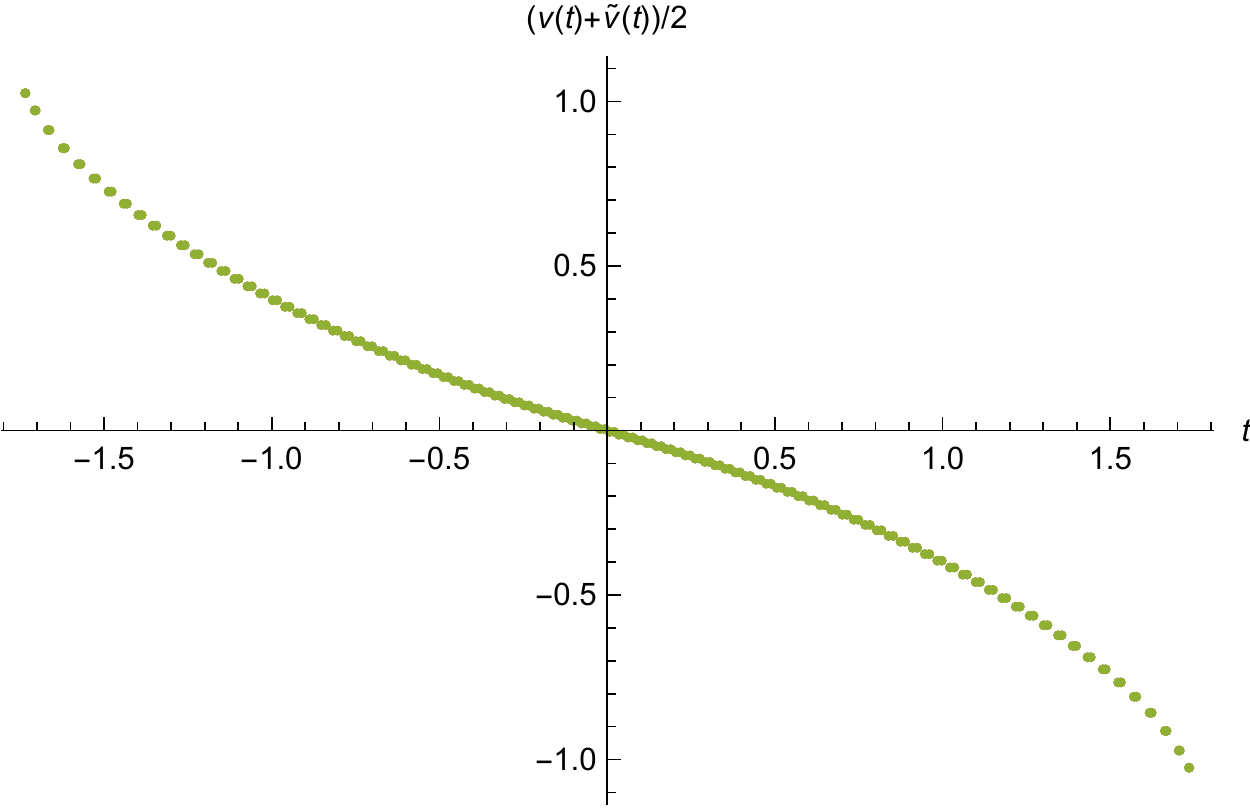}
	        \caption{Real part axis $(v(t)+\tilde{v}(t))/2$}
	    \end{subfigure}
    \end{subfigure}
    \quad %add desired spacing between images, e. g. ~, \quad, \qquad, \hfill etc. 
    \begin{subfigure}[b]{0.2\textwidth}
        \includegraphics[width=\textwidth]{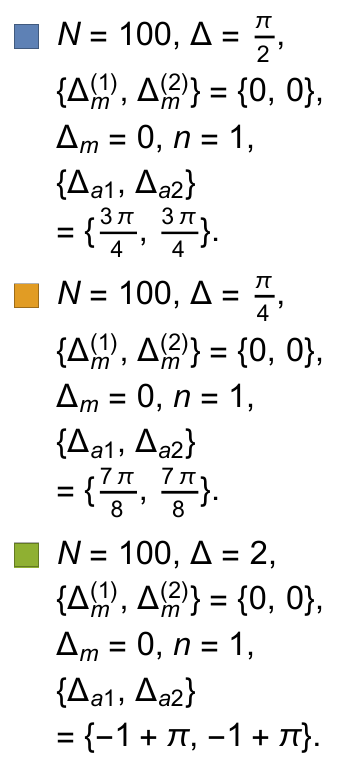}
    \end{subfigure}
    \caption{Eigenvalues for $\Delta_a=\{\frac{\pi}{2}, \frac{\pi}{2}, \frac{\pi}{2}, \frac{\pi}{2}\},\,\{\Delta_{a1}, \Delta_{a2}, \tilde{\Delta}_{a1}, \tilde{\Delta}_{a2}\}=\{\frac{3\pi}{4}, \frac{3\pi}{4}, \frac{3\pi}{4}, \frac{3\pi}{4}\}$ (blue), $\Delta_a=\{\frac{\pi}{4}, \frac{\pi}{4}, \frac{3\pi}{4}, \frac{3\pi}{4}\},\,\{\Delta_{a1}, \Delta_{a2}, \tilde{\Delta}_{a1}, \tilde{\Delta}_{a2}\}=\{\frac{7\pi}{8}, \frac{7\pi}{8}, \frac{7\pi}{8}, \frac{7\pi}{8}\}$ (orange) and $\Delta_a=\{2, 2, \pi-2, \pi-2\},\,\{\Delta_{a1}, \Delta_{a2}, \tilde{\Delta}_{a1}, \tilde{\Delta}_{a2}\}=\{\pi-1, \pi-1, \pi-1, \pi-1\}$ (green) with the same other parameters.}
    \label{fig:Q111_Plot_Dela}
\end{figure}

We set $\Delta_m=\Delta_m^{(2)}-\Delta_m^{(1)}$. The numerical solutions for different values of $N$, $n$, $\{\Delta_m^{(1)}, \Delta_m^{(2)}\}$, $\Delta_m$, $\{\Delta_{a1}, \Delta_{a2}, \tilde{\Delta}_{a1}, \tilde{\Delta}_{a2}\}$ and $\Delta_a=\{\Delta_1, \Delta_2, \Delta_3, \Delta_4\}(\Delta)$ are shown in Figures~\ref{fig:Q111_Plot_N} - \ref{fig:Q111_Plot_Dela}.

Let us briefly summarize the salient features of the exact eigenvalues that we find. Similar to the index of $N^{0,1,0}$ in section \ref{Sec:N010_BAEs}, the eigenvalues are not reflectively symmetric about a particular real axis, except for the case that $(\Delta_m^{(1)}+\Delta_m^{(2)})/2=\pi$. Furthermore, the imaginary part of $u_i$ is not exactly the same as $\tilde{u}_i$ so that there are two numerical results of the eigenvalue density $\rho(t)$, the real part difference $\delta v(t)$ and the real part axis $\left(v(t)+\tilde{v}(t)\right)/2$. In addition, Figure~\ref{fig:Q111_Plot_Delm12} and Figure~\ref{fig:Q111_Plot_Delaq} show that the values of $\{\Delta_m^{(1)}, \Delta_m^{(2)}\}$ keeping the same $\Delta_m$ and the values of $\{\Delta_{a1}, \Delta_{a2}, \tilde{\Delta}_{a1}, \tilde{\Delta}_{a2}\}$ have effects only on the real part axis $\left(v(t)+\tilde{v}(t)\right)/2$.

\subsection{The subleading term of the index at large $N$}

The index should take the form
\be
{\rm Re}\log Z=f_1(n,\Delta,\mathfrak{n},\mathfrak{t})N^{3/2}+f_2(n,\Delta,\mathfrak{n},\mathfrak{t})N^{1/2}+f_3(n,\Delta,\mathfrak{n},\mathfrak{t})\log{N}+f_4(n,\Delta,\mathfrak{n},\mathfrak{t})+\mathcal O(N^{-1/2}),
\ee
where here $\Delta$ represents all of the chemical potentials as above and the functions $f_1$, $f_2$, $f_3$ and $f_4$ are linear in the magnetic fluxes $\mathfrak{n}$ and $\mathfrak{t}$.

The index Eq.~(\ref{Eq:Q111_Index}) and ${\rm Re}\log Z$ can be computed using the numerical solutions. Under the similar decomposition
\be
{\rm Re}\log Z=A+B_1\mathfrak{n}_1+B_2\mathfrak{n}_2+B_3\mathfrak{n}_3+B_4\mathfrak{n}_{a1}+B_5\mathfrak{n}_{a2}+B_6\mathfrak{t}+B_7\tilde{\mathfrak{t}},
\ee
where we have used the marginality condition on the superpotential $\sum_{a=1}^4\mathfrak{n}_a=2$, $\mathfrak{n}_1+\mathfrak{n}_{a1}+\tilde{\mathfrak{n}}_{a1}=2$ and $\mathfrak{n}_2+\mathfrak{n}_{a2}+\tilde{\mathfrak{n}}_{a2}=2$. Then we perform a linear least-squares fit for $A$ and $B_a$ to the function
\be
f(N)=f_1N^{3/2}+f_2N^{1/2}+f_3\log{N}+f_4+\sum_{p=1}^{p_c} f_{p+4}\,\,N^{\left(1-2p\right)/2}.
\label{Eq:N010FitlogZN}
\ee

The results of the numerical fit for ${\rm Re}\log Z$ with $N$ are presented in Table~\ref{tbl:Q111_FitRelogZ}. The analytical leading term computed by the index theorem in \cite{Hosseini:2016ume} and the numerical leading term $f_1N^{3/2}$ match to number of significant digits present in the table. And the leading term is indeed independent of $\{\Delta_m^{(1)}, \Delta_m^{(2)}\}$ keeping the same $\Delta_m$, and $\{\Delta_{a1}, \Delta_{a2}, \tilde{\Delta}_{a1}, \tilde{\Delta}_{a2}\}$. The numerical results indicate that the coefficient $f_3$ of the $\log N$ term is precisely $-1/2$. 

%%%%
\begin{table}[htp]
\renewcommand{\arraystretch}{1}
\setlength{\tabcolsep}{5pt}
\begin{subtable}[h]{1\textwidth}
\centering
\caption{$k=0$;$\,\Delta_a=\{\frac{\pi}{2}, \frac{\pi}{2}, \frac{\pi}{2}, \frac{\pi}{2}\}(\Delta=\frac{\pi}{2})$.}
\begin{tabular}{|c c|c|c|c c||l|l|l|l|}
\hline
$\Delta_m^{(1)}$&$\Delta_m^{(2)}$&$\Delta_m$&$n$&$\Delta_{a1}$&$\Delta_{a2}$&$f_1$&$f_2$&$f_3$&$f_4$\\
\hline
$0$&$0$&$0$&$1$&$\frac{3\pi}{4}$&$\frac{3\pi}{4}$&$-2.41840$&\tabincell{l}{$+1.81381$\\$+0.60460\mathfrak{t}$\\$-0.60460\tilde{\mathfrak{t}}$}&$-0.50033$&$-2.13933$ \\
\hline
$0$&$0$&$0$&$2$&$\frac{3\pi}{4}$&$\frac{3\pi}{4}$&$-3.42013$&\tabincell{l}{$+1.28258$\\$+0.42752\mathfrak{t}$\\$-0.42752\tilde{\mathfrak{t}}$}&$-0.50064$&\tabincell{l}{$-2.03420$\\$+0.00004\mathfrak{t}$\\$-0.00004\tilde{\mathfrak{t}}$} \\
\hline
$0$&$0$&$0$&$3$&$\frac{3\pi}{4}$&$\frac{3\pi}{4}$&$-4.18879$&\tabincell{l}{$+1.04723$\\$+0.34907\mathfrak{t}$\\$-0.34907\tilde{\mathfrak{t}}$}&\tabincell{l}{$-0.50087$\\$-0.00005\mathfrak{t}$\\$+0.00005\tilde{\mathfrak{t}}$}&\tabincell{l}{$-2.23075$\\$+0.00030\mathfrak{t}$\\$-0.00030\tilde{\mathfrak{t}}$} \\
\hline
$\pi$&$\pi$&$0$&$1$&$\frac{3\pi}{4}$&$\frac{3\pi}{4}$&$-2.41840$&$+1.81384$&$-0.50098$&$-2.13483$ \\
\hline
$\pi$&$\pi$&$0$&$2$&$\frac{3\pi}{4}$&$\frac{3\pi}{4}$&$-3.42013$&$+1.28259$&$-0.50097$&$-2.03192$ \\
\hline
$\pi$&$\pi$&$0$&$3$&$\frac{3\pi}{4}$&$\frac{3\pi}{4}$&$-4.18879$&$+1.04725$&$-0.50117$&$-2.22872$ \\
\hline
$\frac{\pi}{2}$&$\frac{\pi}{2}$&$0$&$1$&$\frac{3\pi}{4}$&$\frac{3\pi}{4}$&$-2.41840$&\tabincell{l}{$+1.81383$\\$+0.30230\mathfrak{t}$\\$-0.30230\tilde{\mathfrak{t}}$}&$-0.50082$&$-2.13596$ \\
\hline
$\frac{\pi}{6}$&$-\frac{\pi}{6}$&$-\frac{\pi}{3}$&$1$&$\frac{3\pi}{4}$&$\frac{3\pi}{4}$&\tabincell{l}{$-2.42234$\\$-0.69511\mathfrak{t}$\\$-0.69511\tilde{\mathfrak{t}}$}&\tabincell{l}{$+2.04473$\\$+0.69572\mathfrak{t}$\\$-0.44172\tilde{\mathfrak{t}}$}&\tabincell{l}{$-0.50044$\\$-0.00015\mathfrak{t}$\\$-0.00014\tilde{\mathfrak{t}}$}&\tabincell{l}{$-2.30777$\\$-0.01877\mathfrak{t}$\\$-0.01878\tilde{\mathfrak{t}}$} \\
\hline
$\frac{7\pi}{6}$&$\frac{5\pi}{6}$&$-\frac{\pi}{3}$&$1$&$\frac{3\pi}{4}$&$\frac{3\pi}{4}$&\tabincell{l}{$-2.42234$\\$-0.69511\mathfrak{t}$\\$-0.69511\tilde{\mathfrak{t}}$}&\tabincell{l}{$+1.96819$\\$-0.10269\mathfrak{t}$\\$-0.10269\tilde{\mathfrak{t}}$}&\tabincell{l}{$-0.50101$\\$+0.00008\mathfrak{t}$\\$+0.00008\tilde{\mathfrak{t}}$}&\tabincell{l}{$-2.30374$\\$-0.02007\mathfrak{t}$\\$-0.02007\tilde{\mathfrak{t}}$} \\
\hline
$-1$&$+1$&$2$&$1$&$\frac{3\pi}{4}$&$\frac{3\pi}{4}$&\tabincell{l}{$-2.47964$\\$+1.46576\mathfrak{t}$\\$+1.46576\tilde{\mathfrak{t}}$}&\tabincell{l}{$+3.10442$\\$+0.02602\mathfrak{t}$\\$-0.86801\tilde{\mathfrak{t}}$}&\tabincell{l}{$-0.49893$\\$-0.00314\mathfrak{t}$\\$-0.00336\tilde{\mathfrak{t}}$}&\tabincell{l}{$-3.26833$\\$+0.29060\mathfrak{t}$\\$+0.29211\tilde{\mathfrak{t}}$} \\
\hline
$\pi-1$&$\pi+1$&$2$&$1$&$\frac{3\pi}{4}$&$\frac{3\pi}{4}$&\tabincell{l}{$-2.47964$\\$+1.46576\mathfrak{t}$\\$+1.46576\tilde{\mathfrak{t}}$}&\tabincell{l}{$+2.70455$\\$+0.20706\mathfrak{t}$\\$+0.20706\tilde{\mathfrak{t}}$}&\tabincell{l}{$-0.49839$\\$-0.00226\mathfrak{t}$\\$-0.00226\tilde{\mathfrak{t}}$}&\tabincell{l}{$-3.26854$\\$+0.27889\mathfrak{t}$\\$+0.27889\tilde{\mathfrak{t}}$} \\
\hline
$0$&$0$&$0$&$1$&$-\frac{\pi}{3}$&$-\frac{\pi}{6}$&$-2.41840$&\tabincell{l}{$+1.81381$\\$-0.07557\mathfrak{n}_1$\\$+0.07557\mathfrak{n}_2$\\$-0.15115\mathfrak{n}_{a1}$\\$+0.15115\mathfrak{n}_{a2}$\\$+0.60460\mathfrak{t}$\\$-0.60460\tilde{\mathfrak{t}}$}&$-0.50031$&$-2.13942$ \\
\hline
$0$&$0$&$0$&$1$&$\frac{\pi}{4}$&$-\frac{\pi}{12}$&$-2.41840$&\tabincell{l}{$+1.81381$\\$+0.15115\mathfrak{n}_1$\\$-0.15115\mathfrak{n}_2$\\$+0.30230\mathfrak{n}_{a1}$\\$-0.30230\mathfrak{n}_{a2}$\\$+0.60460\mathfrak{t}$\\$-0.60460\tilde{\mathfrak{t}}$}&$-0.50027$&$-2.13970$ \\
\hline
\end{tabular}
\end{subtable}
\end{table}

\begin{table}[htp]
\ContinuedFloat
\renewcommand{\arraystretch}{1}
\setlength{\tabcolsep}{4.5pt}
\begin{subtable}[h]{1\textwidth}
\centering
\caption{$k=0$;$\,\Delta_a=\{\frac{\pi}{4}, \frac{\pi}{4}, \frac{3\pi}{4}, \frac{3\pi}{4}\}(\Delta=\frac{\pi}{4})$.}
\begin{tabular}{|c c|c|c|c c||l|l|l|l|}
\hline
$\Delta_m^{(1)}$&$\Delta_m^{(2)}$&$\Delta_m$&$n$&$\Delta_{a1}$&$\Delta_{a2}$&$f_1$&$f_2$&$f_3$&$f_4$\\
\hline
$0$&$0$&$0$&$1$&$\frac{7\pi}{8}$&$\frac{7\pi}{8}$&$-2.41840$&\tabincell{l}{$+1.81381$\\$+0.60460\mathfrak{t}$\\$-0.60460\tilde{\mathfrak{t}}$}&$-0.50033$&$-2.13933$ \\
\hline
$-\pi$&$-\frac{\pi}{4}$&$\frac{3\pi}{4}$&$2$&$-\frac{\pi}{4}$&$0$&\tabincell{l}{$-3.42919$\\$+0.55815\mathfrak{t}$\\$+0.55815\tilde{\mathfrak{t}}$}&\tabincell{l}{$+1.47608$\\$-0.34209\mathfrak{n}_1$\\$+0.34209\mathfrak{n}_2$\\$-0.56442\mathfrak{n}_{a1}$\\$+0.56442\mathfrak{n}_{a2}$\\$+0.61625\mathfrak{t}$\\$-0.66691\tilde{\mathfrak{t}}$}&\tabincell{l}{$-0.49693$\\$-0.00524\mathfrak{n}_1$\\$+0.00524\mathfrak{n}_2$\\$-0.00103\mathfrak{n}_{a1}$\\$+0.00103\mathfrak{n}_{a2}$\\$-0.00517\mathfrak{t}$\\$-0.00581\tilde{\mathfrak{t}}$}&\tabincell{l}{$-2.08016$\\$+0.03162\mathfrak{n}_1$\\$-0.03162\mathfrak{n}_2$\\$+0.00615\mathfrak{n}_{a1}$\\$-0.00615\mathfrak{n}_{a2}$\\$-0.08824\mathfrak{t}$\\$-0.08445\tilde{\mathfrak{t}}$} \\
\hline
$\frac{3\pi}{2}$&$0$&$-\frac{3\pi}{2}$&$3$&$\frac{3\pi}{2}$&$\frac{3\pi}{4}$&\tabincell{l}{$-4.22590$\\$-0.62974\mathfrak{t}$\\$-0.62974\tilde{\mathfrak{t}}$}&\tabincell{l}{$+1.26660$\\$+1.34234\mathfrak{n}_1$\\$-1.34234\mathfrak{n}_2$\\$+2.76857\mathfrak{n}_{a1}$\\$-2.76857\mathfrak{n}_{a2}$\\$+0.02775\mathfrak{t}$\\$-0.12140\tilde{\mathfrak{t}}$}&\tabincell{l}{$-0.49950$\\$+0.00018\mathfrak{n}_1$\\$-0.00018\mathfrak{n}_2$\\$+0.00056\mathfrak{n}_{a1}$\\$-0.00056\mathfrak{n}_{a2}$\\$+0.00088\mathfrak{t}$\\$+0.00085\tilde{\mathfrak{t}}$}&\tabincell{l}{$-1.95975$\\$-0.00141\mathfrak{n}_1$\\$+0.00141\mathfrak{n}_2$\\$-0.00445\mathfrak{n}_{a1}$\\$+0.00445\mathfrak{n}_{a2}$\\$+0.25498\mathfrak{t}$\\$+0.25520\tilde{\mathfrak{t}}$} \\
\hline
\end{tabular}
\end{subtable}
\vspace{3mm}

\begin{subtable}[h]{1\textwidth}
\centering
\caption{$k=0$;$\,\Delta_a=\{2, 2, \pi-2, \pi-2\}(\Delta=2)$.}
\begin{tabular}{|c c|c|c|c c||l|l|l|l|}
\hline
$\Delta_m^{(1)}$&$\Delta_m^{(2)}$&$\Delta_m$&$n$&$\Delta_{a1}$&$\Delta_{a2}$&$f_1$&$f_2$&$f_3$&$f_4$\\
\hline
$0$&$0$&$0$&$1$&$\pi-1$&$\pi-1$&$-2.41840$&\tabincell{l}{$+1.81381$\\$+0.60460\mathfrak{t}$\\$-0.60460\tilde{\mathfrak{t}}$}&$-0.50033$&$-2.13933$ \\
\hline
$\frac{\pi}{3}$&$\frac{5\pi}{3}$&$\frac{4\pi}{3}$&$2$&$-\frac{1}{2}$&$-\frac{3}{2}$&\tabincell{l}{$-3.52661$\\$+1.10095\mathfrak{t}$\\$+1.10095\tilde{\mathfrak{t}}$}&\tabincell{l}{$+1.58880$\\$+0.24574\mathfrak{n}_1$\\$-0.24574\mathfrak{n}_2$\\$+0.49147\mathfrak{n}_{a1}$\\$-0.49147\mathfrak{n}_{a2}$\\$+0.40646\mathfrak{t}$\\$+0.40646\tilde{\mathfrak{t}}$}&\tabincell{l}{$-0.49533$\\$+0.00005\mathfrak{n}_1$\\$-0.00005\mathfrak{n}_2$\\$+0.00009\mathfrak{n}_{a1}$\\$-0.00009\mathfrak{n}_{a2}$\\$-0.00387\mathfrak{t}$\\$-0.00387\tilde{\mathfrak{t}}$}&\tabincell{l}{$-2.48477$\\$-0.00038\mathfrak{n}_1$\\$+0.00038\mathfrak{n}_2$\\$-0.00075\mathfrak{n}_{a1}$\\$+0.00075\mathfrak{n}_{a2}$\\$-0.11412\mathfrak{t}$\\$-0.11412\tilde{\mathfrak{t}}$} \\
\hline
$2$&$-1$&$-3$&$3$&$4\pi-12$&$6-2\pi$&\tabincell{l}{$-4.19444$\\$-0.38206\mathfrak{t}$\\$-0.38206\tilde{\mathfrak{t}}$}&\tabincell{l}{$+0.94252$\\$+0.54645\mathfrak{n}_1$\\$-0.54645\mathfrak{n}_2$\\$+1.16441\mathfrak{n}_{a1}$\\$-1.16441\mathfrak{n}_{a2}$\\$+0.06507\mathfrak{t}$\\$-0.49034\tilde{\mathfrak{t}}$}&\tabincell{l}{$-0.50128$\\$+0.00333\mathfrak{n}_1$\\$-0.00333\mathfrak{n}_2$\\$+0.01182\mathfrak{n}_{a1}$\\$-0.01182\mathfrak{n}_{a2}$\\$+0.00218\mathfrak{t}$\\$+0.00163\tilde{\mathfrak{t}}$}&\tabincell{l}{$-2.10032$\\$-0.01903\mathfrak{n}_1$\\$+0.01903\mathfrak{n}_2$\\$-0.07214\mathfrak{n}_{a1}$\\$+0.07214\mathfrak{n}_{a2}$\\$+0.15335\mathfrak{t}$\\$+0.15687\tilde{\mathfrak{t}}$} \\
\hline
\end{tabular}
\end{subtable}
\caption{($Q^{1,1,1}$) Numerical fit for ${\rm Re}\log Z=f_1N^{3/2}+f_2N^{1/2}+f_3\log{N}+f_4+\sum_{p=1}^{p_c=5} f_{p+4}\,N^{\left(1-2p\right)/2}$ and $N$ ranges from $100$ to $200$ in steps of $5$ except for four cases. For numerical stability and accuracy, in the cases $\Delta=\frac{\pi}{2}, \Delta_m=2, \{\Delta_m^{(1)}, \Delta_m^{(2)}\}=\{-1,+1\}$ and $\{\Delta_m^{(1)}, \Delta_m^{(2)}\}=\{\pi-1,\pi+1\}$, $N$ ranges from $100$ to $200$ in steps of $5$ but $p_c=15$. In the case $\Delta=\frac{\pi}{4}, \Delta_m=-\frac{3\pi}{2}, \{\Delta_m^{(1)}, \Delta_m^{(2)}\}=\{\frac{3\pi}{2},0\}, n=3$, $N$ ranges from $100$ to $300$ in steps of $5$ and $p_c=30$. And in the case $\Delta=2, \Delta_m=\frac{4\pi}{3}, \{\Delta_m^{(1)}, \Delta_m^{(2)}\}=\{\frac{\pi}{3},\frac{5\pi}{3}\}, n=2$, $N$ ranges from $200$ to $400$ in steps of $5$ and $p_c=30$.}
\label{tbl:Q111_FitRelogZ}
\end{table}
%%%%

%%%%%%%%%%%%%%%%%%%%%%%%%%%%%%%%%%%%%%%%%%%%%%%%%%%%%%%
\section{One-loop  entropy in eleven dimensional supergravity}\label{Sec:Sugra}
%%%%%%%%%%

Inspired by the seminal work of ABJM  \cite{Aharony:2008ug} who established the now prototypical dual pair of (AdS$_4\times S^7/\mathbb{Z}_k$)/CFT$_3$ where CFT$_3$ stands for the particular Chern-Simons matter theory discussed in section \ref{Sec:TTI},   a plethora of similar examples was  constructed.  A natural way to establish new dual pairs is to consider, on the gravity side, appropriate manifolds that could replace the seven-sphere, $S^7$. The starting point are  Freund-Rubin type solutions of the form  AdS$_4\times\, M^7$ for a certain list of seven-dimensional Sasaki-Einstein spaces, $M^7$     \cite{Freund:1980xh}.  A fairly complete description of solutions of seven dimensional manifolds, providing Freund-Ruben solutions to 11d supegravity, was cataloged by Duff, Nilsson and Pope in \cite{Duff:1986hr} (see a previous discussion in \cite{Castellani:1983yg}).  The list includes further specification about those which are supersymmetric and states what fraction of the supersymmetry is preserved. An exhaustive list of Sasaki-Einstein seven-dimensional manifolds is presented in \cite{Friedrich:1990zg}. Some prominent cases in the list  include  $M^7= \{S^7,	Q^{1,1,1}, M^{1,1,1}, V^{5,2}, N^{0,1,0}\}$ and their quotients by $\mathbb{Z}_k$.  The typical structure of those manifolds is that of toric Sasaki-Einstein manifolds and can be written as a $U(1)$ bundled over a Kaehler-Einstein base.  For example,  $M^{1,1,1}$   is geometrically  a $U(1)$ bundle over $\mathbb{CP}^2\times S^2$ , the dual quiver Chern-Simons matter theory was discussed in \cite{Martelli:2008si,Hanany:2008cd};  $Q^{1,1,1}$   is geometrically a  $U(1)$ bundle over $S^2\times S^2 \times S^2$, the dual theory is an ${\cal N}=2$ supersymmetric Chern-Simons matter quiver gauge theory   \cite{Benini:2009qs,Cheon:2011vi,Benini:2009qs,Jafferis:2009th,Cremonesi:2010ae,Franco:2009sp}. The one  non-toric case in the list  AdS$_4 \times  V^{5,2}$ was addressed in \cite{Martelli:2009ga,Cheon:2011vi}. For all these dual pairs the free energy of the field theory on $S^3$ was shown to agree with the regularized on-shell action on the gravity side largely using techniques presented in \cite{Herzog:2010hf} (see also  \cite{Amariti:2019pky} for recent applications). More recently, the topologically twisted index of a number of these field theories has been computed \cite{Hosseini:2016tor,Hosseini:2016ume,Jain:2019lqb,Jain:2019euv}. 

%%%%%%%%%%%%%%%%%%%%%%%%%

Our goal in this section is to to compute the logarithmic correction to the entropy of the magnetically charged black holes dual to the field theory computations presented in the previous sections and establish that it coincides with the result of the field theory side. To compute such logarithmic corrections one requires only low energy data, that is, only the spectrum of massless fields which in this case would be eleven-dimensional supergravity with background asymptoting to the Freund-Rubin spaces mentioned above plus magnetic flux components. These IR corrections provide a litmus test for the would-be UV complete description of gravity which in our case are simply the Chern-Simons matter field theories discussed in the previous sections. Such powerful IR window into UV physics was studied by Ashoke Sen and collaborators in the case of asymptotically flat string theory black holes \cite{Sen:2011ba,Sen:2012cj}; in this case string theory provides the UV complete result and the IR results are, again, furnished by supergravity theories. In the context of the AdS/CFT correspondence, there have been some developments in matching the gravity computation to the coefficient of $\log N$ term  on the field theory side   \cite{Liu:2017vll,Jeon:2017aif,Liu:2017vbl,Hristov:2018lod, Liu:2018bac,Gang:2019uay, PandoZayas:2019hdb,Hristov:2019xku,Nian:2017hac}. For the cases of AdS/CFT pairs arising from M5 branes wrapping hyperbolic three-manifolds, the field theory results were obtained analytically  and shown to match the gravity result in \cite{Gang:2019uay,Benini:2019dyp}. We are, nevertheless, quite confident in the numerical results presented here and in previous works  \cite{Liu:2017vll,Liu:2018bac,PandoZayas:2019hdb}.

 In this section we compute the one-loop logarithmic correction from the gravity side and confront them with the field-theoretic (UV) results.  Let us start by recalling a number of important facts regarding the one-loop effective actions of supergravity backgrounds. Our setup is 11d supergravity where we assume there is an embedding of the solutions describing magnetically charged asymptotically AdS$_4\times M^7$ black holes. 

We make the assumption that the whole contribution to the one-loop effective action comes from the asymptotic AdS$_4$ region as was the case  in \cite{Bhattacharyya:2012ye} for the AdS$_4$ solution and in \cite{Liu:2017vbl}, for the magnetically charged asymptotically AdS$_4$ black hole case and for black holes described by  M5 branes wrapping hyperbolic 3-manifolds in \cite{Gang:2019uay,Benini:2019dyp}.

On very general grounds of diffeomorphism invariance, it can be argued that in  odd-dimensional spacetimes, the top Seeley-De Witt coefficient $a_{d/2}$ vanishes \cite{Vassilevich:2003xt}. Therefore, the only contribution to the heat kernel comes from the zero modes. Applied to our case, the one-loop contribution due to 11d supergravity  comes from the analysis of zero modes.  As in previous cases  \cite{Bhattacharyya:2012ye,Liu:2017vbl,Gang:2019uay, Benini:2019dyp},   the  gravity computation  performed in 11d sugra is essentially reduced to the contribution of a two-form zero mode in the asymptotically AdS$_4\times M^7$ region.

More explicitly, given that there is a two-form zero mode in AdS$_4$ we need to make sure that there are possible zero modes in $M^7$ that could contribute. In the spectrum of quantum eleven-dimensional supergravity we can have contribution coming from one-form zero modes (ghost), two-form zero modes (ghost) and thre-form zero modes ($C_3$). Other than the two-form zero modes discussed already in \cite{Liu:2017vbl}, there is another potential source of zero modes which could arise if $M^7$
admits a harmonic one-form. This one-form zero mode could contribute to the ghost one-form or it could contribute to the harmonic three-form on AdS$_4 \times M^7$ by
taking the wedge product of a harmonic two-form on AdS$_4$ times a harmonic one-form on
$M^7$.  It is worth pointing that, given the magnetic charges, the space is not really a direct product but there is a fibering of $M^7$ over AdS$_4$. This fibering was studied in detail in \cite{Liu:2017vll} and shown to not affect the counting of two-form zero modes relevant in this section.

We will not reproduce all the details of the computation here, the interested reader is referred to \cite{Liu:2017vbl,Gang:2019uay,Benini:2019dyp} for details. We briefly sketch the derivation of the one-loop effective action. Given that the only zero mode in AdS$_4$ is a 2-form and assuming that the solution is asymptotically of the form   AdS$_4 \times M^7$ we need to decompose the kinetic operator along these two subspaces. For the 2-form zero mode of AdS$_4$ to survive we need to have the corresponding part of the kinetic Laplace-like operator  also vanishing.

When integrating over zero modes there is a  factor of $L^{\pm \beta_A}$ for each zero mode in the path integral. The total contribution to the partition function from the zero modes is
\bea
L^{\pm \beta_A \: n_A^0}, 
\eea
where $n_A^0$ is the number of zero modes of the kinetic operator $A$ and the sign depends on whether the operator is fermionic or bosonic. Typically, zero modes are associated with certain asymptotic symmetries. For example, with gauge transformations that do not vanish at infinity. The key idea in determining $\beta_A$ is to find the right variables of integrations and to count the powers of $L$ that such integration measure contributes.   The scaling exponent for $p$-forms is easily computed \cite{Bhattacharyya:2012ye}, yielding $\beta_p = (d-2p)/2$ in terms of the total dimension $d$ of spacetime. For the case at hand of a 2-form in eleven dimensions, we have $\beta_2=(11-4)/2=7/2$.

Having determined $\beta_2$, the computation of the one-loop effective action reduces to counting the number of 2-form zero modes, $n_2^0$.  A simple way to determine the number of 2-form zero modes is by computing the Euler characteristic of the black hole. In  \cite{Liu:2017vbl,Gang:2019uay}  it was argued that $n_2^0=2(1-g)$ for a black hole of horizon given by a genus $g$ Riemann surface. Note that this number is computed using the non-extremal branch of the solution and that it is independent of the charges of the black holes. Therefore, be it for the magnetically charged or the electrically charged black holes we obtain the same result. 

The full contribution to the logarithmic terms of the one-loop effective action is thus given only by the  2-form zero modes and we have: 
\be
\boxed{
\log Z_{1-loop}=(2-\beta_2)n_2^0\log L= (2-7/2)2(1-g)\log L = \frac{1}{2}(g-1)\log N,} \label{Eq:GravityLog}
\ee
where  according to the AdS/CFT dictionary we have used that for M2 branes backgrounds we have $L^6\sim N$. When restricting to spherically symmetric horizons $(g=0)$  we find perfect agreement with the numerical field theory results in previous sections. The topologically twisted index result  in the previous sections assume $g=0$ but it is easily generalized to arbitrary $g$ and the agreement with Eq.~(\ref{Eq:GravityLog}) remains robust. 

There is a generalization of the above result, obtained in \cite{Gang:2019uay} and \cite{Benini:2019dyp}, for the case where $M^7$ has non-vanishing first Betti number, $b_1$. The generalization takes the form 
\be
	\log Z_{1-loop}=\frac{1}{2}(g-1)(1-b_1)\log N, \label{gravity log N}
\ee
and was shown to match the field theory result for certain $M^7$ constructed as 4-sphere fibration over a hyperbolic 3-manifold \cite{Benini:2019dyp}. The extra contribution proportional to $b_1$ arises from  the supergravity 3-form potential as one can construct a zero mode by combining the 2-form zero mode in AdS$_4$ and a 1-form zero mode in the hyperbolic 3-manifold. 

For complete agreement between gravity and field theory, we need to show the vanishing of the first Betti number for the $M^7$ we considered in this manuscript. This can be shown as follows.  Every  seven-dimensional,  compact Einstein manifold of positive curvature has vanishing first Betti number  (see, for example, \cite{Duff:1986hr}).  This can be seen from the Hodge-de Rham operator acting on one-forms:
\be
\Delta_1 Y_m =\square Y_m+ R_m{}^n Y_n.
\end{equation}
Recall that the Hodge-de Rham operator is defined as
\be
\Delta =d \delta +\delta d, 
\ee
where $d$ is the exterior differentiation mapping $p$-forms to $(p+1)$-forms  and $\delta=(-1)^p \ast d \ast$ is its adjoint where $\ast$ is the Hodge dual operation.  Let us assume that the  Einstein manifold $M^7$ has natural normalization, $R_{mn}=6m^2 g_{mn}$. Considering the eigenvalues 
\be
\Delta_1 V_m=\lambda V_m, 
\ee
it follows immediately that  $\Delta_1 \ge 6m^2$. For one-forms that are co-coclosed $\nabla^m V_m=0$ one can prove an even stronger bound.  Therefore, for the class of Sasaki-Einstein seven-manifolds relevant for our analysis we have vanishing first Betti number and, subsequently perfect agreement of the logarithmic term in Eq.~(\ref{Eq:GravityLog}) with the field theory results in the previous sections. 

Let us finish this section with one important remark. The analysis performed in this section relied only on the asymptotic form of the black hole background. The explicit construction of such black hole backgrounds is,  however, a highly nontrivial problem. In the case of $S^7$ many results exists in the literature for very general black holes. The case of $Q^{1,1,1}$  has been widely discussed with relatively modest results about the near-horizon region presented in \cite{Donos:2008ug,Halmagyi:2013sla,Azzurli:2017kxo,Hong:2019wyi}.

%%%%%%%%%%%%%%%%%%%%%%%%%%%%%%%%%%%%%%%%%%%%%%%%%%%%%%%
\section{Conclusions}\label{Sec:Conclusions}
%%%%%%%%%%

In this manuscript we have numerically studied the topologically twisted index of various Chern-Simons matter quiver gauge theories on the product of a genus $g$ Riemann surface and the circle, $\Sigma_g\times S^1$ and determined that, in all cases, there is a logarithmic contribution of the from $\frac{g-1}{2} \log N$. We are able to explicitly track the contributions to the logarithmic terms coming from different elements of the index including the precise cancellation of $N\log N$ contributions between the vector multiplet and the Jacobian contribution to the topologically twisted index.  We have also provided the dual computation of one-loop quantum supergravity which perfectly matches the field theory result. This gravity computation is quite universal and requires a mild cohomological  property (vanishing of first Betti number, $b_1=0$) on the dual seven-dimensional manifold $M^7$ which is satisfied for most of the examples discussed in this manuscript.

The universality of our result for the topologically twisted index of 3d theories was inspired by the universality of the free energy on $S^3$ discussed in  \cite{Marino:2011eh,Bhattacharyya:2012ye}. This universality also  interestingly resonates with a recent analogous study in four dimensions  which analytically showed that there is a universal logarithmic contribution to the superconformal index of a large class of 4d ${\cal N}=1$ supersymmetric field theories \cite{GonzalezLezcano:2020yeb}. Perhaps similar universal results exist in other dimensions.

We expect that our supergravity analysis  extends to rotating electrically charged asymptotically AdS$_4\times M^7$ black holes beause  the result is independent of the black hole charges and depends only on the dictionary entry relating Newton's contanst, $G_N$, to the rank of the gauge group, $N$, and  the horizon topology. For the case of theories obtained from M5 branes wrapping three-dimensional hyperbolic manifolds, the logarithmic counting for magnetically charged black holes was presented in \cite{Gang:2019uay}; the case of rotating, electrically charged black holes was analyzed in \cite{Benini:2019dyp}. In both cases the logarithmic term in the field theory side was known analytically and the supergravity analysis was essentially the same and the result was independent of the black hole  charges. Indeed, it is clear that the logarithmic computation as presented here and in previous works is independent of the charges. Thus, {\it we claim that our analysis here is also valid for all asymptotically AdS$_4\times M^7$ black holes whether magnetically charged or rotating, electrically charged ones. } It would be interesting to directly verify this claim by analyzing the logarithmic term in the superconformal index  of these theories.

It would be interesting to understand our results from a more analytic point of view. A natural starting point could be  by pursuing the relation between the Bethe Potential ${\mathcal V}$ and the expectation value of the free energy on $S^3$ as pointed out in \cite{Hosseini:2016tor} but beyond the leading order. There are other more formal arguments establishing a relation between the topologically twisted index in $S^2\times S^1$ and the free energy on $S^3$  pointed out in \cite{Closset:2017zgf}.  Namely, the leading in $N$ relations between the free energy on $S^3$ and the topologically twisted index has been well documented \cite{Hosseini:2016tor,Hosseini:2016ume} by explicit computations. Quite remarkably, certain universality of the logarithmic terms in the free energy on $S^3$ of a large class of Chern-Simons matter theories was  established in  \cite{Marino:2011eh}, that is, a universal contribution of the form $-\frac{1}{4}\log N$;  the dual supergravity side was elucidated  in \cite{Bhattacharyya:2012ye} and found to be in perfect agreement.  Our result in this manuscript  --  the universality of $-\frac{1}{2}\log N$, is mostly numerical. It would be interesting to develop a matrix model intuition into some of the crucial subleading in $N$ relations between the free energy on $S^3$ and the topologically twisted index on $\Sigma_g\times S^1$ for this  large class of field theories. It will also be quite natural to include aspects of the superconformal index as presented in \cite{Choi:2019zpz,Nian:2019pxj,Choi:2019dfu} in this universality analysis. We hope to report on these efforts.

We have studied various theories that have M-theory duals. It would be interesting to extend our result to field theories admitting  massive IIA duals  where the growth of the microstates goes as $N^{5/3}$. On the field theory side one focuses on the  topologically twisted index of $SU(N)$  Chern-Simons matter theory at level $k$ whose leading term, of order  $N^{5/3}$,  coincides with the entropy of magnetically charged, asymptotically AdS$_4\times S^6$ black holes  in massive type IIA theory \cite{Benini:2017oxt,Hosseini:2017fjo}.   The black holes in question were presented in \cite{Guarino:2017eag} as a payoff of the arduous work of obtaining AdS$_4$ gauged supergravity from the reduction of massive type IIA theory \cite{Guarino:2015jca,Guarino:2015qaa,Guarino:2015vca,Varela:2015uca}. The log term in this Chern-Simons matter  theory was computed in \cite{Liu:2018bac} using a combination of analytical and numerical techniques, it would be interesting to extend those results to a larger class of theories where a similar universality might be established. The gravity computation of the logarithmic contribution, it merits to say, is quite more complicated due to the dual theory living in an even-dimensional space  leading to a more general type of contributions to the logarithmic term.

Another potentially fruitful avenue would be to  explore the 't Hooft  limit where $N\to \infty$ with $\lambda=N/k$ kept fixed. To the best of our knowledge, there are no results about this limit for the topologically twisted index other than the analysis of  \cite{PandoZayas:2019hdb}. Even for the free energy on $S^3$ we are not aware of systematic numerical explorations beyond the large $N$ leading term. It is worth noticing that in this limit one expects a re-arrangement of the degrees of freedoms as guided by the scaling of the free energy. On the gravity side, subleading corrections are also quite different as the one-loop quantum supergravity computations now depend on more dynamical aspects of the background given that the dual gravity leaves in ten-dimensional type IIA supergravity.

We have not addressed in any detail the subleading $N^{1/2}$ behavior which corresponds to higher curvature corrections  on the gravity side. For the case of the ABJM theory, the $N^{1/2}$ was determine in a combination of numerical and analytical approaches in \cite{Liu:2017vll}.   A number of interesting bottom-up observations regarding the structure of higher curvature corrections in similar classes of theories were made recently in \cite{Bobev:2020egg} and it would be interesting to pursue this entry in the AdS/CFT dictionary more precisely in this context. We hope to report on some explorations along these lines. 

Finally, there is a glaring open challenge to the supergravity community - {\it the problem of missing black holes}. There are some approaches that allow one to determine the entropy of the supergravity dual black holes to certain quiver Chern-Simons matter theory (see, for example,  \cite{Hosseini:2019ddy,Gauntlett:2019roi,Kim:2019umc}). Some progress has also been reported in \cite{Azzurli:2017kxo,Bobev:2017uzs,Bobev:2020pjk}. Our discussion in section \ref{Sec:Sugra} assumes the existence of such black holes and demonstrate that the logarithmic corrections to the entropy precisely matches the field theory results using general aspects of the would-be black hole solution. All these impressive  tests are performed in the backdrop where the explicit construction of the black holes is lacking. It remains a very interesting question to explicitly find those black holes and compute their Bekenstein-Hawking entropy and demonstrate that it agrees with the microscopic prediction of the topologically twisted index.

\section*{Acknowledgments}
We are thankful to Francesco Benini,   Chandramouli Chowdhury,  Marina David,  Dongmin Gang, Jewel Ghosh, Alfredo Gonz\'alez Lezcano, Junho Hong,  Sayed M. Hosseini, Albrecht Klemm,  James T. Liu, Jun Nian, Vimal Rathee, Ashoke Sen,   Wenli Zhao and Shan Zhou.  This work was supported in part by the U.S. Department of Energy under grant DE-SC0007859.

\bibliographystyle{JHEP}
\bibliography{BHLocalization}

\end{document}